\documentclass[lettersize,journal]{IEEEtran}

\AtBeginDocument{%
  \providecommand\BibTeX{{%
    \normalfont B\kern-0.5em{\scshape i\kern-0.25em b}\kern-0.8em\TeX}}}

\usepackage{amsmath,amsfonts}
\usepackage{algorithmic}
\usepackage{array}
\usepackage{textcomp}
\usepackage{stfloats}
\usepackage{url}
\usepackage{verbatim}
\usepackage{tabularx}
\usepackage{multirow}
\usepackage{tikz}
\usetikzlibrary{arrows,shapes,positioning,shadows,trees}
\graphicspath{ {./images/} }
\usepackage{amssymb}
\usepackage{subcaption}
\usepackage{censor}
\usepackage{enumitem}
\usepackage{adjustbox}
\usepackage{ragged2e}
\usepackage{scalerel}
\usepackage{graphicx}

\def\BibTeX{{\rm B\kern-.05em{\sc i\kern-.025em b}\kern-.08em
    T\kern-.1667em\lower.7ex\hbox{E}\kern-.125emX}}
\usepackage{balance}
\begin{document}
\title{Sanitization of Multimedia Content: A Survey of Techniques, Attacks, and Future Directions}
\author{
    \IEEEauthorblockN{Andrea Ciccotelli\IEEEauthorrefmark{1},
                      Hanaa Abbas\IEEEauthorrefmark{2},
                      Roberto Di Pietro\IEEEauthorrefmark{1}}
    \IEEEauthorblockA{\\ \IEEEauthorrefmark{1}
                      King Abdullah University of Science and Technology, Thuwal, Saudi Arabia \\
                      \IEEEauthorrefmark{2}
                      Hamad Bin Khalifa University, Doha, Qatar \\
                      Email: andrea.ciccotelli@kaust.edu.sa, hanaa.abbas.research@gmail.com, roberto.dipietro@kaust.edu.sa}
}

\markboth{Preprint}%
{Sanitization of Multimedia Content: A Survey of Techniques, Attacks, and Future Directions}

\maketitle

\begin{abstract}

The exploding rate of data publishing in our networked society has magnified the risk of sensitive information leakage and misuse, pushing the need to secure multimedia content from unintended exposure to potentially untrusted third parties. Data sanitization---the process of securing multimedia by removing or obfuscating sensitive information such as personally identifiable or confidential data---helps to mitigate the severe impact of security risks and privacy violations related to the published data.

 In this paper, we make several contributions. First, we classify data sanitization methods along two main dimensions: the media type (images, audio, text, and video) and the techniques used to sanitize sensitive regions, which we group into obfuscation-based (e.g., distortion, replacement) and removal-based approaches. Building on this categorization, we present a comprehensive review of technologies designed to protect multimedia content. We then broaden the scope by introducing the attacks that specifically target these technologies, followed by a discussion of potential countermeasures. 
 Each aspect is complemented with critical discussions and lessons learned. Finally, we identify and elaborate on open research challenges in the crucial domain of multimodal multimedia sanitization.\\*
We argue that the systematization provided in this work—together with the highlighted challenges and research directions—offers a valuable blueprint for practitioners, industry, and academia alike, while paving the way for novel research avenues in the field.\\*

\end{abstract}


\begingroup
\renewcommand\thefootnote{}
\footnotetext{This work has been submitted to the IEEE for possible publication. Copyright may be transferred without notice, after which this version may no longer be accessible.}
\addtocounter{footnote}{0}
\endgroup
\section{Introduction}
Over the recent years, the volume of data generated, shared, and stored by healthcare applications, social media platforms, sensor and IoT networks, and video surveillance systems, to cite a few, has drastically increased. However, the exploding amount of available data has increased the risk of privacy breaches. This risk is exacerbated by the availability of affordable hand-held electronic devices (e.g., smartphones, cameras) and high-speed 5G networks, that have made it much easier for the average user to create locally and share globally multimedia files. Such files are often uploaded to untrusted third parties or cloud providers for storage and further processing. Therefore, the need for privacy-preserving technologies is more critical now than ever.

Privacy is a serious concern in our networked societies, and thus several data protection laws and international standards were enacted worldwide. Similarly, most social media networks enforce policies to combat the dire consequences of privacy violations on users, such as social impact (e.g., damaged reputation, identity theft), financial losses, and even physical harm. For over two decades, considerable research was directed towards privacy-enhancing approaches for digital media. The proposed solutions can be categorized into two branches \cite{padilla2015visual}: (a) access restriction; and, (b) controlling media content by modifying the sensitive parts (i.e., multimedia data sanitization). 
In this survey, we consider only the second approach. That is, we do not include works based on hardware-oriented solutions (including intervention and blind vision \cite{sah2017video}), access control (e.g., user authentication and roles management), secure processing (e.g., secure multiparty computation and naïve encryption), and data hiding (e.g., steganography and digital watermarking). 

Data sanitization includes the obfuscation or removal of sensitive information such as personally identifiable information (PII), protected health information (PHI), and payment card information (PCI) to preserve the user's privacy. In some use-cases, some information leakage is needed in the obfuscated media content to maintain its usability and intelligibility; hence, a degree of privacy loss is always inevitable. Throughout the paper, this concept is highlighted as the \textit{privacy-utility trade-off}. Nowadays, research efforts are focused on attaining provable security guarantees that precisely characterize the degree of information leakage about the sanitized multimedia. In the past, the sanitization process was performed manually by trained experts. However, in recent years, with the advent of Machine Learning (ML)-based techniques, there is a widespread trend to semi or fully automate the cumbersome and costly process. 
However,  sanitization is not the only process being automated; attacks against it are also automated using state-of-the-art techniques, such as neural networks, Natural Language Processing (NLP), face recognition \cite{zhao2003face}, etc.. Consequently, in this survey, we additionally review the countermeasures devised to thwart such automated attacks. 

\begin{figure}[ht]
\centering
\includegraphics[scale=1.2]{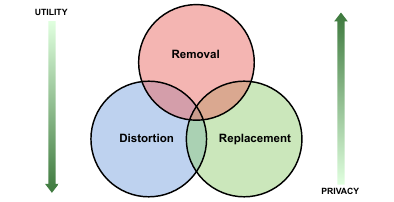}
\caption{The Utility-Privacy Trade-off of Obfuscation Methods}
\label{fig:methods}
\end{figure}

Data sanitization technologies are a double-edged sword. Indeed, we note that many of the to-be-discussed obfuscation techniques can also have malicious use cases. For example, adversaries may utilize obfuscation methods to evade being recognized by biometric verification and video surveillance systems. However, since obfuscation methods should withstand conventional recognition and identification algorithms that the adversary may use, we do not differentiate obfuscation based on its intent. 
Typically, in data sanitization, the region hosting sensitive information is either distorted, replaced, or removed. On the one hand, distortion modifies the original object to render it unintelligible, while replacement substitutes the element with a visually similar one. Object removal ensures complete protection; however, the utility is severely compromised. On the other hand, distortion and replacement methods may leak information about the obfuscated content while achieving higher utility than removal. 
Fig. \ref{fig:methods} depicts the privacy-utility trade-off between the obfuscation categories. The figure also illustrates the possibility of combining the different methods, a technique called  \textit{multimodal obfuscation}. 
As will be shown in the sequel, does not exists a single best approach to sanitization; each redaction solution comes with its benefits and drawbacks. 
Therefore, one of the objectives of this contribution is to help in devising the optimal combinations of techniques to adopt to reach a certain degree of privacy, while maintaining an acceptable value of the utility function.
 
\begin{table*}[b]
\caption{Comparison of Main Surveys on Multimedia Security and Privacy}
\label{tab:survey_comparison}
\centering
\resizebox{0.75\linewidth}{!}{%
\begin{tabular}{|c|c|c|c|c|c|c|} 
\hline
\multirow{2}{*}{\begin{tabular}[c]{@{}c@{}}\textbf{Survey}\\\textbf{\& Year}\end{tabular}} & \multicolumn{4}{c|}{\textbf{Digital Media}} & \multirow{2}{*}{\begin{tabular}[c]{@{}c@{}c@{}}\textbf{Attacks \&}\\\textbf{Solutions}\end{tabular}} & \multirow{2}{*}{\begin{tabular}[c]{@{}c@{}}\textbf{System}\\\textbf{Model}\end{tabular}} \\
\cline{2-5} & \textbf{Images} & \textbf{Audio} & \textbf{Video} & \textbf{Document} &  &  \\ 
\hline
2005 \cite{lopresti2005information} &  &  &  & \checkmark & \checkmark &   \\
\hline
2014 \cite{winkler2014security}& \checkmark &  &  &  &  &   \\
\hline
2014 \cite{vasudevan2014review}&  &  &  & \checkmark &  &   \\
\hline
2015 \cite{padilla2015visual}& \checkmark &  & \checkmark &  & &  \\ 
\hline
2016 \cite{ribaric2016identification}& \checkmark & \checkmark & \checkmark &  &  &  \\ 
\hline
2017 \cite{sah2017video}& \checkmark &  & \checkmark &  &  & \\ 
\hline
2018 \cite{rakhmawati2018image}& \checkmark &  &  &  &  &  \\
\hline
2020 \cite{liu2020privacy}& \checkmark &  &  &  &  &  \\ 
\hline
2023 \cite{zhao2025visual} & \checkmark &  & \checkmark &  & \checkmark &  \\
\hline
2023 \cite{wen2024image} & \checkmark &  &  &  & \checkmark &  \\
\hline
2023 \cite{cao2024face} & \checkmark &  & \checkmark &  & \checkmark &  \\
\hline
2024 \cite{chen2025survey} & \checkmark &  &  &  &  &  \\
\hline
2024 \cite{sun2025ensuring} & \checkmark &  &  &  & \checkmark &  \\
\hline
\textbf{This survey} & \checkmark & \checkmark & \checkmark & \checkmark & \checkmark &  \checkmark  \\
\hline

\end{tabular}}
\end{table*} 
Over the past years, just few attempts were made to survey the literature on sensitive information protection using obfuscation methods. In one of the first surveys in the area, Padilla-López et al. (2015) summarized hardware and software solutions for image and video redaction in the context of privacy-aware intelligent monitoring systems \cite{padilla2015visual}. The authors classified protection methods into five large categories: intervention, blind vision, secure processing, redaction, and data hiding. Intervention and blind vision are mainly hardware-based solutions. In contrast, in our survey, we limit our focus to methods concerning direct modification of the multimedia’s regions of interest (ROI) to obscure information (i.e., redaction). Another study by Vasudevan et al. (2014) \cite{vasudevan2014review} presented an extremely brief review of text sanitization. They discussed the various techniques used to identify and mask the sensitive text. Ribaric et al. (2016) \cite{ribaric2016identification} presented de-identification of non-biometric, physiological, and behavioral biometric identifiers and soft-biometric identifiers found in images, video, and audio. In \cite{sah2017video}, Sah et al. (2017) focused on video redaction, including the problem of object detection and tracking. 
Besides the previously mentioned surveys, there are other notable works in the literature. In \cite{lopresti2005information}, Lopresti et al. (2005) summarized the process of (traditional) document redaction, highlighting the circumstances enabling information leakage, and the detection process of the residual artifacts. In \cite{rakhmawati2018image}, Rakhmawati et al. (2018) presented a very brief survey about the techniques of image privacy protection in a five-page conference paper. Their paper describes the characteristics, evaluation methods, and privacy protection methods that apply to images. The surveys in \cite{winkler2014security} and \cite{liu2020privacy} address two specific applications of image privacy protection in visual sensor networks and image sharing on online social networks, respectively. 
Additional surveys further broaden the scope: Zhao et al. (2025) propose an adversary-centric framework for visual content privacy protection, distinguishing human-vision, computer-vision, and hybrid adversaries and reviewing representative solutions \cite{zhao2025visual}. An image privacy protection survey introduces a three-layer taxonomy (data-, content-, and feature-level) based on privacy-sensitive domains, offering a holistic framework for comparing techniques \cite{wen2024image}. Face-focused surveys analyze de-identification methods in depth: one compares pixel-, representation-, and semantic-level approaches and their empirical performance \cite{cao2024face}, while another concentrates on cloud-based services, grouping methods into image obfuscation-based and adversarial perturbation-based categories and discussing deployment considerations \cite{chen2025survey}. Broader privacy perspectives include a survey on generative data that organizes countermeasures around privacy, controllability, authenticity, and compliance dimensions \cite{sun2025ensuring}. Our survey provides a comprehensive view that combines the sanitization of several media types and addresses attacks and countermeasures concerns in one paper. In comparison to the cited surveys, we include a more detailed discussion of the observed current research trends and possible future research directions. We also provide a general discussion in each multimedia sanitization section reviewing the provided methods, reference applications, key properties, and challenges.

Table \ref{tab:survey_comparison} compares the main existing surveys in the literature (including our contribution) in terms of the addressed media type and whether attacks and countermeasures are explored. As it can be seen from the table, other surveys go back in time and are not as comprehensive as ours. Non-survey papers that are frequently cited are not listed in the table since they lack generality.

\textbf{Contributions.} In this paper, we provide a thorough systematization of sanitization techniques for images, audio, text, and video. We revise the attacks that these techniques are subject to, analyze current countermeasures, and discuss further research directions. Specifically, our main contributions are summarized as follows:
\begin{itemize}
\item Categorization of the sanitization methods according to their underlying mechanism (distortion, replacement, or removal) while highlighting the properties expected to be preserved after sanitization.
\item Summary of the most recent and relevant contributions in the literature about the sanitization techniques for each media type and their known attacks and possible countermeasures.
\item Provisioning of an in-depth discussion for each media type exploring topics such as sanitization challenges, current trends, and future research directions.
\end{itemize}

\begin{figure*}[t]
\centering
\includegraphics[width=\textwidth]{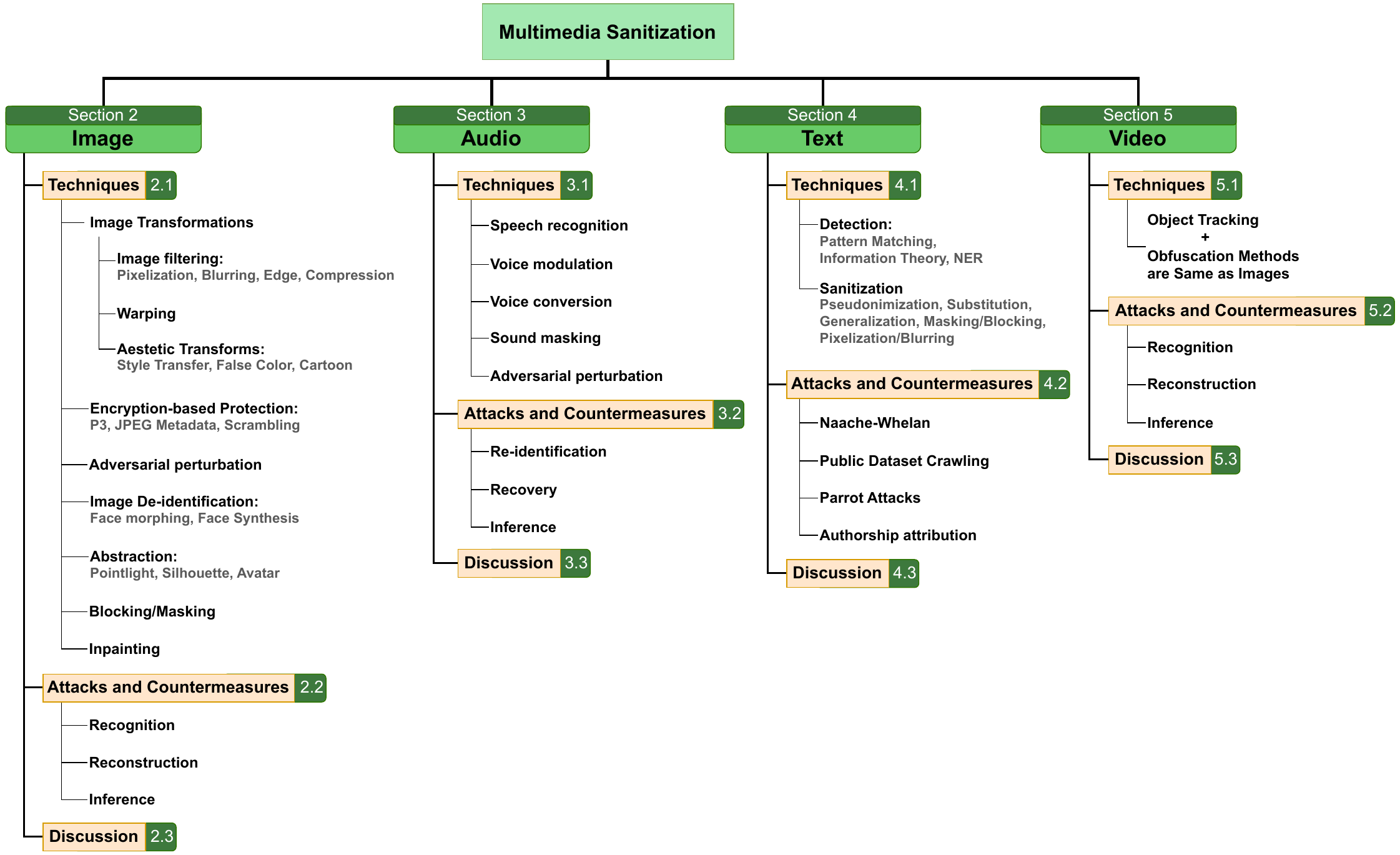}
\caption{Overview of the surveyed domains, techniques, attacks, and provided discussion.}
\label{fig:paper_overview}
\end{figure*}

\textbf{Roadmap.} The remainder of the paper (see paper overview in Fig. \ref{fig:paper_overview}) is organized as follows. We detail the sanitization techniques used for images (Section \ref{image_obfuscation}), audio (Section \ref{audio_obfuscation}), text (Section \ref{document_obfuscation}), and video (Section \ref{video_obfuscation}), where each section also provides a comprehensive discussion about the described techniques in terms of their key properties, the attacks they are subjected to, and possible research directions. 
Section \ref{sec:lessons}
 summarizes the key lessons learned from our exploration of this domain.
  Final conclusions are drawn in Section \ref{conclusion}.  

\section{Image Sanitization}
\label{image_obfuscation}
The number of images captured daily is rapidly increasing due to the availability and affordability of high-quality cameras embedded in personal devices and surveillance systems. A vast percentage of these images are uploaded to online social networks or cloud-based platforms for processing and storage. To put numbers into perspective, in November 2020, Google Photos announced that it had stored more than 4 trillion images \footnote{Google Blog: https://blog.google/products/photos/storage-changes/}. Every day, approximately 300 million photos are uploaded to Facebook and 95 million to Instagram \footnote{https://dustinstout.com/social-media-statistics/}. Thus, the critical regions in the images must be identified then removed or replaced to minimize the risk of disclosing sensitive or personal information.

\begin{figure*}[t]
\centering
\includegraphics[width=\textwidth]{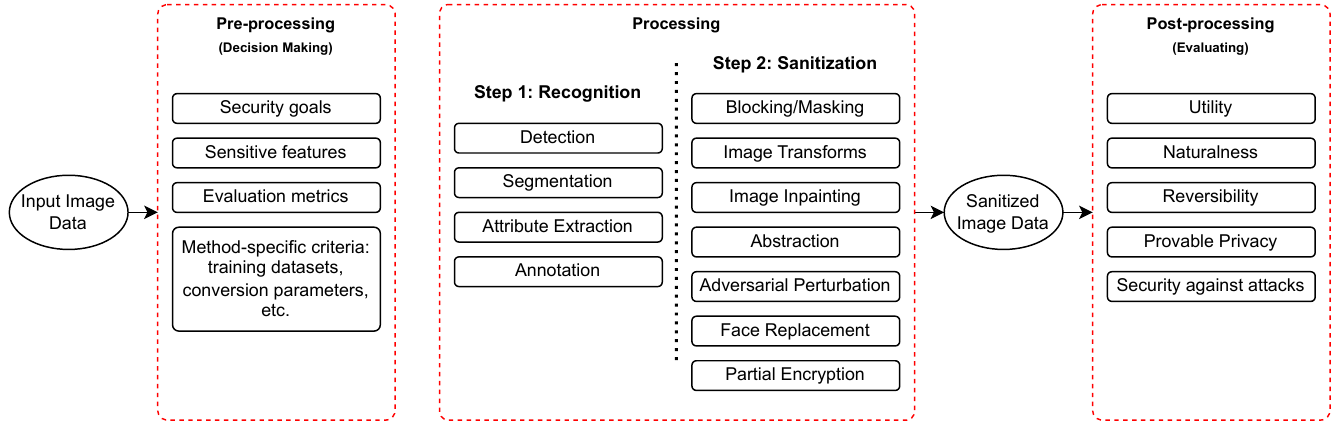}
\caption{Major Steps and Components of an Image Sanitization System}
\label{fig:image_sanitization}
\end{figure*}

An automated sanitization system typically involves multiple steps. The desired security level must be clearly defined in the pre-processing stage, including which sensitive objects must be identified and protected. The second stage involves two further steps: the identification step and the sanitization step. In the identification step, the region of interest (ROI) is detected using well-established \textit{image recognition} techniques (e.g., face recognition). The boundaries of the ROI are then localized (i.e., \textit{image segmentation}) and annotated. An obfuscated image is produced by applying an obfuscation algorithm on the annotated ROI. In the post-processing stage, the output image is evaluated for its security based on the pre-defined security goals, i.e., whether it should withstand human recognition, machine recognition, or both. Fig. \ref{fig:image_sanitization} depicts the main integral components and steps of an image sanitization system.

The authors in \cite{ribaric2016identification} categorized sensitive image attributes as biometric, non-biometric, and soft-biometric. This paper follows the same taxonomy; however, we extend it to include confidential and censored categories to incorporate attributes that are not personal but must be secured. The types and examples of sensitive content in images are listed in Table \ref{tab:sensitive_content_in_obfuscated_images}. Most images have multiple types of objects that need to be identified and protected, known as the \textit{multimodal identification} problem. For instance, even if the face is obfuscated in an image, soft biometrics such as body silhouette, clothing style, and skin color can be strong leads to a person's identity. Even after the obfuscation process, some residual visual features will remain. They are either left intentionally to maintain the utility and intelligibility of the image or missed unintentionally due to detection errors. This is repeatedly referred to as the \textit{privacy-utility trade-off} in the literature. According to the application scenario, the obfuscated image may be required to retain specific characteristics: (a) reversibility, (b) utility, and (c) naturalness. \textit{Reversibility} is a property achieved by the provision of additional translational variables. These variables are expected to remain secret or partially unknown to an attacker; otherwise, the obfuscation security is compromised. The \textit{utility} and \textit{naturalness} of images are highly desirable when images are intended for public use or model training for automatic recognition and classification algorithms. Also, maintaining a natural appearance helps conceal the fact that some identifiers were obfuscated, which can be a security goal on its own. Besides the usability and intelligibility trade-off, another common challenge is providing provable and quantifiable privacy or security guarantees. This section explores the different obfuscation techniques and the possible attacks an adversary may perform to infer the hidden content. Furthermore, we suggest some countermeasures and insights regarding current trends in the research and potential areas of improvement. 

\begin{table}
\caption{Types of sensitive content in images}
\label{tab:sensitive_content_in_obfuscated_images}
\centering 
\resizebox{\columnwidth}{!}{%
\begin{tabular}{|p{0.15\linewidth}|>{\hsize=\hsize\RaggedRight}p{0.4\linewidth}|>{\hsize=\hsize\RaggedRight}p{0.3\linewidth}|}
\hline
\textbf{Type} & \textbf{Description} & \textbf{Examples} \\ 
\hline
Biometric & Unique, measurable and permanent personal identifiers \cite{ribaric2016identification}. & Face, iris, ear, fingerprint \\ 
\hline
Soft biometric & Vague physical, behavioral or adhered personal characteristics that are not necessarily unique or permanent \cite{ribaric2016identification}. & Height, eye color, body shape, age, gender, skin color, tattoos, birthmarks, scars \\ 
\hline 
Non-biometric & Personal identifiers that provide non-physical and non-behavioral contextual information about the individual. Such attributes are either temporary and changeable.  & Clothing, hairstyle, location, license plate, identification cards (e.g. health card, driver's license), credit cards\\ 
\hline
Confidential & Non-personal attributes that should be withheld from public for security and privacy reasons. & QR codes, cheques, keys, classified images (e.g., high-security buildings) \\ 
\hline
Censored & Visual censorship of content due to laws and regulations. & Branded products, pornography, gruesome content \\
\hline
\end{tabular}}
\end{table}

\subsection{Techniques}
We propose the following classification of image sanitization techniques based on the primary mechanism of obfuscation (distortion, replacement) or removal of image features. The list of seven categories is presented in Table \ref{tab:image_techniques_overall}. The categories are \textit{image transformation}, \textit{encryption-based protection}, \textit{adversarial perturbation}, \textit{image de-identification}, \textit{abstraction}, \textit{masking} and \textit{blocking}, and \textit{inpainting} methods. In the following subsections, we provide explanations and further discussions about each technique. In Table \ref{tab:image_techniques_overall}, we rate the main obfuscation techniques based on the protection level against human and machine attackers separately, the utility of the obfuscated image, and the ability to reverse the obfuscation. A low utility rating indicates severe quality degradation and an absence of naturalness and intelligibility. A medium rating indicates a general lack of naturalness or intelligibility. A high rating indicates that naturalness, usability, and quality are preserved. In terms of security, a low rating indicates that the obfuscated image is easily detectable and recognizable either with the naked eye or using image detection and recognition algorithms. A medium rating indicates the existence of known attacks that can leak information; however, the attacks do not have high success guarantees. A high rating indicates that the obfuscated image has a low risk of information disclosure. 

\begin{table*}[ht]
\centering
\caption{Main Image Protection Techniques in the Literature (L: Low, M: Medium, H:High)}
\label{tab:image_techniques_overall}
\resizebox{\linewidth}{!}{
\begin{tabular}{|l|l|l|l|c|c|c|c|} 
\hline
\multirow{2}{*}{\textbf{Category}} & \multirow{2}{*}{\textbf{Main Techniques}} & \multirow{2}{*}{\textbf{Subtype}} & \multirow{2}{*}{\textbf{\textbf{Sub-Subtype}}} & \multicolumn{2}{c|}{\textbf{Security Level}} & \multirow{2}{*}{\textbf{Utility}} & \multicolumn{1}{l|}{\multirow{2}{*}{\textbf{Reversibility}}} \\ 
\cline{5-6}
 &  &  &  & \textbf{Human} & \textbf{Machine} &  & \multicolumn{1}{l|}{} \\ 
\hline
\multirow{11}{*}{Distortion} & \multirow{8}{*}{Image Transformations} & \multirow{4}{*}{Image filtering} & Pixelization & \multirow{4}{*}{L} & \multirow{4}{*}{L} & \multirow{4}{*}{L} & \multirow{4}{*}{\checkmark (lossy)} \\ 
\cline{4-4}
 &  &  & Blurring &  &  &  &  \\ 
\cline{4-4}
 &  &  & Edge &  &  &  &  \\ 
\cline{4-4}
 &  &  & Compression &  &  &  &  \\ 
\cline{3-8}
 &  & Image warping & \hfil - & L & L & L & \checkmark (lossy) \\ 
\cline{3-8}
 &  & \multirow{3}{*}{Aesthetic transforms} & Cartooning & \multirow{3}{*}{L} & \multirow{3}{*}{L} & \multirow{3}{*}{M} & \multirow{3}{*}{\checkmark (lossy)} \\ 
\cline{4-4}
 &  &  & False color &  &  &  &  \\ 
\cline{4-4}
 &  &  & Style transfer &  &  &  &  \\ 
\cline{2-8}
 & \multirow{2}{*}{Encryption-based Protection} & Partial Encryption & P3, TPE & \multirow{2}{*}{H} & \multirow{2}{*}{H} & \multirow{2}{*}{L} & \multirow{2}{*}{\checkmark (lossless)} \\ 
\cline{3-4}
 &  & Scrambling & \hfil - &  &  &  &  \\ 
\cline{2-8}
 & Adversarial Perturbation & \hfil - & \hfil - & L & M & H & $\times$ \\ 
\hline
\multirow{4}{*}{Replacement} & \multirow{3}{*}{Image de-identification} & \multirow{2}{*}{Face synthesis} & k-Same & \multirow{2}{*}{H~} & \multirow{2}{*}{H~} & \multirow{2}{*}{H~} & \multirow{2}{*}{$\times$} \\ 
\cline{4-4}
 &  &  & ML-based &  &  &  &  \\ 
\cline{3-8}
 &  & Face morphing & \hfil - & H & M & M & $\times$ \\ 
\cline{2-8}
 & Abstraction & \hfil - & \hfil - & H & H & M & $\times$ \\ 
\hline
\multirow{2}{*}{Removal} & Blocking/Masking & \hfil - & \hfil - & H & H & L & \checkmark (lossy) \\ 
\cline{2-8}
 & Inpainting & \hfil - & \hfil - & H & H & M & $\times$ \\
\hline
\end{tabular}}
\end{table*}

\subsubsection{Image Transformations}
\label{sec:image_transformations}
Image transformation is an image editing method that refers to mapping an input image to a processed output image using a function. Early attempts at redacting photos relied on image transforms to hide sensitive regions. Image transformation techniques, specifically pixelization and blurring, are typically used for censorship in television and online news \cite{rakhmawati2018image}. Here we consider \textit{image filtering}, \textit{image warping}, and \textit{aesthetic transforms} as types of image transformations used for distorting the appearance of the original image (see Table \ref{tab:image_transformations} for examples). In image filtering and image warping, the obfuscated image is a degraded version, characterized by its information and utility loss. Since the image transforms are based on a parameterized model, this category of obfuscation is highly vulnerable to ML-assisted attacks \cite{mcpherson2016defeating,tekli2019framework}. Such attacks are trained to reconstruct an approximated (i.e., lossy) original from the redacted image or deduce the image’s original using evident correlations from within a set of unobscured images. As per aesthetic transformations, they modify the input image to produce an abstract or a cartoon equivalent which may thwart weak recognition systems. Overall, these methods are ineffective against human and machine attackers.

\begin{table}[ht]
\centering
\caption{Main Image Transformation (Editing) Techniques in the Literature}
\label{tab:image_transformations}
\resizebox{\columnwidth}{!}{%
\setlength{\tabcolsep}{0.5em} 
{\renewcommand{\arraystretch}{3.5}
\begin{tabular}{|>{\hsize=\hsize\RaggedRight}p{0.3\linewidth}|l|c|>{\hsize=\hsize\RaggedRight}p{0.15\linewidth}|} 
\hline
\textbf{Category} & \textbf{Technique} & \textbf{Example} & \textbf{Reference \& Year} \\ [0.5ex]
\hline
\multirow{3}{*}{Image filtering} & Pixelization &
\adjustimage{width=1cm,height=1cm,valign=m}{tools/20px.png}
& 2018 \cite{fan2018image}, 2019 \cite{fan2019practical} \\
\cline{2-4}
& Blurring & 
\adjustimage{width=1cm,height=1cm,valign=m}{tools/blur_20_radius.png}
& 2016 \cite{fan2018image} \\ 
\cline{2-4} & Edge Detection & 
\adjustimage{width=1cm,height=1cm,valign=m}{tools/robertscross.png}
& 2019 \cite{hasan2019can} \\ 
\hline
Image warping & Twirl effect & 
\adjustimage{width=1cm,height=1cm,valign=m}{tools/twirl.png}
& 2013 \cite{korshunov2013using} \\ 
\hline
\multirow{2}{*}{Aesthetic Transforms} & False Color &
\adjustimage{width=1cm,height=1cm,valign=m}{tools/false_color.png}& 2017 \cite{cciftcci2017reliable} \\ 
\cline{2-4} & Cartooning & \adjustimage{width=1cm,height=1cm,valign=m}{cartoon} & 2014 \cite{erdelyi2014adaptive}, 2017 \cite{hasan2017cartooning}, 2018 \cite{larson2018pixel} \\
\hline
\end{tabular}}}
\end{table}

\textbf{Image filtering:} Image filtering has many applications in image processing, including edge detection, smoothing, sharpening, and noise reduction. The image is treated as an $M$ x $N$ matrix. Each cell in the matrix contains an integer from 0 to 255 (0 is black and 255 is white). An image can also have multiple channels such as RGB (red-green-blue), HSV (hue-saturation-value) and YUV \cite{yu2021gan}. Filtering is a mathematical operation where a matrix (aka, kernel) is convoluted with each pixel value and its neighbors within an image.

\begin{figure}[h]
\centering
\begin{subfigure}[b]{.15\textwidth}
\centering
\includegraphics[width=1cm,height=1cm]{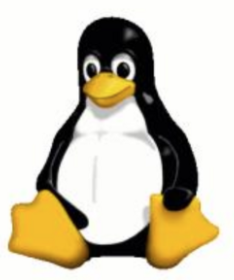} 
\caption{Original}
\end{subfigure}
\quad
\begin{subfigure}[b]{.15\textwidth}
\centering
\includegraphics[width=1cm,height=1cm]{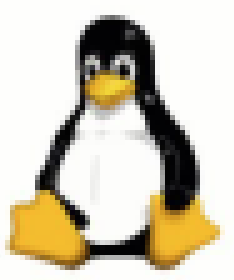}
\caption{5$\times$5}
\end{subfigure}
\quad
\begin{subfigure}[b]{.15\textwidth}
\centering
\includegraphics[width=1cm,height=1cm]{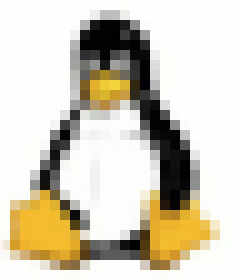}
\caption{10$\times$10}
\end{subfigure}
\quad
\begin{subfigure}[b]{.15\textwidth}
\centering
\includegraphics[width=1cm,height=1cm]{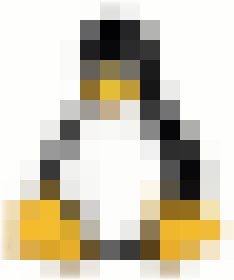}
\caption{20$\times$20}
\end{subfigure}
\quad
\begin{subfigure}[b]{.15\textwidth}
\centering
\includegraphics[width=1cm,height=1cm]{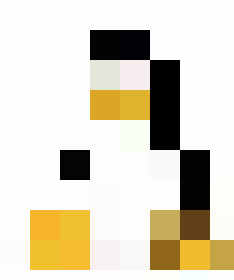}
\caption{30$\times$30}
\end{subfigure}
\caption{Pixelization: 5$\times$5, 10$\times$10, 20$\times$20, and 30$\times$30 grids.}
\label{fig:image2}
\end{figure}

\textit{Pixelization.} Pixelization (aka, mosaicing) is a widely adopted technique where the region of interest (ROI) is divided into a square grid (e.g., 8$\times$8, 16$\times$16) or a rectangular grid. The average color value of the pixels within each box is computed. Then, the value of each pixel within the box is replaced by the computed average. Pixelization can be considered a reduction in the resolution of the obfuscated region \cite{mcpherson2016defeating}. For a pixel box of size \(n\times n\), the resolution is effectively reduced by a factor of \(n^2\) \cite{mcpherson2016defeating}. Increasing the size of the pixel box increases the level of security as more pixels are averaged together. As can be seen in Fig. \ref{fig:image2}, pixelated images of smaller pixel boxes, e.g., 5$\times$5, are still human-recognizable. 

\begin{figure}[ht]
\centering
\includegraphics[scale=0.15]{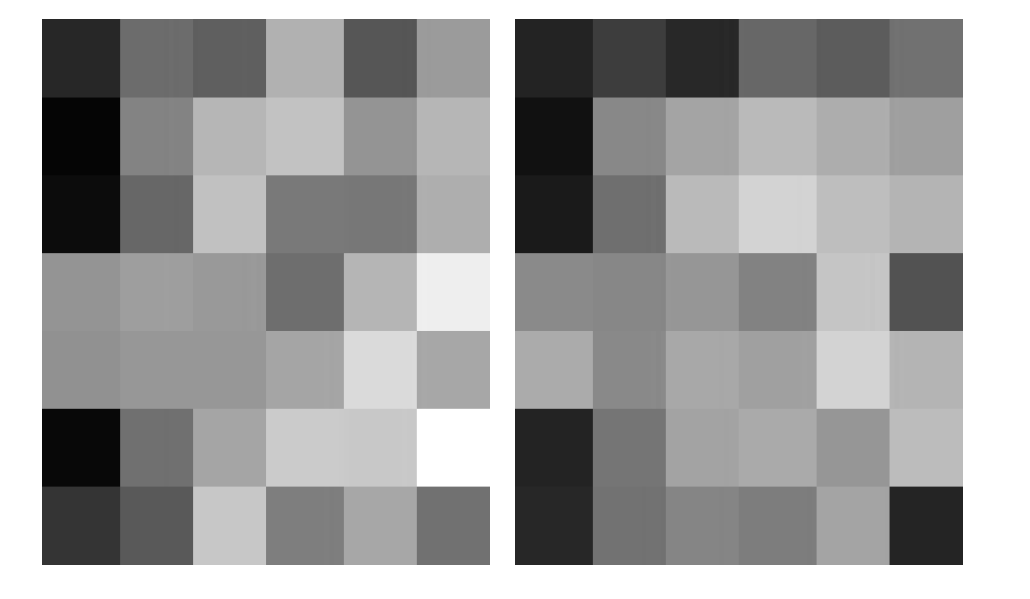}
\caption{Pixelized (left) and DP-pixelized (right) images for $16\times16$ pixel grid, $m = 16$, $\epsilon = 0.5$ \cite{fan2018image}.}
\label{fig:Image_Pix_with_DifferentialPrivacy}
\end{figure}

In \cite{fan2018image}, Liyue Fan (2018) presented an image pixelization method that extends the notion of differential privacy, providing rigorous privacy guarantees. Differential privacy \cite{10.1007/978-3-540-79228-4_1} (aka, statistical  disclosure control) is the state-of-the-art privacy paradigm for statistical database sanitization. It defines a parameter $\epsilon$ that corresponds to the degree of privacy, where the smaller the value of $\epsilon$, the stronger the privacy. The algorithm first performs pixelization on an input image and applies Laplace perturbation to the pixelized image. The Laplace perturbation uses the notion of $m$-neighborhood, which protects up to $m$ pixels. The output image would closely resemble a normally pixelized image except for $m$ grid cells.  The algorithm guarantees the indistinguishability of the output of two neighboring images having the same dimensions and differing by at most $m$ pixels. As $m$ increases, the privacy is increased, but the utility is lowered as the Laplace perturbation noise increases. Fig. \ref{fig:Image_Pix_with_DifferentialPrivacy} shows a pixelized image (left) and its differentially private (DP) alternative (right), where the pixel box size is $16\times16$, $m = 16$, and $\epsilon = 0.5$. The noise added after pixelization protects the image from machine-based recognizers and provides provable and quantifiable privacy guarantees. The author also extended the concept of differential privacy to blurring. The algorithm first adds Laplace noise to each pixel and then applies Gaussian blur to smoothen the output.

In \cite{fan2019practical}, the same author (2019) expanded on DP pixelization to yield higher utility. The output of this method exhibits more resemblance to the original image. Compared to standard pixelization, the accuracy of re-identification using CNN-based attacks is lowered from 96.25\% to 82.5\% for $\epsilon$ = 0.5 and 17.5\% for $\epsilon$ = 0.1. The author used a generalized notion of DP known as \textit{metric privacy}. Metric privacy guarantees a level of indistinguishability proportional to the distance between two inputs. Meaning, the mechanism's output is roughly the same for all visually similar images in a set, and an (informed) adversary cannot infer the exact input image by observing the obfuscated image. The obfuscation algorithm involves two major steps: transformation and random sampling. The first step transforms a sensitive ROI into a k-dimensional feature vector using Singular Value Decomposition (SVD). SVD is used as a (invertible) perceptual image hashing method that captures geometric features and characteristics of the image data. The higher the value of k, the higher the resemblance between the input and obfuscated ROI. Next, the vector is randomly sampled according to a certain probability distribution, satisfying metric privacy. The sampled vector is processed with an inverse transform to produce the obfuscated ROI. The technique was evaluated using Mean Square Error (MSE) and Structural Similarity (SSIM) metrics between the obfuscated ROI and the original ROI. SSIM is a value between 0 and 1 representing the holistic similarity between two images, where 1 indicates identical images \cite{tekli2019framework}.

\textit{Blurring.} Blurring is another image quality degradation technique. There are several types of blur filters. A \textit{Gaussian blur} is where pixels within the ROI are convoluted with a Gaussian kernel to achieve a “smoothing” effect. The blurriness level is controlled by the standard deviation $\sigma$ parameter. Another type of blurring is motion blur. The \textit{motion blur} alters the details of an image by replicating the effect of a synthetic camera’s motion. The degree of blurriness is affected by the length and the angle of the synthesized motion. \textit{Box blur}, also known as a linear box filter, is a form of a low-pass filter where each pixel is replaced with the average value of its neighboring pixels. 

\begin{figure}[h]
\centering
\begin{subfigure}[b]{0.15\textwidth}
\centering
\includegraphics[width=1cm,height=1cm]{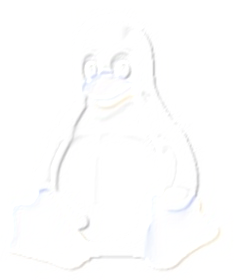} 
\caption{5}
\end{subfigure}
\quad
\begin{subfigure}[b]{0.15\textwidth}
\centering
\includegraphics[width=1cm,height=1cm]{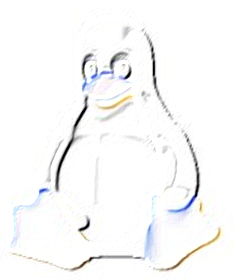}
\caption{25}
\end{subfigure}
\quad
\begin{subfigure}[b]{0.15\textwidth}
\centering
\includegraphics[width=1cm,height=1cm]{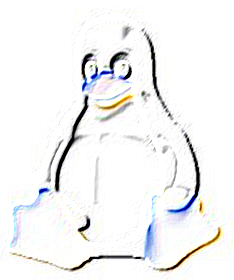}
\caption{50}
\end{subfigure}
\caption{Roberts cross edge detection with varying strengths of (a) 5; (b) 25; and, (c) 50.}
\label{fig:edge_detection}
\end{figure}

\textit{Edge Detection.} In image processing and computer vision, edge detection is a fundamental technique for feature detection and feature extraction. It includes various mathematical methods that identify the points at which the image brightness discontinues or changes sharply. Image edges preserve the shape and some internal details of the ROI \cite{hasan2018viewer}. Unlike pixelization and blurring, edge detection is not a typical obfuscation method and it is considered a form of abstraction. Nonetheless, there were a few mentions of edge detection as an obfuscation method in \cite{winkler2014security,hasan2019can,hasan2018viewer}. The level of obfuscation is controlled by an ``edge filter" parameter which affects the strength and visibility of the edges and the inner details of the image. In Fig. \ref{fig:edge_detection}, we applied Roberts cross edge detection method with increasing strength parameter: 5, 25 and 50, respectively. 

\textit{Discussion.} The methods detailed in this section are regarded as naïve and conventional methods that are no longer secure against machine detectors and that can still leak information to human observers. In \cite{mcpherson2016defeating}, pixelated images of objects belonging to 1 out of 10 classes and smaller pixel boxes, e.g., 4$\times$4, were easily recoverable using neural networks with an identification accuracy of over 70\%. With 16$\times$16 box size, the accuracy of recovery is about 31\%. In \cite{li2017effectiveness} and \cite{vishwamitra2017blur}, the authors investigated the effectiveness of pixelization and blurring against human recognition as well and explored its impact on user satisfaction. The evaluation proved that both methods are ineffective against human recognition. However, considering user satisfaction, they performed better than blocking or masking methods but still are not favorable. 

There are other quality degradation techniques that are sparsely cited as forms of visual privacy protection such as scaling, stretching, rotation, modifying saturation and brightness, grayscaling, and lossy compression \cite{poller2012robust}. However, they are no longer viable due to the robustness of image fingerprinting and image recognition methods that can easily identify matching images even after severe quality degradation. \textit{Lossy compression} (aka, \textit{high compression} or \textit{downsampling}) is an irreversible data encoding that significantly decreases the peak signal-to-noise ratio (PSNR) \cite{poller2012robust,kim2014image} of the image and produces inexact approximation or partial discarding of the image data. It is commonly used to reduce data size during storage and transmission. The most widely used lossy compression scheme is the discrete cosine transform (DCT). Existing compression standards are JPEG and JPEG-2000 for images, and M-JPEG, M-JPEG 2000, MPEG-4, or AVC/H.264 for videos \cite{kim2014image}. Similar to pixelization and blurring, the utility of the image is lowered after compression; however, it can remain easily recognizable by the end-user. Research has shown that compression is also vulnerable to super-resolution ML-trained models \cite{poyser2020impact,wang2020deep,anwar2020deep}. 

\textbf{Image warping:} Image warping is a geometric transformation of an image such that any depicted object is significantly distorted. Warping destroys the details and relationship between neighboring pixels while retaining the general shapes of the image. It is commonly used for animation and artistic purposes. In \cite{korshunov2013using}, Korshunov et al. (2013) proposed an algorithm that uses warping techniques to obfuscate faces. The algorithm works as follows: the face region is detected using face recognition algorithms. Then, a set of key pixels in the image are selected from the eyes, nose, mouth, and sides of the face. These pixels are shifted (i.e., their coordinates changed) by a random shifting distance that is decided using a pseudo-random algorithm. The next step is to estimate the transformation matrix based on the original and destination coordinates of the shifted pixels. The transformation is applied to the remaining pixels using the estimated matrix, while any gaps are interpolated using a bicubic algorithm. The discussed process is reversible if the inverse transformation matrix can be estimated \cite{cavedon2011getting}. However, due to the interpolation of some pixels, an unwarped image is only an approximation of the original. The strength of the warping effect, and thus the privacy, is based on the shifting distance. Ideally, a warped face would be detectable by face detection algorithms but unidentifiable by recognition algorithms. Security is enforced by using a secret key for seeding the pseudo-random algorithm and encrypting the selected key pixels. 

\textbf{Aesthetic Transforms:} Artistic transformations on images are mainly intended for improving the utility of obscured images. It can be beneficial for online social networks, where sensitive information can be hidden without degrading the user experience. In \cite{hasan2019can} and \cite{hasan2018viewer}, the authors conducted an objective evaluation to study aesthetic transforms as a visual obfuscation method and their impact on viewer satisfaction. The study applied cartooning and false colors to conventional obfuscation methods (pixelization, edge detection, and masking) known for their high utility loss and low visual perception.  Despite the applied beautifications, the results show that user satisfaction did not significantly increase.

\textit{False Color.} False colors are typically used as a visualization aid in image processing. They are used to depict invisible features that our eyes cannot normally see (e.g., invisible spectral emissions and sensitivities, line of sight analysis). An RGB input image is first transformed into grayscale. Then, the grayscale value is mapped to an RGB value based on a predefined color palette. The mapped RGB value then replaces the original pixel value. The primary advantages of this technique are that the intelligibility of the image is not compromised, the selection of an ROI is not required, and it can be used in conjunction with other obfuscation techniques \cite{cciftcci2017reliable}. However, false colorization is partially reversible (i.e., lossy) because: (a) the color palette may not be one-to-one; and, (b) the initial RGB to grayscale conversion is a lossy operation---the color information is not retained. {\c{C}}ift{\c{c}}i et al. (2017) \cite{cciftcci2017reliable} presented a reversible false color-based obfuscation for the JPEG standard, where the entire image is colorized, and the restorative information is encrypted and embedded in the file's metadata. The output of the system is the false-color image (FI), whose JPEG APP Markers contain the encrypted and compressed values of the difference image (DI), sign image (SI), the histograms of the original image, and the color palette. To ensure the reversibility of the scheme, instead of using direct RGB-to-gray conversion, a false-color value is computed for each color value in the original image. Since most color palettes are not one-to-one, a pseudo-inverse image is computed using the inverse table lookup. The resulting FI is then encoded to the JPEG standard. DI is obtained by subtracting FI from the pseudo-inverse of the original image. SI is included since the difference can be negative. 

\begin{figure}[ht]
\centering
\includegraphics[scale=0.2]{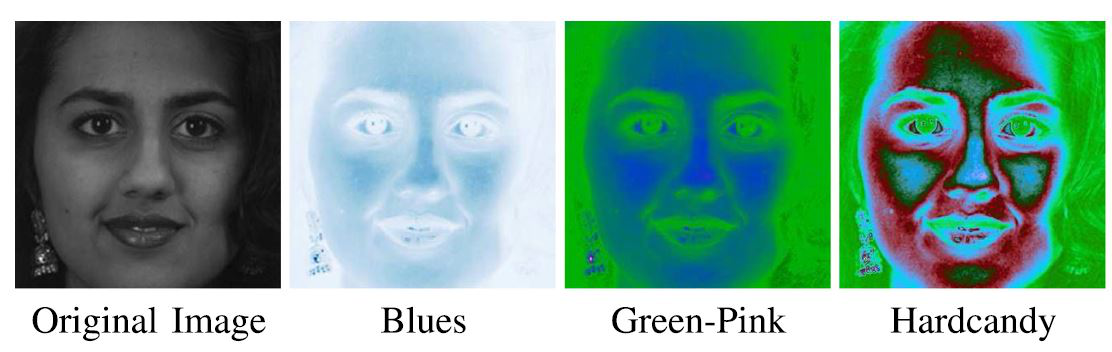}
\caption{False color-based protection using three different palettes \cite{cciftcci2017reliable}.}
\label{fig:false_color}
\end{figure}

\begin{figure}[ht]
\centering
\includegraphics[scale=0.1]{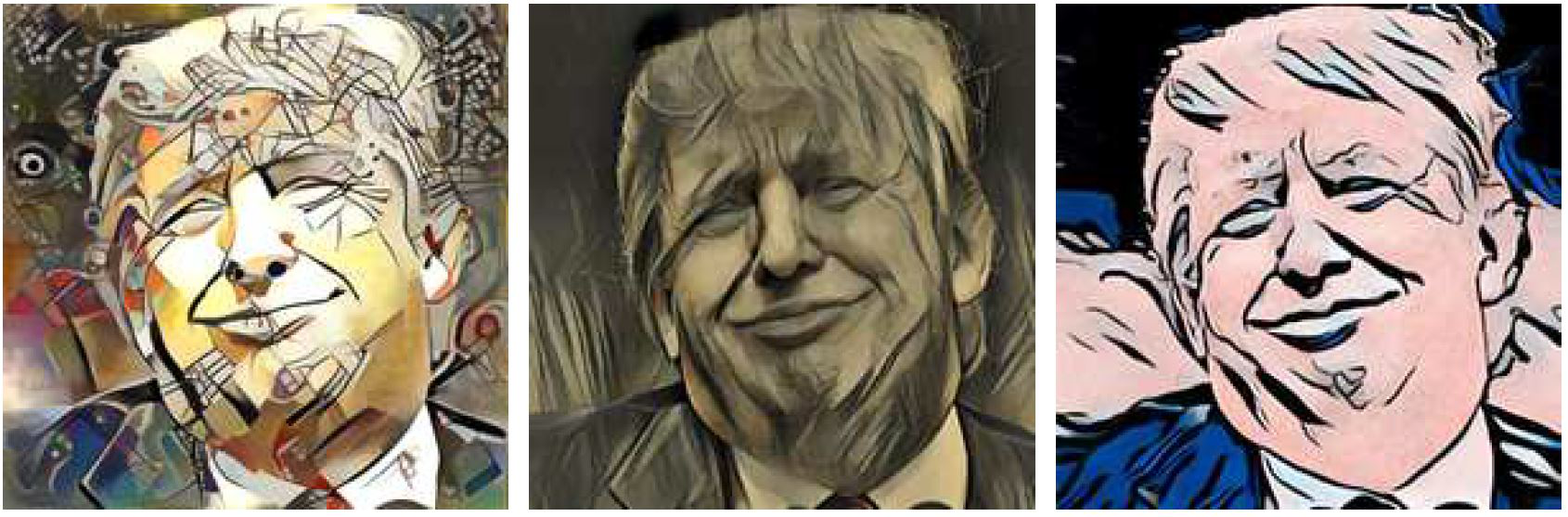}
\caption{Examples of aesthetic CNN-based style transfer for face obfuscation \cite{li2019anonymousnet}.}
\label{fig:aesthetic_transforms}
\end{figure}

\textit{Cartooning.} Cartooning achieves an effect similar to watercolor painting, where an ROI is converted into an aesthetically pleasing abstracted frame. For images containing faces, cartooning maintains a high utility since it preserves general properties like behavioral information, gender, emotion, and at the same time obscures facial identity \cite{erdelyi2014adaptive,hasan2017cartooning}. In  \cite{erdelyi2014adaptive}, Erdelyi et al. (2014) proposed an adaptive mechanism where the intensity of the cartooning filter can be changed according to the level of protection and utility required for specific scenes. The method uses two key techniques: color filtering and edge enhancements, i.e., smoothing the areas with moderate color variations to single-colored areas \cite{erdelyi2014adaptive}. Their evaluation metrics were the structural similarity (SSIM) index and peak signal-to-noise ratio (PSNR). The results show that cartooning achieves higher privacy and utility than blurring and pixelization. Hasan et al. (2017) in \cite{hasan2017cartooning} proposed a cartooning algorithm having two major components: the first is abstracting out the whole scene using image processing, and the second is using computer vision to detect the sensitive objects and replace them with visually similar and randomly selected clip-art images of the same class. Their method is intended to preserve semantic scene information while obfuscating fine-grained details. The authors note several challenges in the accuracy of identification and localization of objects, mainly: the selection of suitable clip-art, and the alignment of the clip art with the removed object in terms of scale, position, and orientation in an aesthetically pleasing manner. The results showed a general acceptance by the participants of the images' semantics, aesthetics, and privacy. However, the major downside is that the replacing clip-art is selected randomly from a dataset and, therefore, some participants were not pleased with the quality of the clip arts.    
ML-based algorithms are being utilized for cartooning and aesthetic \textit{style transfers}. For example, in \cite{larson2018pixel}, Larson et al. (2018) applied to the entire image a cartooning model based on Generative Adversarial Network (GAN). They intend to confuse image classifiers not to predict the correct category of the image while maintaining its visual aesthetics.  Aesthetic style transfer (illustrated in Fig. \ref{fig:aesthetic_transforms}) can be achieved using Convolutional Neural Networks (CNN), where a model is trained to generate an output combining the content of an image with the style of another \cite{li2019anonymousnet}. 

\subsubsection{Encryption-based Protection}
A proper encryption of the photos (e.g., using AES in CBC mode) provides complete security. Still, it destroys the information that is not privacy-breaching, therefore, hindering the ability to perform any image processing task without the secret key. Several technologies solve this problem by providing what is referred to as \textit{conditional access} \cite{padilla2015visual} and significantly distorting only the sensitive region of interest in the image while leaving the rest as is. The main advantage of this type of protection is that the original data can be perfectly recovered using the secret key. The drawbacks are that a secure channel to exchange the key and high computing power are needed, making it inefficient for real-time applications. Encryption and scrambling can be performed in different domains: image pixels, bitstream, or Discrete Cosine Transform (DCT) coefficients \cite{yuan2017image}.

\textit{P3.} Ra et al. (2013) \cite{ra2013p3} proposed P3 encryption for JPEG-compliant images, which splits the image based on an AC coefficient threshold into a JPEG public part and a private part. The two parts are shared independently, and a secret key is needed to uncover the private part. Their combined size does not significantly exceed the size of the original image.  P3 assumes that most information about an image is carried in the 0\textsuperscript{th} DC coefficient and the remaining 63 AC coefficients. Therefore, P3 encrypts the most significant bits of the significant DCT coefficients in the private part; the remaining coefficients and least significant bits are left in the public part. The image can be reconstructed by decrypting the private parts and positioning them back in the public part. According to the authors, P3 was explicitly designed to protect against automated recognition technologies \cite{ra2013p3} since the public image does not resemble the original image. Later on, it was uncovered that P3  is not effective in privacy preservation against artificial neural networks\cite{mcpherson2016defeating,fan2018image}. When the splitting threshold is not small enough, some visual information could be leaked from the public image. For a threshold of 1, the highest privacy setting of P3, Mcpherson et al. (2016) \cite{mcpherson2016defeating} achieved an 83\% re-identification accuracy of obfuscated faces using a simple neural network, and 97\% for a threshold of 20.  

\textit{JPEG Metadata Embedding.} Several works in the literature leveraged the metadata (application segments or APPn marker) in the JPEG standard to fully reverse the obfuscation. In \cite{yuan2015privacy}, an JPEG sub-image is constructed from the DCT coefficients of the original image, where the coefficients outside of the identified ROI are set to zero. A mask matrix is built from the shape, size, and position parameters of the obfuscated ROI. The sub-image is encrypted using symmetric encryption. The mask matrix and other metadata are also encrypted and embedded in the image file. A similar approach in \cite{farrugia2014reversible} proposed a two-level reversible watermarking scheme used in conjunction with any obfuscation process where the residual information needed to reverse the obfuscation process is compressed, authenticated, encrypted, and finally embedded within the obfuscated image. The authors implemented their scheme with $k$-same face replacement, where the reversing information is the difference between the original and obfuscated image. Other JPEG-specific partial encryption methods include \textit{Cryptagram} \cite{tierney2013cryptagram} and \textit{Thumbnail Preserving Encryption (TPE)} \cite{wright2015thumbnail}. TPE scheme allows the reconstruction of a low-resolution thumbnail from an encrypted image while preventing the extraction of any further data. 

\textit{Scrambling.} Image scrambling (or image \textit{permutation}) is the process of rearranging  pixel tiles (or DCT coefficients for JPEG images) randomly in the spatial domain to break the correlation between the neighboring pixels and make the image unintelligible \cite{poller2012robust,yuan2015secure,mondal2018cryptographic}. It can also be performed in transform domain or codestream domain depending on the media format \cite{rakhmawati2018image}. For a useful and comparative review of the various scrambling techniques proposed in the literature, refer to the survey given in \cite{mondal2017comparative} by Mondal et al. (2017).  

\begin{figure}[ht]
\includegraphics[scale=0.15]{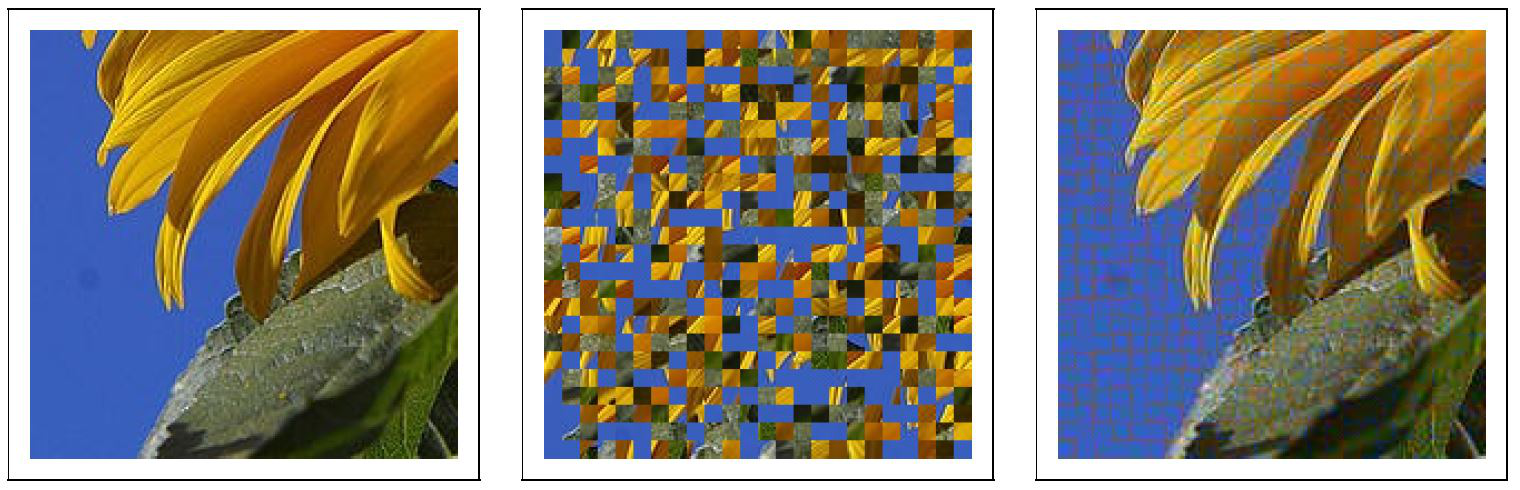}
\centering
\caption{Conventional Spatial-Domain Permutations \cite{poller2012robust}: (left: original, center: permuted, right: de-permuted with blurred grid borders).}
\label{fig:permutation}
\end{figure}

Permuting DCT coefficients or image tiles is insufficient to achieve secure scrambling. Scrambling can be solved through a brute force attack whose complexity is $t!$ for $t$ tiles \cite{sharma2018image}. An adversary can further reduce the attack's complexity by analyzing the boundary, color, content shape, and texture of neighboring blocks in the image. Additionally, scrambling is impacted by lossy compression \cite{korshunov2013using,poller2012robust}. When scrambling a compressed image, a blurring effect appears at the borders of the permuted pixel grids. Fig. \ref{fig:permutation} shows an example of a spatial-domain permutation and de-permutation on a compressed JPEG image. In \cite{poller2012robust}, Poller et al. (2012) proposed a secure jigsaw scrambling method that works with compressed JPEG images. The border blurring effect is reduced by replacing the pixel values at each grid's margins with the pixel values at the border during de-scrambling. To achieve higher security, the authors increased the number of possible permutations by shifting, rotating, and mirroring pixel grids randomly based on a secret key, and then by pseudo-randomly modulating the pixels' channel intensity \cite{poller2012robust}. The authors suggested salting the encryption key for additional security and decreasing the grid size to increase the number of permutations. Sharma et al. (2018) \cite{sharma2018image} applied block-wise Randomized Multidimensional Transformations (RMT) to the pixel matrix of the image. Acting like the encryption key, a random orthogonal (i.e., rotation) or a projection matrix multiples the image's matrix. Then, random additive noise selected uniformly from a known range is added to the multiplication output. 

\subsubsection{Image de-identification}
de-identification is the process of hiding, eliminating, or substituting personal identifiers (aka, personally identifiable information or PII) with surrogate identifiers to prevent the disclosure and use of private data for unintended and potentially harmful purposes. It aims to hide personal information from human recognizers, automatic recognition techniques, or both. The terms anonymization and de-identification are often used interchangeably in the literature. Nonetheless, the terms can be differentiated based on whether the obfuscation process is reversible (de-identification) or irreversible (anonymization) \cite{ribaric2016identification}. Table \ref{tab:image_deidentification-features} lists the main attributes and features (biometric, soft-biometric, and non-biometric) that can be extracted from facial and body landmarks in images. The required level of protection and utility determines which of the listed features to obfuscate and which ones to preserve. In Table \ref{tab:image_deidentification-features}, we refer to works in the literature investigating recognition and de-identification (and, in some cases, the preservation) of the listed features. The human face is the most prevalent biometric identifier in photos \cite{ribaric2015overview,meden2018k}. A trivial solution to de-identifying faces involves blacking out the eyes, mouth, or the entire face, where the first two fail to thwart currently robust face recognition solutions \cite{newton2005preserving}. Nowadays, most research focuses on achieving face obfuscation that is photo-realistic while maintaining standard facial characteristics. In this section, we look at image de-identification from the perspective of face de-identification. Nonetheless, image de-identification also includes the obfuscation of non-biometric identifiers such as license plates \cite{mcpherson2016defeating}, QR codes \cite{zhang2019protection}, etc..     

\begin{table}[ht]
\centering
\caption{Image de-identification: Balancing Privacy-Utility in Facial \& Body Features}
\label{tab:image_deidentification-features}
\resizebox{\columnwidth}{!}{%
\begin{tabular}{|l|l|c|l|} 
\cline{2-4}
\multicolumn{1}{c|}{} & \multicolumn{1}{c|}{\textbf{Feature}} & \begin{tabular}[c]{@{}c@{}}\textbf{Semi-, Non-, or}\\\textbf{biometric}\end{tabular} & \multicolumn{1}{c|}{\textbf{Reference}} \\ 
\hline
\multirow{5}{*}{\textbf{Body}} & Fingerprint & Biometric & \cite{yoon2012altered,lugini2014removing} \\
\cline{2-4}
& Skin tone & Semi-biometric & \cite{du2014garp,qin2017automatic,karras2019style}  \\ 
\cline{2-4}
& \begin{tabular}[c]{@{}l@{}}Tattoos, Scars, \\Birthmarks\end{tabular} & Semi-biometric &  \cite{marvcetic2014experimental,hrkac2016deep} \\ 
\cline{2-4}
& Clothing style & Non-biometric & \cite{brkic2017face,brkic2017know,prinosil2018clothing} \\ 
\cline{2-4}
& Body Pose, Activity & Semi-biometric & \cite{ren2018learning,nambiar2019gait} \\ 
\hline
\multirow{6}{*}{\textbf{Face}} 
& Iris, eyes & Biometric & \cite{thavalengal2014iris,zhang2018biological} \\
\cline{2-4}
& Ear & Biometric & \cite{emervsivc2017ear,jiddah2018fusion}  \\ 
\cline{2-4}
& Gender & Semi-biometric & \cite{du2014garp,gross2005integrating,othman2014privacy} \\ 
\cline{2-4}
& Age & Semi-biometric & \cite{du2014garp} \\
\cline{2-4}
& Hair style, hair color & Non-biometric & \cite{qin2017automatic,brkic2017face,prinosil2015automatic} \\
\cline{2-4}
& Facial Expression & Semi-biometric & \cite{gross2005integrating} \\
\hline
\end{tabular}}
\end{table}

\textbf{Face de-identification:} Table \ref{tab:image_deidentification-methods} shows a breakdown of the reviewed solutions for face de-identification: face morphing and face synthesis using $k$-same and adversarial learning methods.

\begin{table*}[ht]
\centering
\caption{The Main Methods of Face De-identification in the Literature}
\label{tab:image_deidentification-methods}
\resizebox{\linewidth}{!}{%
\begin{tabular}{|l|p{0.3\linewidth}|p{0.4\linewidth}|} 
\hline
\textbf{Main Methods} & \textbf{Year \& References} & \textbf{Description}\\ 
\hline
Face Morphing & 2013 \cite{korshunov2013using}, 2015 \cite{nakashima2015facial}, 2017 \cite{mahajan2017swapitup}, 2019 \cite{li2019identification}  & Synthesizing a face that maintains a likeness to an input and a target face using various techniques. \\ 
\hline
Face Synthesis & 2005 \cite{newton2005preserving}, 2018 \cite{sun2018natural}, 2019 \cite{karras2019style,li2019anonymousnet}, 2020 \cite{le2020anonfaces}  & Synthesizing realistic faces that preserve generic facial characteristics using $k$-same algorithms that compute an averaged face representing a cluster of $k$ similar faces or using DL-based synthesis. \\ 
\hline
\end{tabular}}
\end{table*}

\begin{figure}[h]
\centering
\includegraphics[scale=0.15]{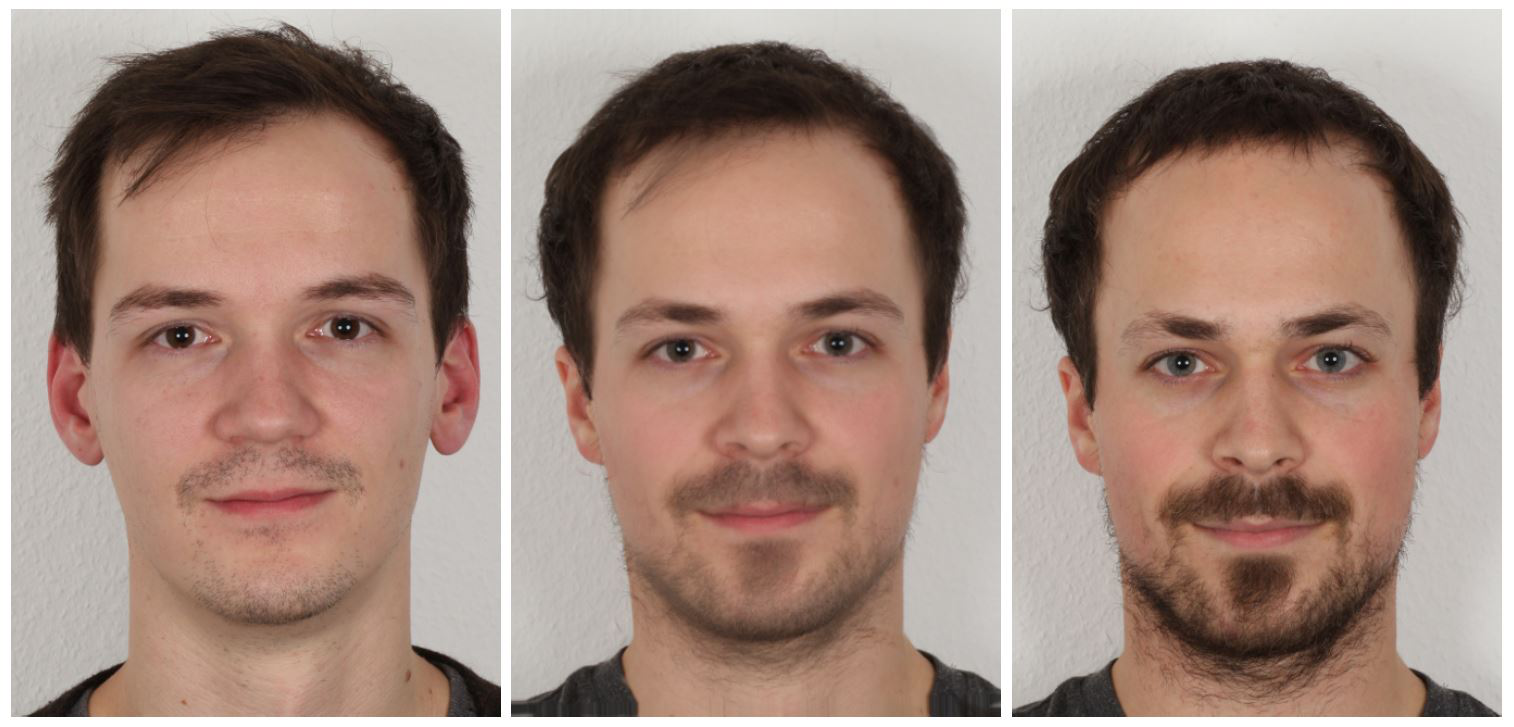}
\caption{Face Morphing \cite{seibold2017detection}: (left, right: original faces, center: morphed face).}
\label{fig:morphing}
\end{figure}

\textit{Face Morphing.} Face morphing (illustrated in Fig. \ref{fig:morphing}) is a geometrical transformation involving two face images to create a manipulated image that maintains a likeness to each of the original identities. In \cite{korshunov2013using}, Korshunov et al. (2013) proposed a partial and randomized face morphing method that works by interpolating the position and intensity of selected key pixels in the input face to match a target face \cite{rakhmawati2018image,korshunov2013using}. The interpolation of the key pixels is achieved somewhere in-between the input and target images, and the pixel intensities are weighted. First, a face landmark localization algorithm automatically selects pixels around the detected eyes, nose, and mouth from both images. The original and target images are divided into triangles using Delaunay triangulation, with the selected points being the vertices of the triangles. For corresponding pixels in both images, the final pixel coordinate is determined using a given level of interpolation value. For an interpolation level and intensity strength equal to 0, the original pixels are generated, and for a level of 1, the target image is produced. Recovery of the original image is possible by applying the inverse morphing transformation, given knowledge of the target image, selected key points, interpolation values, and intensity values. Security can be ensured by using a secret key for seeding the pseudo-randomization of the interpolation and intensity weights and encrypting the key points used for the triangulation. 

The previous method lacked naturalness due to partial morphing. To achieve a higher utility, Nakashima et al. (2015) in \cite{nakashima2015facial} proposed a method to preserve facial expression based on image-melding. Compared to the previous method, this technique leverages only a smaller number of corresponding points, specified manually by the user, in target and source images of similar orientation. The corresponding points in the source image are roughly matched to the target's by a transformation process, applying a weighted mask. The mask adjusts the scale, orientation, and position of the source image's facial features. Mahajan et al. (2017) \cite{mahajan2017swapitup} developed a system for \textit{face swapping}. The system works by directly fitting a secondary image over the input image by rotating and scaling it. Color balance adjustment and blending of the facial landmarks from the second image onto the input are made to achieve a more natural appearance. Li et al. (2019) \cite{li2019identification} created a Facial Attribute Transfer Model (FATM) using autoencoders. The DNNs blend the target's facial attributes to those of the original faces. 

\textit{Face Synthesis:} Here, we discuss $k$-same and GAN-based methods. Both methods rely on different mechanisms to effectively produce a new face image, unlike face morphing that modifies an input face to match a target face. 

\textit{$k$-same Family.} The basis of $k$-same algorithms is \textit{$k$-anonymity}, an extension from differential privacy for micro-aggregated data. Other data anonymization techniques exist for categorical data; the two most popular among those are $l$-diversity and $t$-closeness \cite{le2020anonfaces}. $k$-same algorithms compute an averaged face to represent a cluster of $k$ similar faces. The first work to introduce this concept was by Newton et al. (2005) \cite{newton2005preserving} to enable secure sharing of image data with provable guarantees. 
The algorithm first determines the similarity between faces and clusters them based on a distance metric. The authors proposed a pixel-wise averaging ($k$-Same-Pixel) or eigenvectors averaging ($k$-Same-Eigen) \cite{newton2005preserving}. Images produced by $k$-Same-Eigen have a blurred effect because only a minimal number of important facial characteristics is retained. Even though the averaged face may appear similar to one image than another, because the same averaged face replaces all images in the cluster, the method guarantees that the correlation success rate of the $k$ faces is  $\frac{1}{k}$ at most. One of the biggest challenges is determining the appropriate value for $k$ to achieve the highest privacy and finding the optimal groups of closest faces, which is an NP-hard problem  \cite{newton2005preserving}. The main limitation of \cite{newton2005preserving}'s method is its lack of naturalness and information loss during the de-identification process. 

Many additions to the $k$-same family have been proposed to overcome some of the previously mentioned limitations. Gross et al. (2005) \cite{gross2005integrating} presented $k$-Same-Select, which guarantees the utility of the data by preserving facial expressions or gender. However, $k$-Same-Select works in the pixel domain and sometimes suffers from alignment mismatch in the de-identified face. $k$-Same-M by the same authors \cite{gross2006model} addressed this limitation by using statistical models instead. Meng et al. (2014) \cite{meng2014face} proposed $k$-Same-furthest, which aims to achieve perfect privacy for any source face regardless of the value of $k$. Sun et al. (2015) proposed $k$-Diff-furthest \cite{sun2015distinguishable} to solve the tracking problem of individuals in $k$-Same de-identified videos, where $k$ similar faces are replaced with the same averaged face. More recently, in 2018-2019, two studies incorporated the concept of $k$-Same algorithms in neural networks and GAN-based face synthesis in \cite{meden2018k} and \cite{pan2019k}, respectively. 

\textit{GAN-based Methods.} Images produced by the early solutions of $k$-Same may exhibit poor visual quality and ghosting artifacts due to inexact alignment of facial features \cite{le2020anonfaces,croft2019differentially,bursic2021quantitative}. Generative Adversarial Networks (GANs) are more successful in producing natural-looking face images \cite{kim2020selective} that preserve many facial characteristics and cannot be reliably linked to the original identity. However, face detection is problematic when the subjects appear in varied backgrounds, positions, and activities. Most generative face models are suited for frontal and strictly aligned faces. To tackle this problem, Sun et al. (2018) \cite{sun2018natural} devised a two stages solution. Depending on the visibility of the face pixels, facial landmarks are either detected or generated. They are detected when the original face image is clear, but hypothetical realistic facial structures and poses are generated when the face has been obfuscated. An advantage of the second scenario is that it allows upgrading weak obfuscations, including blocking and blurring, to use realistic synthetic faces that blend naturally into the context. 

Le et al. (2020) \cite{le2020anonfaces} designed a system based on StyleGAN (style-based GAN) \cite{karras2019style} and $k$-same algorithms to achieve provable privacy, which synthesizes a face for a given dataset while simultaneously and iteratively tuning its privacy-utility trade-off. The system uses a two-steps cluster analysis component for feature vector extraction and clustering all input images into fixed-sized sets. The authors report that low-dimensional feature vectors produce better clustering and ensure high naturalness. The synthesized faces preserve properties such as age, gender, skin tone, and emotional expressions. Finally, the authors used a measure of the average Euclidean distances between images in a cluster and the corresponding synthesized image (i.e., measuring total information loss). StyleGAN \cite{karras2019style} architecture is superior to traditional GAN in terms of established quality metrics and the unsupervised separated learning of high-level attributes (e.g., pose, freckles, hair). Another noteworthy work is AnonymousNet (2019) \cite{li2019anonymousnet} which uses GANs to produce realistic synthetic faces but notably adds adversarial perturbations to reduce the recognition accuracy and provides provable privacy metrics through a $k$-anonymity-based attribute selection.

\textit{Diffusion-based face de-identification.}
Recent work increasingly relies on diffusion models to anonymize faces by synthesizing identity-safe content while preserving non-identity attributes. Compared to GAN pipelines, diffusion approaches often dispense with explicit landmarks or masks, and can better preserve pose, gaze, lighting, and background. These methods fit naturally in the “replacement/synthesis” family, aiming to suppress identity cues without sacrificing downstream utility. They also enable controllable anonymization, where identity is removed but task-relevant appearance is maintained. A representative research work, conducted by Kung et al. (2024) \cite{kung2025face} introduce a reconstruction-only diffusion framework that removes identity without auxiliary detectors, preserving expression, pose, gaze, and background while achieving strong identity suppression and high perceptual quality. Complementarily, Shaheryar et al. (2024) \cite{shaheryar2024iddiffuse} propose a dual-conditional diffusion model that maps real identities to synthetic ones while conditioning on non-identity “texture” features, improving controllability and consistency when the same subject appears across multiple images.
Similarly, He at al. (2025) \cite{he2024diff} explore diffusion-based privacy protection that balances identity removal with retention of semantic attributes important for downstream tasks.

\subsubsection{Adversarial Perturbation}
Studies have shown that advanced ML systems can, in partially obfuscated images, make correct malicious inferences about the person, such as religion, age, and occupation, thus deeming visual obfuscation methods insufficient when adopted in isolation  \cite{oh2016faceless}. Adversarial perturbation mechanisms were developed to change the prediction of deep neural networks, so the adversary draws incorrect or specific intended inferences from the image data without making perceptible changes to the input image. Xiao et al. (2020) \cite{xiao2020evade} extend adversarial techniques to the domain of deep‑hashing image retrieval. Rather than visibly altering photographs, they inject imperceptible perturbations that push images away from their hash‑space neighbours, effectively ‘stashing’ private images in distant regions. Their approach increases the effort required for an attacker to retrieve protected images by two to seven orders of magnitude and transfers to unseen models in 30–60\% of cases under a black‑box threat model, while leaving perceptual quality intact. Another advantage of this technique is that it can be applied to an already obfuscated image to strengthen the obfuscation further. Recent works have developed the concept of Universal Adversarial Perturbations, which can be added to any image and cause the target ML model to misclassify it. Universal perturbations are more practical to deploy than per-instance perturbations since minimal computation and knowledge (black-box attack) are needed \cite{chaubey2020universal}. 

Some approaches to crafting adversarial face images lack quality and take an unreasonable amount of time to be generated. Deb et al. (2019) \cite{deb2019advfaces} proposed, AdvFaces, an automated adversarial face synthesis method that learns to generate minimal perturbations in the facial regions via GANs. AdvFaces produces imperceptible perturbations that evade recognition with high success rates up to 97.22\% for obfuscation attacks (falsely reject a genuine subject) and 24.30\% for impersonation attacks (falsely accept an imposter) \cite{seibold2017detection,scherhag2019face}. To contrast image classification tasks, Li et al. (2019) \cite{li2019scene} proposed an algorithm that generates adversarial noise using a novel adaptation of Fast Gradient Sign Method (FGSM) \cite{sanchez2020exploiting}, which limits the probability of inferring the true class of a distorted image successfully. Different variations of FGSM such as iterative FGSM, private FGSM, random FGSM, and least-likely FGSM may select the true class as the target class. The authors designed the new FGSM approach to pick the target class from an adaptive subset of classes that most likely does not include the class to be protected. A recent survey by Chaubey et al. (2020) in \cite{chaubey2020universal} provides a comprehensive summary of the existing adversarial models and defense mechanisms against them. Other examples of works based on image adversarial methods are referred in \cite{hukkelaas2019deepprivacy,raval2017protecting,oh2017adversarial,rajabi2021practicality}. 

\subsubsection{Abstraction}
\begin{figure}[!ht]
\includegraphics[scale=0.3]{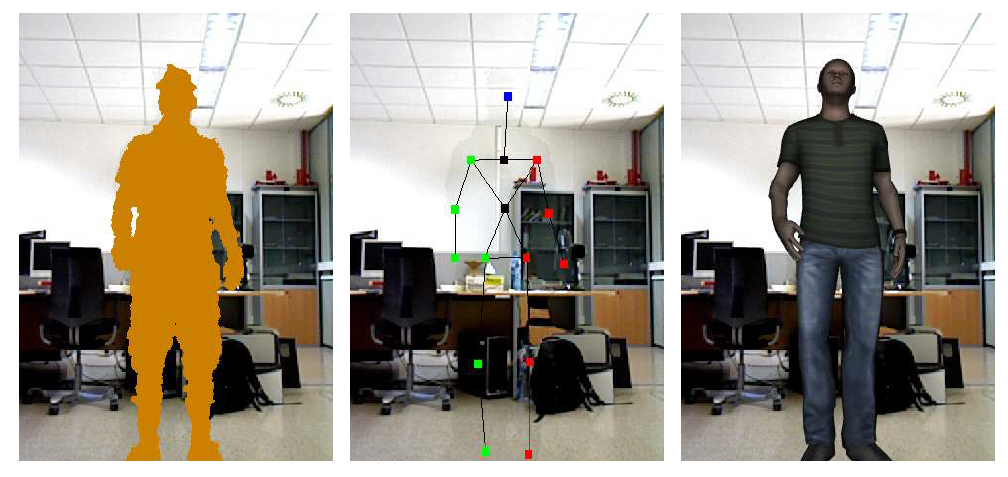}
\centering
\caption{Abstraction techniques \cite{padilla2015visual}: (left: solid silhouette, center: point-light, right: avatar).}
\label{fig:abstraction}
\end{figure}

Abstraction removes an object of interest and replaces it with a graphical representation that somewhat resembles the removed content. Fig. \ref{fig:abstraction} shows three examples of object abstraction found in the literature: \textit{silhouette}, \textit{point-light} (or \textit{skeleton}), and \textit{avatar}. Since abstraction is a utility-and-privacy-preserving method, it is widely used in online social networks. An abstraction algorithm applies image inpainting methods to remove the segmented object of interest, renders a visually similar object, and places the abstracted form over the exact pixel locations of the removed ROI. 

The effectiveness of the following abstraction methods was investigated by Li et al. (2017) in \cite{li2017effectiveness}. The point-light or skeleton method replaces an ROI with points and lines corresponding to the ROI's shape outline. A human point-light can only illustrate a person’s activity and height, but completely hides his identity. The silhouette method is where an ROI is replaced with a (monochrome) object that mirrors its shape. A human silhouette can preserve the subject's body pose, activity, clothing style, gender, health condition, or hair style, but still completely removes biometric identifiers. A human avatar can preserve, in addition to all previously listed features, finer details about the person such as facial expressions and age, and still protect the identity from disclosure. The privacy-utility balance is vividly exhibited in the previous examples. Unlike face replacement methods, abstraction does not aim to preserve naturalness; however, it seeks to generate realistic representations of the removed sensitive region. Nonetheless, abstraction is still considered one of the most effective methods for visual obfuscation today. 

\subsubsection{Blocking / Masking}
\begin{figure}[ht]
\centering
\begin{subfigure}[t]{40pt}
\centering
\includegraphics[width=1cm,height=1.5cm]{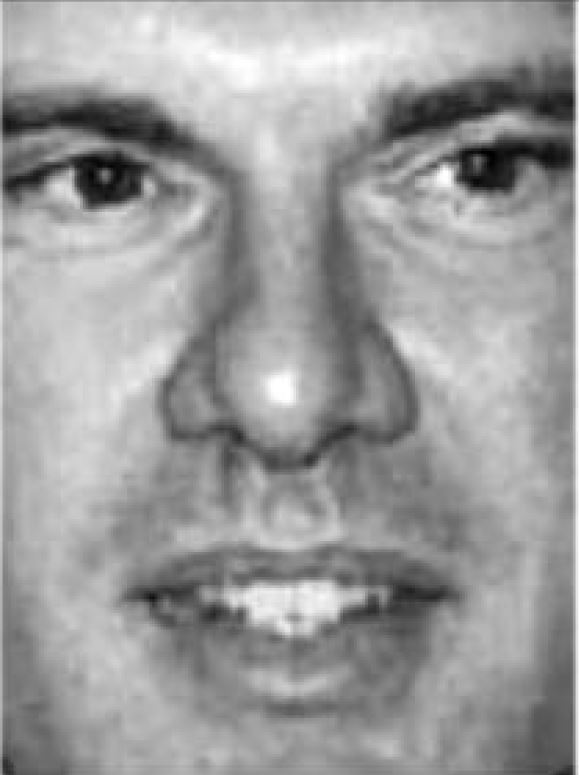} 
\caption{\centering Original}
\end{subfigure}
\quad
\begin{subfigure}[t]{30pt}
\centering
\includegraphics[width=1cm,height=1.5cm]{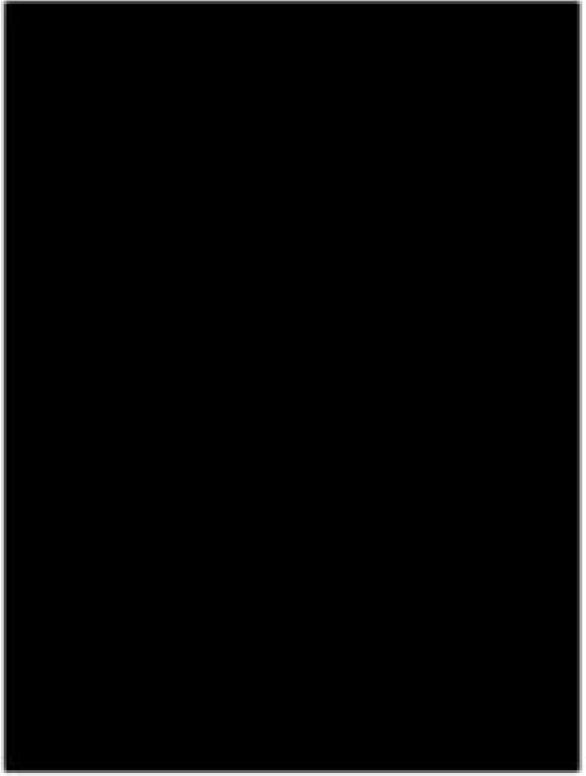}
\caption{\centering Block}
\end{subfigure}
\quad
\begin{subfigure}[t]{30pt}
\centering
\includegraphics[width=1cm,height=1.5cm]{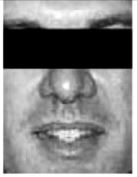}
\caption{\centering Mask}
\end{subfigure}
\quad
\begin{subfigure}[t]{35pt}
\centering
\includegraphics[width=1cm,height=1.5cm]{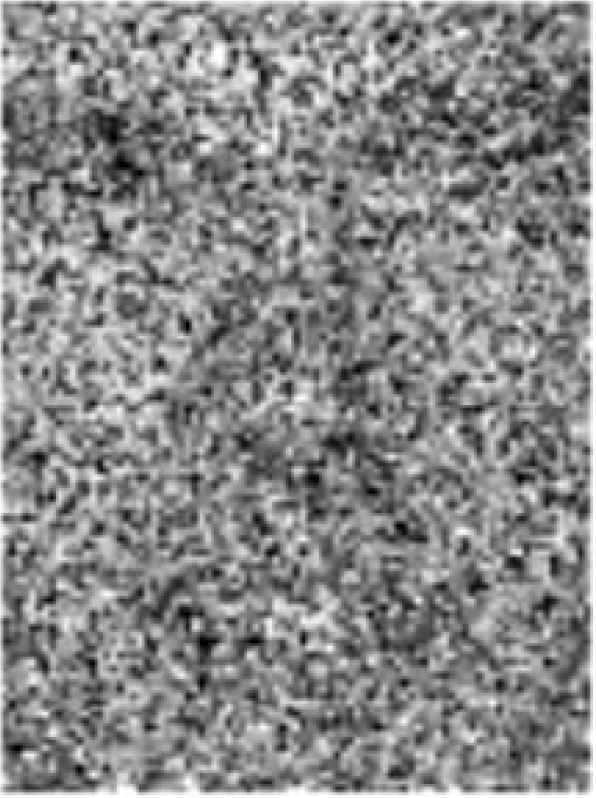}
\caption{Grayscale\\\centering Noise}
\end{subfigure}
\quad
\begin{subfigure}[t]{30pt}
\centering
\includegraphics[width=1cm,height=1.5cm]{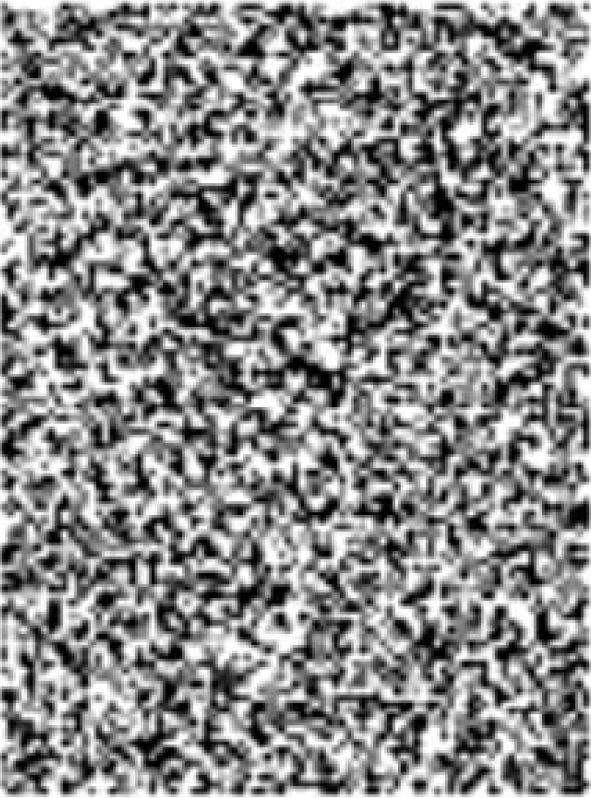}
\caption{\centering Black- white noise}
\end{subfigure}
\caption{Examples of Blocking/Masking \cite{newton2005preserving}}
\label{fig:blocking}
\end{figure}

\textit{Blocking} or \textit{blacking} (see Fig. \ref{fig:blocking}-b) is where the sensitive bounding box is detected and replaced with a shape (e.g., square, rectangle, ellipses) in a solid color, typically black. Whereas the term \textit{masking} is similar to blocking and is frequently used interchangeably in the literature. However, we find that it commonly refers to less intrusive covering of the sensitive regions, e.g., the bar mask in Fig. \ref{fig:blocking}-c, where the eyes are determined as the sensitive area of interest. It may also refer to covering the sensitive regions with other objects, e.g., covering a face area with a clip-art or a sticker \cite{yuan2015privacy}. \textit{Noise addition} can be considered a subtype of image masking \cite{newton2005preserving}. In noise addition, random image pixels are replaced by black and/or white pixels (referred to as \textit{salt-and-pepper noise} \cite{seibold2017detection}) as seen in Fig. \ref{fig:blocking}-e. Black/white images are modified by flipping random pixels, and in gray-scale images (see Fig. \ref{fig:blocking}-d), a random value between 0 and 255 replaces random pixels. However, additive random noise does not thwart automated recognition except when roughly 50\% of the pixel values have been flipped in black/white images or 75\% of the pixels are changed in gray-scale images \cite{newton2005preserving}. Moreover, the black bar mask around the eyes is ineffective against image inpainting attacks, e.g. \cite{barnes2009patchmatch,yeh2017semantic,zhao2019guided}. Blocking and masking are the most utility-reducing among all other ad-hoc obfuscation methods \cite{vishwamitra2017blur}. Nonetheless, blocking offers high protection against human and automated object recognition \cite{tekli2019framework}. 

\subsubsection{Inpainting}
\begin{figure}[ht]
\centering
\includegraphics[scale=0.15]{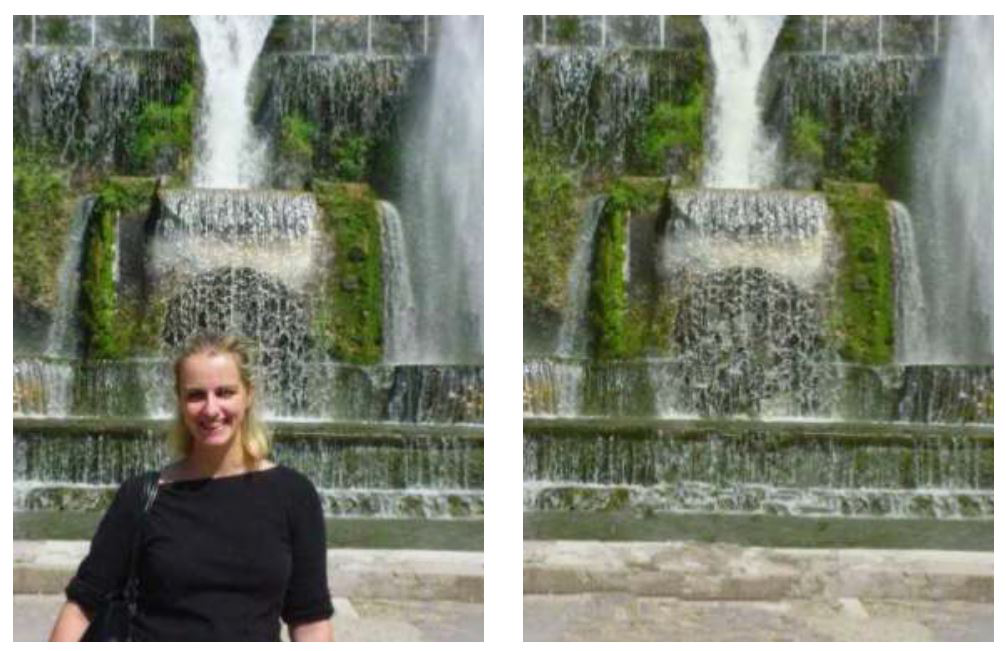}
\caption{Image Inpainting \cite{padilla2015visual}: (a) original image (left), (b) inpainted image (right).}
\label{fig:inpainting}
\end{figure}

Inpainting methods are generally used to repair damaged or missing parts of photos. However, inpainting as an obfuscation method (see example in Fig. \ref{fig:inpainting}) refers to completely removing sensitive content (e.g., full-body region) of an image. The cleared space is then filled to blend naturally in a visually consistent manner, appearing as part of the background \cite{li2017effectiveness}. Based on the study conducted in \cite{li2017effectiveness}, inpainting is the most effective technique for achieving true privacy. In the cited study, it produced the smallest percentage of identification success rate amongst 13 other obfuscation methods. The output of inpainting algorithms is becoming more seamless and photo-realistic with the use of deep learning and adversarial training approaches. Once the image is inpainted, a visual abstraction of the removed sensitive information (e.g., \cite{padilla2015visual,blavzevic2015towards}) or synthetic faces (e.g., \cite{sun2018natural}) can be added to prevent complete information and utility loss. A major limitation of inpainting algorithms is their computational complexity \cite{rakhmawati2018image}.

\subsection{Attacks and Countermeasures}
Image obfuscation alters or removes features from the images to hide sensitive information and retains some visual features to keep the image intelligible and suitable for processing. However, these visual features can still be used to identify or reconstruct the protected objects through different attack styles. The attacks against image obfuscation can be classified as human-based, machine-based, recognition-based, reconstruction-based, or inference-based. In this section, we focus separately on recognition (or re-identification) attacks,  reconstruction (or restoration) attacks, and inference attacks. 

\textbf{Recognition-based Attacks.} Recognition-based attacks are where an attacker attempts to re-identify obfuscated objects or the person(s) from their obfuscated faces. Human-based recognition is based on the degree of information loss in the obfuscated image and the attacker's familiarity with the scenes and objects before the obfuscation. Removal methods such as blocking, masking, and inpainting may completely prevent human recognition \cite{li2017effectiveness,vishwamitra2017blur}. Conversely, image distortion methods (e.g., image filters, warping, false colors, cartooning) preserve the general shape and intensities of the objects, and therefore, are not very resilient to human detectors \cite{li2017effectiveness}. There are several ways to qualitatively measure an identity disclosure risk. A definitive privacy leak is when an image can be correctly linked to an individual with high certainty. Potential privacy risks exist when private information is linked to several individuals. The highest level of privacy is when an image cannot be related to any individual. 

In terms of machine-based recognition, an attacker would use machine learning algorithms for image recognition. Three different kinds of automated attacks involve the matching of input and output images in various arrangements: (a) original to obfuscated; (b) obfuscated to original; and, (c) obfuscated to obfuscated \cite{newton2005preserving}. In the first type of attack, the obfuscated images are run directly through the image recognition software. In this context, the recognition algorithm uses a training set (aka, gallery images \cite{newton2005preserving}) composed of the original images, while the obfuscated images (aka, probe images) are used as input. The second type is the dual of the first; the training set comprises obfuscated images, and the original images are used as input. This attack assumes that the attacker has the complete collection of original images. According to \cite{newton2005preserving}, the trained model is expected to have decomposed and dispersed the alterations on the obfuscated images. The work in \cite{guo2017countering} serves as an example of this attack, where input transformations on perturbed images counteract the effect of adversarial perturbations. 
In the third attack, the attacker invokes the same obfuscation technique on a set of probe images. The recognition is done by matching an obfuscated probe to an obfuscated gallery. This attack assumes that the adversary has access to the training set obfuscated by a given method. This is also called \textit{parrot attack} or \textit{imitation attack}; a process of comparison and elimination to recognize the obfuscated image. In \cite{padilla2015visual}, pixelization and blurring were found to be vulnerable to parrot recognition which can achieve recognition rates close to 100\%, despite looking somewhat de-identified to human recognizers \cite{padilla2015visual}. It is shown that parrot attacks drastically reduce the level of privacy protection and improve the recognition rate \cite{ribaric2016identification,ribaric2015overview}. 

To counter image recognition attacks, adversarial perturbations can be added to alter the machine-learning based prediction, while having the introduced perturbation not detectable by the human eye. However, there are some defense mechanisms suggested in the literature for detecting and rectifying such perturbations. Two of such defenses are considered: the first one is to pre-process images at the inference time \cite{guo2017countering}, and the second one relies on increasing the robustness of the model against adversarial training (i.e., training with perturbed images) \cite{chaubey2020universal}. In \cite{akhtar2018defense}, Akhtar et al. (2018) presented a rectification method called Perturbation Rectifying Network (PRN) that pre-processes the images to be classified before passing them to the target recognition model. The output of PRN is a rectified image minus the perturbation. A detector network, separately trained on the DCT of the input-output difference of the PRN, performs a binary classification to decide if an adversarial perturbation is present. If perturbed, a rectified image is passed to the target model instead of the probe image. PRN is shown to have a success rate up to 97.5\% against adversarial perturbations. Another possible countermeasure to recognition attacks is pseudo-randomization (based on a secret key) in the de-identification algorithm. The randomization will impede an attacker's ability to replicate the de-identification process exactly. When attaining high privacy is the main concern, we recommend not to use standard image transformations (such as pixelization, blurring, masking, lossy compression, and other transforms), even though they have lower computational complexity, since they are ineffective against ML attacks.

\textbf{Reconstruction-based Attacks.} Reconstruction-based attacks aim to recover the original image before its transformation. Today, with the advent of deep neural networks, this task has become much less challenging. Several studies have shown that conventional image transformations, namely pixelization, blurring, masking, and compression, are ineffective against bicubic interpolation, statistical modeling, and DL-assisted attacks \cite{mcpherson2016defeating,tekli2019framework,hill2016effectiveness}. 

For pixelized and compressed artifacts, a study in \cite{ledig2017photo} used neural networks to achieve image super-resolution \cite{wang2020deep,anwar2020deep}, upscaling images up to a factor of 4. In further works, GAN-based models were used to solve super-resolution problems \cite{wang2018esrgan}, and the study in \cite{yu2016ultra} achieved ultra-resolution for images of lower resolution (e.g., 16$\times$16). Similarly, for canceling the impact of blurring and restoration of sharper images, several image-deblurring methods were proposed in the literature using neural networks \cite{koh2021single}, and GANs \cite{kupyn2018deblurgan}.

Image inpainting can be used to infer missing pixels for blacked-out sections in an image. Yeh et al. (2017) \cite{yeh2017semantic} trained a deep generative (GAN-based) model to find an encoding in the latent space that is closest to the corrupted image. The generator then uses the encoding to reconstruct the missing parts. The major advantage of their method is that it works with arbitrarily structured and large masked regions. Another method in the literature was also successful in recovering blacked image regions, however, relying on the presence of similar patches and structures in the remaining parts of the input image \cite{barnes2009patchmatch}. Other solutions used statistical modeling such as total variation and low rank and machine learning such as nearest neighbor (NN) and autoencoders \cite{yeh2017semantic}. In comparison, semantic image inpainting produced much more realistic images with sharper edges, having higher PSNR and SSIM in randomized masks, even when 80\% of the pixels are missing. Prediction of the missing pixels can be done using external data (e.g., guidance images \cite{zhao2019guided}) or internet-based inpainting \cite{whyte2009get}. 

\textbf{Inference Attacks.} As the name suggests, an adversary can infer hidden attributes or information about the obfuscated image in an inference attack. Inference includes determining that an image itself has been manipulated or synthesized. This is specifically valid for naturalness-preserving obfuscation technologies, such as face morphing and face synthesis. In addition, inference can target the learning process itself: Shokri et al.~\cite{shokri2017membership} demonstrated that an attacker can determine whether a specific image was used during the training of a model, through membership inference. Such leakage may occur even when only obfuscated or partially modified images are exposed, revealing sensitive associations despite the presence of sanitization.

In face de-identification, indistinguishability is an important property where the synthetic face should look realistic and not be easily differentiated from a natural one. Examples of works dealing with image forgery exist; detection through analysis of reflections and shadows or binary statistical features, detection of adversarial images \cite{hendrycks2016early}, or detection of deep fake images \cite{tolosana2020deepfakes}.

\subsection{Discussion}
\textbf{Applications.} The need for image obfuscation is urgent and evident, given the amount of images that  are circulated and shared daily on the Internet or over On-line Social Networks. Due to the privacy-intrusive nature of the Internet, the community was driven to advance the research in privacy-preserving and privacy-aware systems and their applications. Examples of such systems and applications are: (a) image processing, search, and retrieval on the cloud; (b) image capture in video camera surveillance, health monitoring applications, IoT sensor networks, Internet of Multimedia Things (IoMT), and commercial applications such as Google Street View; (c) image sharing and publication on online social networks (OSNs); and, (d) image classification and recognition for automated tasks, such as autonomous driving and supply chain optimization \cite{tekli2019framework}. 

\textbf{Challenges and Open Research.} Undesirable visual exposure and malicious inference of high-level sensitive information are standard image privacy problems \cite{liu2020privacy}. According to previous studies, implicit information such as health conditions, social class, and relationship status can be deduced from images using ML models, leading to privacy infringement. However, protecting against malicious inference remains challenging due to the multi-modality of the underlying features, which are typically tricky to recognize using conventional image recognition. The field of multimodal attribute detection and obfuscation in images is still in its infancy; only a few studies addressed the perturbation of hairstyle, clothing, race, and gender in images. In image de-identification, face and body detection significantly impact the achieved protection--- failure to locate several soft-biometric identifiers results in a high re-identification risk. Some face detection challenges are partial visibility of the face regions due to occlusions, accessories, sunglasses, and variations in head pose, lighting, and backgrounds, which can obstruct the detection algorithm's ability to recognize face regions correctly. Current studies still try to mitigate those recognition challenges. In addition, other studies aim to generate realistic faces that maintain a high degree of similarity in terms of facial expressions and head pose and mimic body poses to relay an accurate representation of the human activity being performed \cite{le2020anonfaces}. There is also a need for quantifying privacy and utility through objective evaluation.

\textbf{Key Properties.} The key properties to consider when selecting an optimal obfuscation method for protection against information leakage are \cite{li2019scene}: (a) \textit{naturalness and utility}: the distortion should be unnoticeable; (b) \textit{(ir)reversibility}: (in)ability to undo the transformation; (c) \textit{robustness and security}: thwarting most human and machine-based attacks; (d) \textit{adjustable/adaptable}: protection with different degrees of strength; (e) \textit{provably secure}: provides measurable security metrics based on differential privacy or distance metrics; and, (f) \textit{computational complexity}: it should be easy to implement in terms of resource usage. The study field of image obfuscation has reached a mature level, where the latest techniques are based on adversarial learning. Therefore, rather than developing new obfuscation methods, most research is now focused on improving the utility of the adversarial training output while keeping security/privacy a priority. When one refers to the ``utility'' of an obfuscated image, they are typically concerned with three things: its naturalness (i.e., Does it look realistic?), its intelligibility (i.e., Does the distortion impact the ability to infer important attributes?), and other context-specific properties (e.g., image quality, generated obfuscations for images with faces belonging to different people should be distinguishable). The first three properties introduced at the beginning of this paragraph (naturalness, reversibility, and security) are addressed in most image obfuscation methods. However, the remaining ones are selectively implemented based on the user requirement. In terms of provable privacy guarantees, the main way to achieve it is by applying statistical and aggregated data protection methods derived from differential privacy such as $k$-anonymity. However, most existing protection solutions do not provide measurable privacy guarantees. Moreover, selective obfuscation (i.e., applying the obfuscation on a per-feature basis), adjustable privacy filters, and automatic evaluation of image privacy and utility, are still ongoing research topics.

\textbf{Review of Methods.} In Table \ref{tab:image_techniques_overall}, we have given low, medium, and high ratings for the main categories of image obfuscation regarding their utility and the ability to reverse the obfuscation and protection level against human and machine attackers. We further explain the rating choice here. Simple ad-hoc image transformation methods (blurring, pixelization, lossy compression, edge, warping, style transfer, false color, cartooning) distort sensitive ROIs in a way that results in a high degree of information loss. Still, at the same time, their output maintains a strong resemblance to the original general shapes of the image. However, aesthetic transforms (false colors, style transfer, and cartooning) provide a better visual experience than the others. These methods are no longer viable due to robust ML recognition \cite{newton2005preserving} that can deduce the original input before its transformation. 

On the one hand, encryption and scrambling completely hide the ROI and, therefore, adversely impact the image’s utility. On the other hand, they allow full recoverability of the original content using a secret key. The security of the encryption-based protection depends on the key strength and the permutations’ strength. Possible solutions to break the protection are based on brute force and are computationally complex. 

Suppression methods such as masking, blocking, and inpainting altogether remove the sensitive ROIs, resulting in the highest degree of information loss. Although, inpainting is less visually intrusive and produces a more natural effect. These methods are robust against human and machine recognition. Studies have shown that image inpainting methods can be used to fill in missing (masked) areas. 

Abstraction entirely replaces an ROI, thus protecting its identity with a visually similar representation. The utility of the replacement depends on the replacing object, its quality, and its resemblance to the removed ROI.  Overall, inpainting and abstraction may be most useful as obfuscation methods because they are both effective at increasing privacy for elements of an image and providing a good viewer experience \cite{li2017effectiveness}.  This shows that to achieve a higher privacy-utility balance, more processing time and computations are needed. Machine learning can provide efficient obfuscation models; however, their accuracy depends on how extensive the training set is.

In terms of advanced obfuscation, the solutions are mainly based on face synthesis and adversarial perturbation. Adversarial perturbations are not perceivable by humans; they add little visual distortion to the image. Their purpose is to deceive automated machine recognition. However, some studies in the literature have shown that adversarial perturbations are detectable and can be countered using ML methods. Though, the cited countermeasures assume an attacker with access to the perturbed training set, which is not always possible in real-world scenarios. Face synthesis is a non-reversible method of generating new faces by averaging a cluster of similar faces ($k$-same family algorithms) or adversarial learning. The method effectively reduces the re-identification risk and typically generates natural-looking faces that preserve gender, facial expressions, age, or race while protecting the identity.

\section{Audio Sanitization}
\label{audio_obfuscation}
Audio data is another significant dimension of big data with several applications in surveillance, voice-based verification, lifelog services, telephone-based services, and speech-based services such as Apple's Siri, Google Assistant, and Amazon Alexa. Voice data plays another role in call centers and help-desks where customer phone calls are stored for auditing and training purposes. Such audio recordings may contain sensitive information, also known as \textit{Private to a Person (P2aP)} information, such as names, demographics, locations, ID numbers, etc. \cite{sheikh2019fuzzy}. Traditionally, audio segments containing sensitive information were first manually identified and then muted \cite{vatanparvar2020adapting}. Alternatively, speech-to-text (STT) conversion is used to transcribe audio conversations automatically. Text sanitization methods are then applied to the transcriptions, where patterns of sensitive text are identified. Then, the corresponding audio clips are censored by muting, or replacing the segment of interest with a beep (aka, \textit{bleep censor}), or another sound effect \cite{sah2017video}. This latter approach has been used in television, radio, and some video-sharing platforms for many years. 

However, redacting PII in the voice data alone is far from enough to prevent security and privacy risks; given that, any utterance includes the speaker's voiceprint. Exposing an individual's voiceprint can cause significant security risks since it is used by voice-based authentication and identification systems. A voiceprint is a distinct and stable biometric identifier that uniquely represents a hybrid of physiological and behavioral characteristics \cite{sah2017video,ribaric2016identification}. From the speaker's voice, personal attributes can be inferred, such as gender, age, emotion, health (e.g., presence of coughing, dysarthria), education (vocabulary choice), and origin (using accents). Malicious attackers, given background knowledge of a victim speaker, may link the anonymized speech back to the source (\textit{linkage attack}). Another security risk is \textit{spoofing attacks} (i.e., identity theft) and \textit{reputation attacks}. That is, an attacker may impersonate an individual using his or her voiceprint to bypass authentication systems, commit fraud, or release controversial statements. Reputation attacks have become more prevalent; voice conversion in conjunction with DeepFake technology \cite{tolosana2020deepfakes} was used to create controversial videos of celebrities and politicians causing public outrage, e.g., the Fake Obama Speech \footnote{Fake Obama Speech, https://goo.gl/pnR3VK}.
Therefore, audio obfuscation is essential and has real-world applications in censorship and speaker de-identification. Speaker de-identification alters the identity information without affecting the textual content of the audio recording. Moreover, censorship is widely used in television or radio to remove profanity or hide the speaker's identity by using conventional transformation or modulation of the speech signals.

\begin{figure*}[t]
\centering
\includegraphics[width=\textwidth]{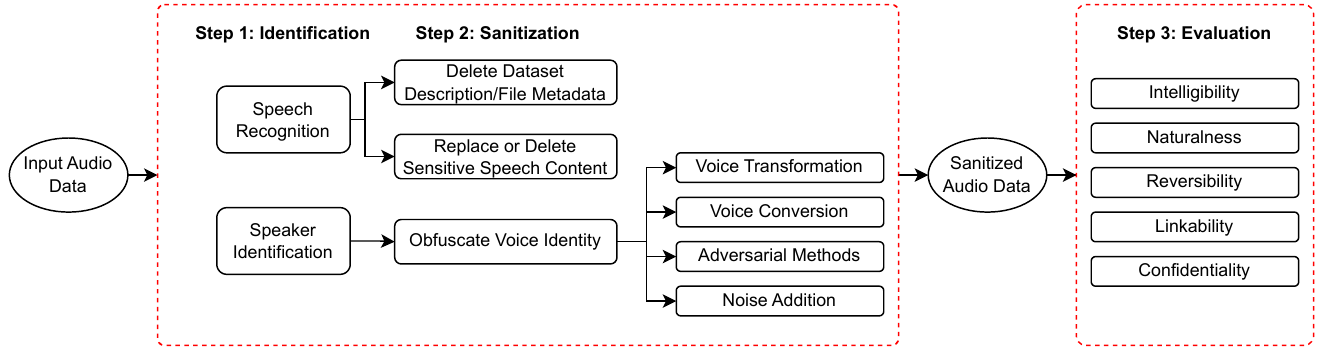}
\caption{Major Steps and Components of an Audio Obfuscation System}
\label{fig:audio_system}
\end{figure*}

Audio sanitization is a multimodal identification challenge; it entails the obfuscation of multiple items in the audio signal, which are the sensitive speech, the speaker's voiceprint, and any available dataset description or file metadata. File metadata can reveal private information about the included speakers. For example, the filename containing the keywords ``boys" and ``high school" will allow the classification of the speakers based on age, gender, and possibly geographical regions which in itself can be a privacy concern. Speech obfuscation methods can be categorized into two types: \textit{general} where the obfuscation is applied to the whole audio; and \textit{context-aware} where the algorithm applies the obfuscation to a targeted class of the audio segments \cite{vatanparvar2020adapting}. Fig. \ref{fig:audio_system} depicts the integral components and steps of an audio sanitization system. There are likely three steps to sanitize an audio file thoroughly: the identification step, the obfuscation step, and the evaluation step. In the last phase, some systems attempt to evaluate the effectiveness of the obfuscation before publicly releasing the file. Based on the use-case, obfuscated audio should have the following properties: (a) intelligibility; (b) naturalness; (c) reversibility; (d) unlinkability; and, (e) confidentiality. \textit{Intelligibility} means that the speech utterances should be clear and make sense to the listener. \textit{Naturalness} indicates that the machine-generated voice should sound as human-like as possible and possibly carry normal voice qualities such as loudness, pitch, hoarseness, etc.. Ensuring both intelligibility and naturalness of an audio enforces the \textit{indistinguishability} property, such that an attacker cannot distinguish between natural and synthetic utterances.  In some cases, \textit{reversability} of the obfuscation is desirable. When it is an undesirable property, any used conversion parameters must be kept secret. The de-identified audio should be \textit{unlinkable} to its source speaker(s) and should not reveal the voiceprints. Lastly, any \textit{confidentiality}-breaching content should be removed or replaced with generalized terms. In the following subsections, we explore speech data sanitization techniques and possible attacks against the presented methods. Finally, we conclude the section with a discussion of current trends and research directions in the  field.

\subsection{Techniques}
\label{audio_techniques}
The research topic of privacy-preserving speech data publication is a relatively new and growing field of study \cite{zhang2020enhancing,han2020voice,qian2018towards}, especially now with the prevalence of machine learning technologies. In this section, we first review voice representation used in audio processing. Then, in the following subsections, we investigate the main techniques for sanitizing speech data,  categorizing them as shown in Table \ref{tab:audio_obfuscation}: speech recognition techniques (speech-to-text conversion), voice conversion, audio perturbation, and adversarial voice. Voice modulation is also a traditional approach to hide voice qualities, such as gender and age, by altering the pitch and timbre (frequency warping). This can be achieved by applying filters to alter the spectral characteristics or using vocoders to modify the features of input speech data during synthesis. However, the original speech can be recovered using inverse filters, therefore, security is ensured by hiding the transformation parameters \cite{han2020voice}. Furthermore, the audio quality is usually low \cite{yoo2020speaker}. We do not review the just mentioned category of obfuscation in more detail because it is now considered outdated. Examples of frequency warping methods in the literature are referred to in \cite{qian2019speech, wu2013blind,magarinos2016piecewise}. 

\textbf{Voice Representations.} Raw audio signals are converted into digital representations to facilitate secure, efficient storage and processing. Human speech can be discriminated (aka, \textit{speech segmentation}) using specific spectral and temporal features that are distinguishable from other non-speech or background noise segments. These voice features have thresholds defined in controlled environments, therefore, feature extraction is impacted heavily when the training data is noisy. However, recently, Deep Neural Networks have been used in speaker recognition and identification systems, as summarized in a comprehensive survey in \cite{tirumala2016review}. Each layer in the deep (unsupervised) network learns from the previous layer, i.e., the output of one layer acts as an input to the next layer, making it well-equipped to handle noisy data efficiently. Table \ref{tab:audio_features} gives an overview of some of the features typically used in speech recognition systems. Usually, a set of features (vectors) are selected for a complete and optimal audio representation. Additionally, various schemes and models, such as, heuristic rules, Gaussian Mixture Model (GMM), Hidden Markov Model (HMM), vector quantization, neural networks, k-nearest neighbor (kNN), decision trees, and Support Vector Machine (SVM) \cite{Lu2009} use the extracted vectors to infer semantic labels associated with the audio data for the purpose of speech segmentation and voice authentication.

\begin{table}[ht]
\centering
\caption{An overview of some audio features used for speech segmentation.}
\label{tab:audio_features}
\resizebox{\columnwidth}{!}{%
\begin{tabular}{|>{\hsize=\hsize\RaggedRight}p{0.3\linewidth}|>{\hsize=\hsize\RaggedRight}p{0.6\linewidth}|} 
\hline
\textbf{Main Features} & \textbf{Description} \\ 
\hline
Short-time energy & It is a measure of the total spectral power of a frame and represents the loudness of the signal. Short-time energy can be used to distinguish between speech and music signals. \\ 
\hline
Zero-Crossing Rate (ZCR) & ZCR is defined as the
relative number of times the audio signal crosses the zero-line within a frame. It is used in characterizing sound in terms of voiced and unvoiced categories. \\ 
\hline
Mel-frequency cepstral coefficient (MFCC) & MFCC is a spectral feature obtained from the transformation of the mel-scaled spectrum using the Discrete Cosine Transform (DCT) \cite{Lu2009}. \\ 
\hline
Linear Predictive Coding (LPC) & LPC is a coding method used to represent the spectral envelope of an audio signal based on a linear predictive model. LPC is widely used in speech analysis and speech synthesis. \\
\hline
\end{tabular}}
\end{table}

\begin{table}[ht]
\centering
\caption{Main Methods of Audio sanitization in the Literature}
\label{tab:audio_obfuscation}
\resizebox{\columnwidth}{!}{%
\begin{tabular}{|l|>{\hsize=\hsize\RaggedRight}p{0.15\linewidth}|>{\hsize=\hsize\RaggedRight}p{0.2\linewidth}|>{\hsize=\hsize\RaggedRight}p{0.3\linewidth}|} 
\hline
\textbf{Category} & \textbf{Main Methods} & \textbf{Year \& References} & \textbf{Description}\\ 
\hline
Removal & Speech-to-Text & 2015 \cite{justin2015speaker} & Transcribing the speech using speech-to-text algorithms and then synthesizing a new audio. \\ 
\hline
\multirow{3}{*}{\vtop{\hbox{\strut Distortion \&}\hbox{\strut Replacement}}} & Voice Conversion & 2008 \cite{stylianou2008voice}, 2009 \cite{jin2009voice}, 2014 \cite{pobar2014online}, 2016 \cite{magarinos2016piecewise}, 2018 \cite{gao2018voice}, 2019 \cite{qian2019speech}, 2020 \cite{han2020voice,yoo2020speaker,srivastava2020evaluating} & Mapping the speaker's input voice to a different target speaker's voice to hide the voice identity by modifying the magnitude response of the vocal tract system or applying pitch, time-scale and energy modifications. \\ 
\cline{2-4}
& Noise Addition & 2019 \cite{tung2019exploiting,vatanparvar2019generative}  & Adding random acoustic noise or randomizing the audio signal to reduce its intelligibility. \\ 
\cline{2-4}
& Adversarial Perturbation & 2019 \cite{neekhara2019universal,abdullah2019practical,vatanparvar2019generative}, 2020 \cite{espinoza2020speaker} & Generation of audio samples that result in incorrect recognition by ML-classifiers. \\ 
\hline
\end{tabular}}
\end{table}

\subsubsection{Speech-to-Text Conversion}
A naïve way to protect voiceprint is by converting the speech records into texts and generating synthetic voice using text-to-speech (TTS) tools \cite{han2020voice}. In \cite{justin2015speaker}, Justin et al. (2015) designed a text-independent speaker anonymization system that disguises the speaker’s identity and preserves the intelligibility of the speech content. The system consists of two submodules, a speech recognition module and a speech synthesis module. The first module uses language-dependent diphone recognition to identify speech and convert it into a phonetic transcription. The authors proposed two different techniques for the speech synthesis module: HMM-based and the Time Domain Pitch Synchronous Overlap and Add (TD-PSOLA)-based. The system was tested using i-vector/PLDA automated speaker verification approaches; the success metric of the test is defined as the percentage of executions where speaker re-identification was successful. A limitation of this technique is that it depends on the precision of the automatic speech recognition (e.g., speech-to-text, speech-to-phoneme) and text-to-speech modules for a given language \cite{han2020voice}. The synthesized speech has a degree of utility and information loss based on the level of recognition errors. The major advantage of this scheme is that it ensures complete anonymity since the speech synthesis and speech recognition models are independent; in addition, all source utterances are converted to the same target speaker \cite{ribaric2016identification}.

Qian et al. (2019) \cite{qian2019speech} designed a speech data sanitization system with two goals in mind: speech content desensitization and voice anonymization. Speech content is sanitized via three processes: sensitive keyword determination, keyword spotting, and keyword substitution. Sensitive keywords are determined through TF-IDF-based and vote aggregation algorithms. Voice anonymization is done using a compound frequency warping function. 

\subsubsection{Voice Conversion}
Voice conversion (VC) has become a major research topic in recent years. It refers to the mapping of a speaker’s voice to a target voice, altering the speaker identity, without changing the linguistic content. The technique is used in various tasks, e.g., communication aids for the speech-impaired, voice dubbing, voice disguise, to cite a few \cite{sisman2020overview}. Voice disguise refers to evading voice authentication and verification systems through missed detections (aka, \textit{speaker de-identification}) or false acceptance (aka, \textit{spoofing} or \textit{impersonation}) \cite{alegre2014evasion}. There are numerous studies about VC in the literature. In the following, we review a selected few that address major improvements. For a more comprehensive overview of voice conversion, it is suggested to refer to a recent and very useful survey by Sisman et al. (2020) \cite{sisman2020overview} that reviews the challenges of voice conversion and its evolution from statistical modeling to deep learning. Deep learning and adversarial learning revolutionized voice conversion in recent years; they allowed the mapping of more speaker characteristics and improved the voice quality and similarity to the target speaker tremendously. Another highly-cited survey on the topic, by Mohammadi et al. (2017), is presented in \cite{mohammadi2017overview}. 

Voice conversion modifies the speaker characteristics, such as age and gender, of a given speech regardless of the spoken language. VC strategies are based on: (a) simple codebook-based conversion; (b) linear estimators; or, (c) neural network models \cite{gao2018voice}. The conversion process typically has three main steps: speech analysis to extract voice features, speech mapping, and speech synthesis. 
VC methods can be text-dependent (\textit{parallel}) or text-independent (\textit{non-parallel}). Text-dependent approaches require a perfectly time-aligned parallel dataset of the source and target speakers speaking the same text. The conversion can also be \textit{one-to-one}, where each source speaker is mapped to a different target speaker, \textit{many-to-one}, where all source speakers are mapped to the same target speaker, or \textit{many-to-many} where the model learns the mapping of different speech attributes. A drawback of one-to-one conversion is that it is easy to identify which of the converted utterances belong to the same source speaker \cite{he2020voice}. However, there is an advantage of having a single target speaker, since the collection of additional recordings for the target speaker training set is not required.  

Early methods of VC were based on spectral mapping of parallel training data using various models, such as vector quantization, GMM, non-negative matrix factorization, and artificial neural networks, to name a few \cite{sisman2020overview}. Jin et al. (2009) \cite{jin2009voice} proposed four reversible transformation strategies (using a secret key) to generate intelligible voices that disguise the speakers’ identities. In the learning phase, pairs of utterances for the exact spoken text by the source and target speakers are used. Then, in the transformation phase, all source voices are transformed to the same target voice. In \cite{jin2009speaker}, the same authors expanded on their previous work to achieve distinguishability when multiple speakers are speaking simultaneously in the same recording. There is a higher probability that the resulting de-identified voices can sound the same when many speakers are present.  In both works, the authors used automatic speaker identification approaches, a GMM-based and a phonetic approach for testing the effectiveness of the VC methods. Generally, GMM-based identification systems are more susceptible to voice transformations than phonetic-based identification, which relies on gathering linguistic identifiers (i.e., idiosyncrasies) from the speech \cite{jin2009voice}. 

Obtaining parallel recordings presented a challenge, especially in terms of time aligning the utterances. Therefore, studies shifted their focus to non-parallel voice conversion for speaker identity protection. However, the conversion mapping is more challenging to obtain with non-parallel methods, therefore, they are usually more limited in terms of the quality and conversion effect compared with parallel methods \cite{kameoka2018stargan,fang2018high}. Kinnunen et al. (2017) \cite{kinnunen2017non} transformed the speaker characteristics in the i-vector space of an input speech towards a target speech using the probabilistic linear discriminant analysis (PLDA) model, where the i-vector of an utterance is a fixed low-dimensional representation of its GMM. Fang et al. (2019) \cite{fang2019speaker} proposed a speaker anonymization system that extracts the x-vector (representing speaker identity), sequence of phonemes (representing linguistic content), and the fundamental frequency (F0) from the input speech. The x-vector is anonymized by replacing it with the average of a set of pre-trained x-vectors. The output is synthesized from the combination of F0, the extracted linguistic content, and the anonymized x-vector.

Variational autoencoders (VAEs), consisting of encoder and decoder networks, are used for one-to-one voice conversion in \cite{hsu2016voice}. A probabilistic variant of VAEs, Conditional Variational Autoencoders (CVAEs), was used in \cite{saito2018non}. CVAEs feed an auxiliary input $c$ into the decoder, where $c$ is the attribute labels of the acoustic features to be converted. The network learns the mapping conversion of an attribute of source speech to a target attribute according to $c$. For testing CVAEs, $c$ must be known in advance and fed into the encoder of the trained CVAE. However, a limitation of VAE-based methods is that the converted speech tends to be of poor quality. In \cite{hsu2017voice}, VAEs were combined with GANs (VAE-GAN) to enhance the quality of the converted speech, where the decoder of the VAE is the generator of the GAN \cite{yoo2020speaker}. The GAN discriminator derives the adversarial loss and incorporates it into the training loss to make the decoder output’s indistinguishable from natural speech. In \cite{tobing2019non} and \cite{yook2020many}, the VAE-GAN is extended to include cycle-consistency loss, known as CycleVAE-GAN. The encoder extracts the speech content (latent vector) from the source speech. Then, the target speaker identity vector and the latent vector are fed into the decoder to generate the converted speech. Finally, the converted speech goes back through the encoder with the source speaker identity vector to recover the original input. After optimizing the VAE module, the Wasserstein  GAN network trains the CycleVAE-GAN model. The cyclic conversion drives the non-parallel training of voice conversion and improves the voice quality further. Another method based on CycleVAE-GAN was proposed by Yoo et al. (2020) \cite{yoo2020speaker}. The authors generated anonymized identity vectors to be fed into the decoder instead of the conventional one-hot encoded vectors using various approaches; uniform and non-uniform value vectors. Their method achieved a speaker identification accuracy of 0.1-9.2\% and speech recognition accuracy of 78.2-81.3\%. Espinoza et al. (2020) \cite{espinoza2020speaker} proposed a speaker-characteristics-invariant method based on Domain-Adversarial Training (DAT) and autoencoders. The goal is to hide the speaker’s identity to a greater extent by suppressing other characteristics present in the speech, such as gender and accent. The encoder-decoder autoencoder branch generated an anonymized x-vector. Then, in each adversarial branch of the Domain Adversarial Neural Network (DANN), the speaker characteristics present in the latent features are removed via backpropagation.

In \cite{gao2018voice}, Gao et al. (2018) used Generative Adversarial Nets (GANs) to produce natural-sounding converted voice. It follows the concept of voice impersonation, where the resultant voice mimics a target speaker’s identifiers and his style (temporal pitch, speaking rate, energy patterns) and voice quality (nasality, roughness, breathiness). Voice quality and style, although identifiable, are not quantifiable and therefore presented a challenging research topic. Both the generators and the discriminators of the GAN are modeled using DNNs. As the learning process is based on discrimination between data with the desired features instead of input-output mapping, a parallel corpus of utterances is not required (i.e., it is a text-independent approach). Given the spectral representations of the speakers’ voices and sufficient training data, the model learns to transform the target speaker’s voice style and quality. The training process is also independent of the linguistic content of the utterances and outputs a spectrogram reconstructed from the audio signals using the Griffin-Lim method. Kaneko et al. (2017) \cite{kaneko2017parallel} designed a non-parallel voice conversion named Cycle-GAN, cycle-consistent GAN. The generator network learns one-to-one acoustic feature mapping and its inverse mapping. The discriminator uses training loss, combining adversarial loss and cycle consistency loss, to distinguish converted speech from original speech. For many-to-many conversion tasks, a naïve way is to train different mapping and inverse mapping pairs for all attributes.  However, this is an ineffective approach since there are common latent features shared across the attribute domains. Also, parameter training becomes challenging as the number of attribute domains increases. During testing, the same source attributes used in training must be used. Kameoka et al. (2018) \cite{kameoka2018stargan} proposed an extension to CycleGAN-VC to allow many-to-many voice conversion using only a single StarGAN (an encoder-decoder type GAN) network. The model is applicable for real-time implementations and also uses a smaller training sample. StarGAN-VC offers the benefits of both CVAE-VC and CycleGAN-VC. The generator is trained using an adversarial loss (derived by the discriminator) to learn the many-to-many attributes mapping according to an auxiliary input $c$. 
The converted utterances are indistinguishable from real speech and have higher quality than VAE-GANs. Also, StarGAN-VC does not require knowing $c$ of the input speech at test time. 

In the previous examples, StarGAN and Cycle-GAN, the conversion is limited by the speakers present in the source training set. Zero-shot conversion methods were proposed to mitigate this limitation \cite{chou2019one,qian2019autovc}. A zero-shot VC converts an input using only a single utterance of a possibly unknown speaker (i.e., not seen during training) without finetuning. The VC scheme based on autoencoders consists of a content encoder, which extracts the content information from the first input ($t$), a speaker encoder, which embeds the speaker characteristics from the second input ($x$) as a latent vector, and a decoder, which produces a spectrogram combining the outcome of both encoders \cite{huang2021defending}. 

Voice anonymization via VC requires disentangling linguistic content from speaker identity while keeping prosody and naturalness intact. Diffusion-based VC brings stronger generative priors and flexible conditioning, helping match target timbre without copying a real individual. As in images, the aim is to suppress identity cues while preserving intelligibility and paralinguistic utility. DiffVC+, the work by Huang Fan et al. (2024) \cite{huang2024diffvc} adapts diffusion-based VC to anonymization by combining a self-supervised content encoder with a speaker encoder and an embedding generator for non-existent pseudo-speakers. Two variants (light/decoupled) address edge vs. server constraints, with evaluations showing competitive privacy (ASV/EER) together with improved intelligibility and naturalness.

Most of the studies previously discussed do not provide measurable and provable privacy guarantees. For this purpose, Han et al. (2020) \cite{han2020voice} designed a formal privacy metric for a model-level framework, called $\epsilon$-voice-indistinguishability, extending metric privacy (a generalized notion of differential privacy). For each utterance, the speech content and voiceprint are separated, where x-vectors (DNN-based embeddings) are used to represent the voiceprint component. The mechanism randomly perturbs the input x-vector by selecting an x-vector from a dataset according to certain probability distributions. This method guarantees that an attacker cannot identify the original voiceprint from other voiceprints regardless of the attacker’s knowledge. The degree of indistinguishability ($\epsilon$) depends on the distance (i.e., similarity) between two x-vectors. For objective evaluation, the similarity between the original and converted voice was tested using the PLDA model and the difference using Mean Squared Error. Speech recognition and character error rate were used to assess the utility of the speech. 

\subsubsection{Noise Addition}
In statistical data security, perturbation adds some noise or performs randomization to the data to achieve confidentiality and privacy. This is also the case with voice perturbation. In \cite{chen2008audio}, Chen et al. (2008) proposed a method that preserves environmental sounds (e.g., gunshots, yelling, glass breaking, crying, falling objects) while significantly reducing speech intelligibility and word recognition rate to 7\%. Their method is useful for applications  interested in processing non-speech sounds, for example, in health applications that analyze the user's breathing or coughing or for Voice Activity Detection applications (e.g., detecting whether the user is in a car) \cite{liaqat2017method}. The entire speech signal is considered sensitive to mitigate user privacy concerns. Their key idea is to apply a \textit{vowel replacement} algorithm that renders speech unintelligible and non-reconstructable from non-speech sounds \cite{chen2008audio}. The  algorithm identifies regions with human voices using Linear Predictive Coding (LPC) coefficients and LPC gain. The speech filtering process relies on the fact that vowels play a prominent role in forming and recognizing words.  Thus, replacing vowels with noise will significantly reduce the intelligibility of speech. The speech segments are identified by detecting if the estimated pitch is within range for an adult and the voicing ratio is above a given threshold. The speech is obfuscated by replacing the input LPC coefficients with random coefficients from a set of pre-recorded vowel sounds, resulting in randomized and unintelligible speech (while maintaining the original gain and pitch information). 

Tung et al. (2019) \cite{tung2019exploiting} presented an application-level \textit{sound masking} \cite{caprolu2020short} method that adds a masking sound to prevent unauthorized eavesdroppers from understanding the speech content of the recorded audio. The masking sound is composed of two components, the masking interference and the masking noise. The masking interference comprises several pre-recorded spoken sentences, and the noise is a Gaussian noise filtered using a 16kHz low-pass filter. Only the intended receiver can remove the masking sound to retrieve the original signal. Vatanparvar et al. (2019) \cite{vatanparvar2019generative} used Generative Adversarial Network (GAN) to segment human speech in a recording and replace it with a random speech. The generated audio samples cannot be traced back to any private information. More importantly, the generated audio matches the distribution of human speech and so is indistinguishable from human speech and preserves naturalness. Vatanparvar et al. (2020) \cite{vatanparvar2020adapting} developed more accurate speech segmentation in the presence of noisy settings on smartphones. Their method is characterized by a reduced memory footprint and execution time. 

Abdullah et al. (2019) \cite{abdullah2019practical} developed a hidden command model for a black-box attacker (i.e., an ignorant attacker). The system generates noise-like audio that is resilient to identification by human listeners but not by ML-classifiers. That is, the perturbed audio is transcribed the same as the original audio by ML-based recognition systems. A known limitation is that the attack audio will have reduced effectiveness under noisy conditions, as noise impacts the accuracy of the ML detectors. 
\subsubsection{Adversarial Perturbation}
Adversarial Perturbations are (generally subtle and human-imperceptible) noise introduced in audio samples to deceive ML-based adversaries, namely automated speech recognition (ASR) and automated speaker verification (ASV) systems, into incorrectly recognizing speech content or speaker identity. The perturbations can be applied on the waveforms, spectrograms, or MFCC features of the audio \cite{huang2021defending}. In malicious contexts, adversarial perturbations are referred to as \textit{adversarial attacks} or \textit{hidden voice commands} \cite{neekhara2019universal}. This subsection highlights some notable studies that look into obfuscating speech content and obfuscating speaker identity using adversarial perturbations.

Huang et al. (2021) \cite{huang2021defending} proposed a novel targeted adversarial attack that prohibits VC models from producing voices with similar speaker characteristics to those of a defended speaker. The second design goal is that the added noise should be human-imperceptible, i.e., perturbed utterances are indistinguishable from original ones. The authors applied three different attack approaches (end-to-end, embedding, and feedback) on zero-shot encoder-decoder-based VC models. The perturbations, subjected to a constraint $\epsilon$ to make it subtle, are added to the output of the decoder (end-to-end attack) or the speaker encoder (embedding attack). Meanwhile, the feedback approach combines the previous two attacks, where the spectrogram of the defended speaker is perturbed and fed back to the speaker encoder. The effectiveness of the three attacks in terms of the two design goals was evaluated and compared for white-box and black-box scenarios. Based on the results, it was noted that the embedding approach produced the best results and is more time-efficient. 

Universal perturbations are more efficient and applicable in real-time than per-instance perturbations that output specific mistranscriptions. Neekhara et al. (2019) \cite{neekhara2019universal} aimed to find an audio-agnostic universal perturbation that causes ASRs to produce erroneous transcriptions in a white-box setting. The authors formulated the algorithm as an optimization problem that iteratively builds a minimum perturbation vector that pushes most data points in the training dataset outside their correct transcription region. The success metric requires the character error rate to be higher than a predetermined threshold. In the cited case, the adversarial transcription is incomprehensible, and the original transcription cannot be inferred. The generated universal perturbation is transferable; the output of the network can be used to deceive other network architectures.  The results also show that generated perturbations are quasi-imperceptible and achieve a higher success rate than a random noise perturbation having the same distortion magnitude.

\subsection{Attacks and Countermeasures}
As stated and depicted earlier in Fig. \ref{fig:audio_system}, there are two features to obfuscate in audio: speech content and speaker’s voiceprint. Therefore, an attack against audio obfuscation is expected to utilize speech recognition systems to learn confidential content, possibly containing hints that could deduce the speaker’s identity. Or, the attack would use speaker identification systems to identify the voiceprint and its corresponding speaker. Generally, the attacks could be white-box (informed attacker) or black-box (uninformed attacker). A white-box attacker has full access to the obfuscation model, the architecture, and all trained parameters. However, a black box attacker cannot directly access the model to be attacked and its parameters, and the architecture is unknown.

\textbf{Recovery Attacks.} In this category of attacks, we specifically consider voice transformation schemes. Srivastava et al. (2020) \cite{srivastava2020evaluating} proved that simple VTLN-based VC methods, with appropriate target selection strategy, provide protection against attackers with partial knowledge. An attacker that has extensive knowledge of the type of transformation and its parameters can reverse the transformation to recover the original audio \cite{srivastava2020evaluating}. An adversary could utilize DNN-based speech recognition systems to infer conversion parameters. They considered three types of attackers classified based on their knowledge (informed, semi-informed, and ignorant).

Qian et al. (2019) \cite{qian2019speech} used a \textit{reversing attack} against invertible frequency warping functions. The attacker can reverse VTLN and partially recover the original frequency axis from the warped one. As a countermeasure, the conversion parameter is chosen randomly from the proper range. Then, the speeches from the same user would turn out sounding differently. The authors prove that applying voice conversion twice to a speech with two conversion parameters successively, yields exactly the same output as applying the voice conversion once with the sum of the conversion parameter which has much weaker distortion strength than the originally sanitized speeches. Consequently, an informed attacker can achieve better recognition accuracy using this type of \textit{reversing attack}.

\textbf{Re-identification Attacks.} Re-identification attacks aim to learn the identity of the source speaker. There are two ways this objective can be achieved: using weaknesses in the content obfuscation or weaknesses in the voice conversion. Another name for this attack is the \textit{linkage attack} since it links the content or voiceprint to a speaker (or speaker set) \cite{srivastava2020evaluating}. In the first attack, attackers can determine the speaker’s real identity by analyzing the speech content and matching quasi-identifiers or leaked PII with attributes that were publicly acquired in advance before audio obfuscation \cite{zhang2020enhancing}.  Using the disclosed information, the attackers can narrow the search range and get a candidate subset from a speech data dataset. The success metric of the attack depends on the randomness and the size of the candidate list. Attackers can also exploit inconsistencies in the applied voice conversion methods, specifically between the audio content and the speaker’s voice. For example, if the speaker’s voice was changed from child to adult to mask the age identity, but the speech content contains contradicting references inferring a different age group, this would leak information about the obfuscation technique. As a result, the attackers could search a source dataset and obtain a filtered candidate list matching the disclosed age criteria. Zhang et al. (2019) in \cite{zhang2020enhancing} designed an ML model that learns the correlation between speech content and speakers voice while achieving an optimal data sanitization by weighing the risk of privacy loss and data utility loss. 
A possible countermeasure to consider is further perturbing the obfuscated audio, making it more resilient to speech recognition. The transcriptions would contain indecipherable or misleading phrases. If the phrases are indecipherable, no knowledge could be gained. Alternatively, misleading phrases would intentionally lead to incorrect re-identification.  
The second type of the attack attempts to identify the speaker from the voiceprint rather than the content. Re-identification attacks use automated ML-based speaker identification systems to determine the speaker with the highest similarity measure from the dataset. The success of this type of attacks highly depends on the knowledge of the attacker \cite{han2020voice,srivastava2020evaluating}. 

He et al. (2020) \cite{he2020voice} proposed two new attacks on GMM-based (parallel and many-to-one) voice conversion. The typical speaker re-identification attack computes the similarity score between a converted utterance and the utterances of possible source speakers, and selects the source speaker with the highest similarity score. The first new attack trains a VC model that transforms the de-identified utterance into an utterance of the source speaker. The second attack assumes a stronger attack model; the attacker has an utterance of the target speaker. If the target speaker's utterance is part of a public text-to-speech dataset, the re-identification rate increases further. The similarity score between the resulting converted utterance and the speaker’s original (i.e., non-converted) utterance is computed; the speaker with the highest similarity score is selected. The attacks achieved a higher speaker re-identification rate (19\% - 26\%) than only applying the typical re-identification attack (4\% success). Following a successful re-identification, the attacker can extract private information from the speech and construct the speaker’s profile. 

\textbf{Inference Attacks.} It is crucial to detect and defend against voice conversion artifacts, especially since VC can be used to evade authentication systems. In this attack, an attacker can infer that the audio sample is either transformed or synthesized using VC methods. Synthetic and converted voice can be detected using estimates of dynamic speech variability and natural speech phase \cite{alegre2014evasion,wu2015spoofing}. Depending on the VC approach, the converted speech may exhibit less dynamic variability and limited natural phase compared to natural speech. In \cite{alegre2014evasion}, the authors applied local binary pattern (LBP) analysis of speech spectrograms for finding the natural spectro-temporal variability as a generalized approach  for detecting speaker obfuscation. In \cite{wu2013blind}, the authors used SVM classifiers on a training database composed of an original voice set and a disguised voice set using frequency warping. According to $K$ different disguising factors, the disguised voice set is divided into $K$ subsets. The model outputs $K$ SVM classifiers, and each of them is used to identify whether a testing voice is disguised or not. Similar techniques can be used to distinguish adversarial audio from natural speech. 

\subsection{Discussion}
\textbf{Applications.} As stated earlier, audio obfuscation has several applications in video surveillance, voice-based verification and lifelog and telephone-based services. With the growth of deep learning, particularly adversarial learning, the interest in the field is growing as attacks are boosting in complexity and the need for more natural and human-like voices is ever-increasing.

\textbf{Challenges and Open Research.} Some of the existing speaker de-identification challenges are related to attaining measurable and provable privacy-utility balance, mitigation of background noise, and achieving intelligibility and distinguishability in the presence of multiple overlapping speakers. Recent efforts are made to achieve the naturalness of the synthetic voice using style-based adversarial learning. This includes impersonating a target speaker's voice quality and speech patterns and preserving his emotional expression. Some studies also look into gender de-identification in speech data. In terms of future research directions, we speculate that voice conversion methods are here to stay in the foreseeable future, and investing in adversarial learning has a higher impact than optimizing other traditional or simple obfuscation methods. Also, since adversarial perturbation is an evolving paradigm in deep neural networks, it will remain an indispensable protection method against automated speaker identification.  

\textbf{Key Properties.} The key properties to consider when selecting an optimal obfuscation methods are: (a) \textit{naturalness and intelligibility}: the semantics of the speech should be preserved and voice quality should be acceptable; (b) \textit{(ir)reversibility}: (in)ability to undo the transformation; (c) \textit{robustness and security}: thwarting machine-based speech and speaker recognition including linkage and inference attacks and/or human recognition; and, (d) \textit{provably secure}: to provide measurable security metrics. The first three properties are addressed in most audio obfuscation methods. However, provable security remains an open research topic. 

\textbf{Review of Methods.} We have considered four of the main approaches for speech data sanitization: voice transformation (also referred to as voice conversion), diphone recognition-based (i.e., speech-to-text conversion), perturbing voice through noise addition, and adversarial perturbation to deceive neural networks. In Table \ref{tab:audio_techniques_overall}, we rate the utility and security of the obfuscation methods. Traditional speech-to-text methods produce unnatural synthetic voices; however, when combined with voice conversion, the output can preserve high utility. In terms of security level, the voiceprint and sensitive utterances are completely removed from the audio, so it is rated high. Depending on the type of the voice conversion, the security level ranges from medium to high where some are susceptible to inference and linkage attacks. With current generative models, the utility of the generated voices is extremely high compared to traditional methods. Adversarial perturbations deceive automated attacks, and so have no impact on the utility or on human recognition. Noise addition methods distort the audio for human recognition, however some algorithms can be applied to cancel the added noise. 

\begin{table*}[ht]
\centering
\caption{The Main Audio Protection Techniques in the Literature (L: Low, M: Medium, H:High)}
\label{tab:audio_techniques_overall}
\resizebox{0.7\linewidth}{!}{%
\begin{tabular}{|c|l|l|l|c|c|} 
\hline
\multirow{2}{*}{\textbf{\textbf{Category}}} & \multirow{2}{*}{\textbf{\textbf{Main Methods}}} & \multicolumn{2}{c|}{\textbf{Security Level}} & \multirow{2}{*}{\textbf{Utility}} & \multicolumn{1}{l|}{\multirow{2}{*}{\begin{tabular}[c]{@{}l@{}}\hfil \textbf{Reverse}\\\textbf{Conversion}\end{tabular}}} \\ 
\cline{3-4}
 &  & \textbf{Human} & \textbf{Machine} &  &  \\ 
\hline
Removal & Speech-to-Text & \multicolumn{1}{c|}{H} & \multicolumn{1}{c|}{H} & \multicolumn{1}{c|}{M} & \multicolumn{1}{c|}{$\times$}\\ 
\hline
\multirow{4}{*}{\begin{tabular}[c]{@{}l@{}}Distortion\textbackslash{}\\ Replacement\\\end{tabular}} & \begin{tabular}[c]{@{}l@{}}Voice Conversion\end{tabular} & \multicolumn{1}{c|}{M-H} & \multicolumn{1}{c|}{M-H} & \multicolumn{1}{c|}{H} &  \multicolumn{1}{c|}{\checkmark} \\ 
\cline{2-6}
 & Noise Addition & \multicolumn{1}{c|}{H} & \multicolumn{1}{c|}{L} & \multicolumn{1}{c|}{L} & \multicolumn{1}{c|}{$\times$} \\ 
\cline{2-6}
 & Adversarial Perturbation & \multicolumn{1}{c|}{L} & \multicolumn{1}{c|}{H} &  \multicolumn{1}{c|}{-} & \multicolumn{1}{c|}{$\times$} \\
\cline{2-6}
 & \begin{tabular}[c]{@{}l@{}}Diffusion-based\end{tabular} 
   & \multicolumn{1}{c|}{M-H} 
   & \multicolumn{1}{c|}{H} 
   & \multicolumn{1}{c|}{H} 
   & \multicolumn{1}{c|}{\checkmark} \\
\hline
\end{tabular}}
\end{table*}

\section{Text Document Sanitization} 
\label{document_obfuscation}
Data can take either form: structured (e.g., databases) or unstructured (e.g., audio and video). Plain textual data is the most common form of unstructured data found in documents, emails, social media posts, financial records, medical records, or webpages. In this section, we focus on the redaction of free-text documents. We note that documents may also contain images; however, the topic of image sanitization is covered separately in Section \ref{image_obfuscation}.

Documents are publicly released for a variety of reasons. Nonetheless, the sensitive information contained therein must be kept confidential and protected for security and ethical reasons. Governments release previously classified documents to the public to comply with laws such as the Freedom of Information Act \cite{ateniese2005sanitizable}. In addition, organizations may share their text data with potentially untrusted third parties for mining and storage. For instance, textual data analysis aids the advancement of research in recommender systems, information retrieval systems, and the medical and pharmaceutical industries. In fact, research on de-identification was primarily driven for the privacy protection of medical records \cite{ribaric2016identification}. Official regulations and data protection laws (e.g., HIPAA and GDPR) were developed to ensure compliance and guide the process of desensitizing documents. 
The HIPAA (Health Insurance Portability and Accountability Act) mandates that medical records are sanitized before publication. It identifies a total of eighteen private attributes that must be removed to achieve ``Safe Harbor" de-identification, such as: (a) names; (b) addresses; (c) social security numbers; (d) telephone and fax numbers; (e) vehicle identifiers; (f) medical record numbers; (g) URLs; (h) IP addresses; (j) biometric identifiers; (k) all dates (except year), etc. \cite{mamede2016automated}. Meanwhile, the GDPR (General Data Protection Regulation) is an EU bill that obligates data controllers and processors to implement the necessary protection safeguards, including obtaining explicit consent from data subjects and anonymizing personal identifiers \cite{hassan2019automatic}. 

{\em Document redaction} or document sanitization is the process of identifying sensitive information and removing or hiding such information. Both terms are often used interchangeably in the literature. However, redaction is specifically the blacking out of text in the document \cite{domingo2020privacy}. As per sanitization, it can utilize various methods other than blocking, such as generalization, random substitution, etc.. Document sanitization should also include removing the file metadata that leaks information, such as user comments, bookmarks, or other file attributes. Traditionally, the redaction of printed material was a manual and tedious endeavor. Trained experts recognized sensitive information and then used black marker pens, opaque adhesive tape, or cutters to remove the paper's text \cite{lopresti2004quantifying}. Today, robust commercial software solutions for document redaction are available and built-in by default into word processors and PDF viewers such as Microsoft Word and Adobe Acrobat Pro \cite{raskind2019primer}. The software works by entirely removing the selected sensitive terms and replacing them with black rectangles or placeholder text according to user-specified configurations. Still, the major pitfall untrained users incur is when they place a rectangle overlay without removing the text underneath, making it available upon copy-paste \cite{dixon_2019}. Most importantly, the alterations done in the anonymization process should not render the document's context unintelligible or unreadable and, at the same time, should ensure minimum information leakage. This intricate process is known as privacy-utility balance. The sequel of the section summarizes text sanitization and attacks that result in information leakage. We conclude with a discussion of the research domain and a detailed comparison among the summarized techniques. 

\subsection{Techniques}
Redaction is a two-step process: detecting sensitive attributes and replacing those identified attributes. We discuss both steps separately below. 

\subsubsection{Detecting Sensitive Terms}
Due to the increasing volume of documents, manual redaction has become impractical. There has been plenty of research and proposed models to semi-automate or automate the process of detecting sensitive information. Automated sanitization of unstructured text is relatively an underdeveloped field of study compared to structured text \cite{hassan2019automatic}. Confidential information could be grouped into two classes, namely, identifiers and quasi-identifiers. An identifier is an attribute whose value directly links to an individual (e.g., name, IP address, social security number). A quasi-identifier is an attribute that alone does not allow re-identification but a combination of which (e.g., date of birth and postal code) may be used to reveal the identity of a subject. Identifying sensitive terms in unstructured text, specifically quasi-identifiers that are more diverse and spread out, is more complex since there is no fixed list of attributes as in databases. Semi-automated or automated methods of detecting sensitive entities are presented in Table \ref{tab:text_detection_methods}. They are classified into the following categories: (a) pattern matching and dictionary-based lookup; (b) information-theoretic; (c) Named Entity Recognition (NER) using statistical methods (e.g., Hidden Markov Model and Conditional Random Fields); and, (d) NER based on ML or Natural Language Processing (NLP) techniques \cite{vasudevan2014review,hassan2018anonymization}.

\begin{table*}[ht]
\centering
\caption{The Main Methods of Sensitive Text Detection in the Literature}
\label{tab:text_detection_methods}
\resizebox{0.7\linewidth}{!}{%
\begin{tabular}{|l|l|} 
\hline
\textbf{Method} & \textbf{Reference \& Year} \\ 
\hline
Pattern matching, and dictionary lookups & 1996 \cite{sweeney1996replacing}, 2008 \cite{chakaravarthy2008efficient}~ \\ 
\hline
Information-theoretic approaches~ & 2012 \cite{sanchez2012detecting}, 2013 \cite{sanchez2013automatic}  \\
\hline
NER using statistical models & 2018 \cite{hassan2018anonymization,madan2018redaction} \\
\hline
NER using machine learning & 2015 \cite{mcdonald2015using}, 2018 \cite{madan2018redaction}, 2019 \cite{hassan2019automatic} \\ 
\hline
LLM-based zero-shot detection \& substitution & 2023 \cite{albanese2023text} \\
\hline
\end{tabular}}
\end{table*}

\textbf{Pattern Matching.} One of the first proposed systems by L. Sweeney (1996) \cite{sweeney1996replacing} for automated redaction is named scrub. Scrub provided multiple detection algorithms running in parallel, each one dedicated to a specific type of entity (e.g., names, phone numbers, locations), that used pattern matching and dictionary-based lookup. The sensitive text is replaced with pseudo-values. The format of the pseudo-value is based on the recognized class of the sensitive term. The system achieved a high detection rate, 99-100\%, of personal identifiers, but it could not efficiently detect quasi-identifiers. Manual intervention was also needed to assess the extent of residual identifiers, including matched words unrelated to the data subject that were preserved to maintain the text's utility (i.e., false positives). Similarly, the authors in \cite{chakaravarthy2008efficient} developed a system named ERASE which relied on a public database of entities (e.g., locations, person names, diseases). Each category contained a set of related terms where some may be tagged as protected. Protected entities are then removed from the document. 

\textbf{Information Theory.} Using information theory principles, Sanchez et al. (2012-2013) \cite{sanchez2012detecting,sanchez2013automatic} presented a framework that evaluates the sensitivity degree of entities using the notion of information count (IC). Formally, the IC of a term is inversely proportional to the probability of encountering it in a corpus. Specifically, less common words carry higher IC and are considered more sensitive than commonly used ones. The detection of sensitive terms is based on finding nouns or noun phrases corresponding to concrete concepts such as diseases or individual names. The nouns are detected using NLP tools such as sentence detection, tokenization, part of speech tagging, and syntactic parsing. The authors computed the IC by finding the number of times a specific term is queried on web search engines. If the IC value exceeds a given threshold, the term is considered sensitive and is replaced with its generalized class label to retain a high level of utility. 

\textbf{Named Entity Recognition (NER).} Unstructured text data must first be refined and represented in a well-defined format before the de-identification task. NER is the process of locating and classifying private entities, such as names, locations, dates, and phone numbers,  within a document \cite{mamede2016automated,hassan2018anonymization,hakkani2006document}. Hassan et al. (2018) \cite{hassan2018anonymization} used NER based on Conditional Random Fields (CRF), a statistical modeling method often used in pattern recognition. Their proposed system model consisted of three steps: token generation; feature extraction; and, the CRF tagger. First, sentences are split into tokens. The stem, shape, length, and part-of-speech tag for the words are then identified.  Finally, using the extracted features, a tag is decided for each word based on probabilistic measures. Madan et al. (2018) \cite{madan2018redaction} compared the performance of two NER models, bi-directional LSTM and CRFs, for the prediction of PII in health records. CRFs are discriminative undirected probabilistic graphical models that require manual feature engineering. As per LSTMs (a variant of recurrent neural networks RNNs), they extract context information from word representations without extensive prior knowledge about the dataset. The pre-processing phase includes converting sentences into a stream of tokens; later, meaningful sentence boundaries are detected using rule-based or machine learning-based methods. The next step is tagging the named entities using an appropriate scheme, then feeding them into the de-identification algorithm. For the CRF model, the following features are used: token windows, parts of speech (PoS) tags, structurally special features (e.g., capitalization, punctuation, contained digits), word representation features, token labels, and word shapes. The authors showed that LSTM models achieve better F1 scores than CRF-based models; this is attributed to the powerful fitting capabilities of LSTMs as compared to CRF's standard log-linear models \cite{madan2018redaction}. However, the authors indicate that CRFs can still outperform LSTMs when the input training data is limited and more structured.

\textbf{LLM-based zero-shot redaction and substitution.}
Recent work leverages large language models (LLMs) to flag and sanitize sensitive spans without domain-specific training. The approach identifies low-probability tokens under the model’s contextual distribution (high information content), redacts them, and then substitutes semantically close alternatives to preserve coherence and utility. Unlike fixed-label NER pipelines, zero-shot LLMs can capture out-of-schema identifiers and quasi-identifiers across domains and languages. The work by Albanese et al. (2023) \cite{albanese2023text} presents a generic, configurable implementation with a privacy threshold driving the privacy–utility trade-off.

In \cite{mcdonald2015using}, Mcdonald et al. (2015) presented an approach that uses the sensitivity load of Part-of-Speech (POS) N-grams to identify sequences of sensitive text. Text classifiers (e.g., using deep neural networks or Random Forest) can then use the POS N-grams as features. Their work does not assume that all named entities are sensitive. It is also able to identify sequences of sensitive text and not only individual terms. Hassan et al. (2019) \cite{hassan2019automatic} proposed an unsupervised method that does not require classifiers pre-trained on manually tagged datasets, not restricted to predefined types, and is language agnostic. Neural networks can extract vectors from words through the continuous bag of words or skip-gram approaches \cite{hassan2021utility}. In the skip-gram method, the neural network predicts neighboring words based on the current term. The authors chose to train a skip-gram word embedding model to extract word vectors, capturing the semantic relationship between words. The disclosure risk is directly related to the semantic relatedness between textual terms co-occurring in the document; hence the entities whose vectors are similar to a private entity's vector are removed. 

Beigi et al. (2019) \cite{beigi2019privacy} proposed a text representation learning framework that preserves both the utility and the privacy of the learned representations, which are to be shared publicly. The framework uses an auto-encoder to extract the text representations and then adds Laplacian (random) noise based on $\epsilon$-differential privacy. The noise added to the original representations minimizes the risk of re-identification, leakage of private attributes, and recovery of the original text. It also guarantees that an adversary cannot predict whether the shared dataset contains a specific target representation. However, noise addition can adversely impact the utility of the text; therefore, the framework utilizes two semantic and private attribute discriminators to automatically learn the optimal noise amount. 

\subsubsection{Sanitization of Sensitive Terms} Anonymization methods of structured data generally provide provable privacy metrics based on differential privacy or $k$-anonymity models. These privacy models have been adapted also to unstructured data, additionally, the de-identification process is commonly achieved by deleting identifiers and replacing quasi-identifiers with surrogate values. In Table \ref{tab:text_sanitization}, some ways of sanitizing unstructured text in documents are listed. Moreover, examples and comparison of the sanitization techniques are shown in Table \ref{tab:text_sanitization_comparison} in the discussion section.

\begin{table}[ht]
\caption{The Main Text Sanitization Techniques in the Literature}
\label{tab:text_sanitization}
\centering
\resizebox{\columnwidth}{!}{%
\begin{tabular}{|l|l|l|} 
\hline
\textbf{Category} & \textbf{Technique} & \textbf{Reference \& Year} \\ 
\hline
Distortion & Pixelization/Blurring & 2016 \cite{hill2016effectiveness} \\ 
\hline
\multirow{4}{*}{Replacement} & 
Generalization & 2012 \cite{anandan2012t}, 2014 \cite{vasudevan2014automatic} \\ 
\cline{2-3} & Pseudonymization & 2018 \cite{kleinberg2017netanos} \\ 
\cline{2-3} & Substitution & 2016 \cite{mamede2016automated} \\
\cline{2-3} & LLM-guided substitution & 2023 \cite{albanese2023text} \\
\hline
Removal &  Blocking/Masking &  2013 \cite{sanchez2013minimizing}, 2016 \cite{maeda2016fast}  \\
\hline
\end{tabular}}
\end{table} 

\textbf{Pixelization and Blurring.} Pixelization and blurring are image filtering methods (see Section \ref{sec:image_transformations}) that can be applied when a text is represented as an image. Indeed, text documents can be captured as an image, and then computer vision and optical character recognition are used to identify the document's text \cite{stahl2018deeppdf}. In \cite{hill2016effectiveness}, Hill et al. (2016) studied the effectiveness of pixelization and blurring for document redaction. They developed an HMM-based model and tested it against multiple common typefaces, font sizes, colors, pixel grid sizes, and blurring strengths. The tests were conducted with real-life examples ranging from bank cheques, email addresses, and natural language. Compared to brute-force strategies based on template matching and exhaustive search, HMMs provide a much higher performance boost as a probabilistic data-driven model. The authors recovered both short and arbitrarily long instances of redacted text regardless of the used font metrics. A few  studies reported that pixelization and blurring are not recommended for text redaction due to the ease of text recovery using deep learning and statistical methods \cite{mcpherson2016defeating,hill2016effectiveness}. 

\textbf{Generalization.} Generalization replaces sensitive entities with a label or placeholder text. The label would indicate the class of the entity, which is obtained by querying external knowledge bases (KB) \cite{mamede2016automated}. For example, if the sensitive text is `Paris', it will be replaced by its class label `[CITY]'. Therefore, generalization maintains both intelligibility and privacy; however, it is limited by the availability and precision of the KB. Other sanitization methods should be used if the entity's match is not found in the KB. Generalization based on $k$-anonymity guarantees that the obfuscation is indistinguishable from at least $k$-1 other records in the same set \cite{raskind2019primer}. 
    
\textbf{Pseudonymization.} Pseudonymization is a reversible anonymization technique where personal identifiers are given a unique temporary ID. For example, if the named entity is `Paris', it is replaced with its indexed class value `City\_1' or `Location\_1' \cite{kleinberg2017netanos}. The ID mapping is known only to the owners of the document. If an attacker obtains access to the pseudonymization mapping, the security of the obfuscation is  compromised. The cited technique allows the de-identification of individuals while maintaining the readability of the text. Unlike the generalization method, it keeps the co-referential integrity of the same entity appearances throughout the text \cite{mamede2016automated}. 

\textbf{Substitution.} Similar to generalization, substitution requires access to knowledge bases. An entity is replaced by another random entity of the same class. This substitution provides higher utility than other methods (e.g., masking) since its output appears as natural text. However, it can cause semantic drifts because the replacing entities are selected at random. The replacing entities may not preserve the semantics and grammatical context, e.g., gender and case, of the text. In practice, LLM guidance can improve substitution quality by proposing semantically compatible replacements that retain grammatical agreement and task-relevant context, reducing drift compared with random swaps. A recent zero-shot framework acts this idea by coupling probabilistic redaction with embedding-based substitution, maintaining utility on downstream tasks while increasing privacy \cite{albanese2023text}. Other data sanitization methods that we can consider as forms of substitution are \textit{swapping} and semantic \textit{noise addition}. The swapping process exchanges private entities (relatively similar) based on a distance metric within the same document or between documents in the same set \cite{vasudevan2014review}. In contrast to swapping, noise injection is the addition of unrelated terms or substituting entities that are not present in the document set. 

\textbf{Blocking/Masking.} Blocking is the most common technique used in document redaction. It entails using black rectangles to cover words, paragraphs, or even pages containing sensitive entities. As per masking, it replaces the original text with a predefined set of characters, for example, `***,' `XXXX' or `[REDACTED]'. These characters may or may not match the length of the hidden text. In some works, blocking and masking are referred to as \textit{suppression} \cite{mamede2016automated,maeda2016fast}. Both masking and blocking achieve anonymization by completely removing the sensitive terms; however, they reduce the document's utility as the content around the removed entities may be insufficient for a clear understanding of the text's context. 

\subsection{Attacks and Countermeasures}
This section reviews the potential limitations of sanitized text that are exploited in the attacks described below. 

\textbf{Naache and Whelan Attack.} The first documented attack on document redaction by blocking was by David Naache and Claire Whelan in 2004 \cite{lopresti2004quantifying}. Though they did not publish an academic paper to detail their attack, it still gained widespread attention at the time. They were able to recover instances of a single redacted word from previously classified U.S. intelligence reports. The attack consists of four main steps: (a) image
processing; (b) font metrics identification; (c) dictionary-based word guessing; and, (d) natural language processing. Image processing, namely Optical Character Recognition (OCR), is used to obtain a digitized copy of the redacted document. Font metrics (e.g., typeface, size) are identified with high probability through optical font recognition, nowadays a well-developed process. Font recognition can be performed using pattern matching or ML-based methods. Examples of font metrics identification using ML include the use of State Vector Machine (SVM) classifiers \cite{bharath2017font}, and Convolutional Neural Networks (CNN) \cite{zia2019novel}. The third step of the attack is to measure the obfuscated word's width and produce a potential list of dictionary words (having the same font metrics) that would fit the same width when redacted. Lastly, the options are narrowed down using NLP to determine predictions of the redacted word(s). The success of the attack is measured based on the number and randomness of the final potential predictions. This attack was not considered a large-scale threat because it is not applicable when large regions are redacted and requires manual effort to narrow down the potential list of words.  

Lopresti et al. (2005) \cite{lopresti2005information} compiled a list of countermeasures against this attack type. The idea is to ensure that the predictions are as numerous and random as possible and to hide the actual word width \cite{deshpande2015mask}. The countermeasures can be summarized as: (a) remove several unimportant words surrounding the sensitive text (non-minimal redaction); (b) avoid using mono-spaced fonts since they leak information via word widths---for instance, the width of a character can be revealed through the width of the gap between adjacent texts; (c) use ``secure-for-redaction" fonts in which the character width is varied randomly; (d) follow known governmental document redaction guidelines, e.g., by NIST and NSA \footnote{https://fas.org/sgp/othergov/dod/nsa-redact.pdf}; and, (e) re-typeset text to hide the width of the redacted regions.  

\textbf{Public Dataset Crawling.} In this type of attack, an attacker crawls the internet or public knowledge sets (e.g., archives of original and redacted document pairs) to infer private information in the redacted documents, e.g., hidden attributes or author of the document \cite{hakkani2006document}. The authors in \cite{nettleton2015information} used a similar information retrieval approach to measure the information loss and risk of information disclosure. 

\textbf{Context-based entity inference.}
Even if tokens are fully masked, the surrounding sentence often leaks the hidden entity type. RedactBuster, the work proposed by Beltrame et al. \cite{beltrame2024redactbuster} trains transformer encoders to classify the type of the redacted span from context, achieving high accuracy on the TAB benchmark \cite{pilan2022text}; the authors also propose a practical “character evasion” countermeasure using homoglyph swaps to degrade model performance.

\textbf{Measuring residual risk via re-identification attacks.}
Beyond span-recall against masks, TRIA models disclosure as a text re-identification attack that links anonymized documents to individuals using neural language models; the associated TRIR score (attack accuracy) quantifies empirical risk under configurable attacker resources (background breadth/depth, compute) \cite{manzanares2024evaluating}. This reveals discrepancies between recall and true re-identification resistance and highlights limitations of purely NER-based anonymization.

\noindent\textbf{LLM-assisted deanonymization.}
Foundation models can act as powerful “motivated intruders”: in a zero-shot setting, an LLM substantially outperforms human participants at recovering the subject of anonymized bios, and the study argues for LLM-aware defenses (e.g., context-aware substitutions and misleading cues) and LLM-oriented metrics \cite{van2025redacted}.

\textbf{Parrot Attacks.} Text classification systems sometimes overlook a small percentage of PII known as ``residual PII." For human observers, the sanitized PII and leaked PII are generally indistinguishable. However, machine-assisted attacks known as parrot attacks can expose leaked PII \cite{carrell2019machine}. The attacker is then able to infer patterns of residual PII. This attack applies to small datasets, since  it requires the attacker to annotate private attributes, which is a costly and tedious task for larger datasets.

\textbf{Authorship Attribution Attacks.} 
These attacks aim to identify the author of a document based on an analysis of the writing style. Authorship sanitization methods were proposed to counter attribution attacks. Some of the cited obfuscation methods  are style transfer-based, and some are based on identifying the words that most likely reveal an author’s identity and obfuscating them. Researchers studied the style transfer of an input text to a generic or a simplified style or a specific target style using GANs. However, these obfuscation methods themselves leave behind signatures that are detectable using neural language models. The obfuscations degrade the text's smoothness in a way that is distinguishable from natural text. In \cite{mahmood2020girl}, Mahmood et al. (2020) proposed stealthy authorship obfuscation solutions that modify a text’s style while preserving its semantics and escaping current obfuscation detection methods.

\begin{table*}[t]
\centering
\caption{Examples and Comparison of Text Sanitization Methods}
\label{tab:text_sanitization_comparison}
\resizebox{\linewidth}{!}{%
\begin{tabular}{|l|l|l|} 
\hline
\textbf{Method} & \textbf{Example} & \textbf{Pros \& Cons} \\ 
\hline
Original & Joe Biden issued a statement yesterday. &  - \\ 
\hline
\multirow{2}{*}{Blocking} & \censor{Joe} \censor{Biden} issued a statement yesterday. & \begin{tabular}[c]{@{}l@{}}\begin{tabular}{@{\labelitemi\hspace{\dimexpr\labelsep+0.5\tabcolsep}}l}Achieved anonymization\\Width can leak information\\Utility loss\end{tabular}\end{tabular} \\ 
\cline{2-3}
 & \censor{Joe Biden} issued a statement yesterday. & \begin{tabular}[c]{@{}l@{}}\begin{tabular}{@{\labelitemi\hspace{\dimexpr\labelsep+0.5\tabcolsep}}l}Achieved anonymization\\Implies a single redacted word\\Robust against~Naache \& Whelan attack~\\Utility loss\end{tabular}\end{tabular} \\ 
\hline
\multirow{2}{*}{Masking} & [REDACTED] issued a statement yesterday. & \begin{tabular}[c]{@{}l@{}}\begin{tabular}{@{\labelitemi\hspace{\dimexpr\labelsep+0.5\tabcolsep}}l}Achieved anonymization\\Utility loss\end{tabular}\end{tabular} \\ 
\cline{2-3}
 & *** ***** issued a statement yesterday. &  \begin{tabular}[c]{@{}l@{}}\begin{tabular}{@{\labelitemi\hspace{\dimexpr\labelsep+0.5\tabcolsep}}l}Achieved anonymization\\Character length can leak information\\Utility loss\end{tabular}\end{tabular} \\ 
\hline
Generalization & [Person] issued a statement yesterday. &  \begin{tabular}[c]{@{}l@{}}\begin{tabular}{@{\labelitemi\hspace{\dimexpr\labelsep+0.5\tabcolsep}}l}Achieved anonymization\\Preserved utility~\end{tabular}\end{tabular} \\ 
\hline
Pseudonymization & Person12345 issued a statement yesterday. & \begin{tabular}[c]{@{}l@{}}\begin{tabular}{@{\labelitemi\hspace{\dimexpr\labelsep+0.5\tabcolsep}}l}Achieved de-identification, reversible \\Preserved utility~\end{tabular}\end{tabular} \\ 
\hline
Substitution & John Doe issued a statement yesterday. & \begin{tabular}[c]{@{}l@{}}\begin{tabular}{@{\labelitemi\hspace{\dimexpr\labelsep+0.5\tabcolsep}}l}Achieved anonymization\\Preserved a degree of utility~\end{tabular}\end{tabular} \\ 
\hline
Pixelization & \raisebox{-.5\height}{\includegraphics[scale=0.25]{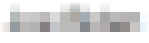}} issued a statement yesterday. & \begin{tabular}{@{\labelitemi\hspace{\dimexpr\labelsep+0.5\tabcolsep}}l}Ineffective against ML-assisted attacks~\\Utility loss\end{tabular} \\ 
\hline
Blurring & \raisebox{-.5\height}{\includegraphics[scale=0.25]{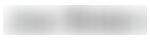}} issued a statement yesterday. & \begin{tabular}{@{\labelitemi\hspace{\dimexpr\labelsep+0.5\tabcolsep}}l}Ineffective against ML-assisted attacks~\\Utility loss\end{tabular} \\
\hline
\end{tabular}}
\end{table*}

\subsection{Discussion}
\textbf{Applications.} While the authenticity of redacted documents does not fit our topic of discussion per se, we believe it is worth mentioning since recent and interesting research efforts are still emerging in the area. 
Digital signatures were created to guarantee the integrity and authenticity of data, including text documents, and to ensure its non-repudiation. To ascertain the source and the integrity of the redacted document, the original document signer must re-sign it, which is not always feasible. For this reason, cryptographic constructions were introduced for redactable signatures in Johnson et al. (2002) \cite{johnson2002homomorphic} and sanitizable signatures in Ateniese et al. (2005) \cite{ateniese2005sanitizable}. Redactable signatures allowed anyone to generate a valid signature of the redacted document. Sanitizable signatures allowed authorized censors to modify parts of a signed document in a controlled fashion without interacting with the original signer. More recently, research efforts have been directed to the development of \textit{Redactable Signature Schemes (RSS)}, also known as authenticated document redaction \cite{ma2017authenticated,ma2020authenticated}.

In some multilevel security applications, documents are sanitized such that one's access rights determine the amount of redaction that the document should be subjected to before disclosure. Subjects with differing security clearances will receive different versions of the same redacted document. In \cite{gehres2010toward}, a model is proposed where information is redacted at multiple classification levels such that it is securely stored and made accessible only to the authorized users. This is achieved using intermediary authorization and protection layers between the storage location and the document viewing service. The documents are redacted following a bottom-up strategy; that is, the sensitivity of the redaction is increased incrementally.

\textbf{Challenges and Open Research.} The main challenge of text redaction is establishing an acceptable privacy-utility balance, and this includes maintaining text semantics \cite{hassan2021utility} and avoiding malicious over-sanitization \cite{ma2020authenticated}. The AI sub-fields of optical character recognition and natural language processing are relatively mature and well-studied domains, including other relevant NLP tasks such as intent detection, relationship extraction, and co-reference resolution. Even though attacks against text sanitization can be automated, they are still limited in scope with the correct identification and generalization of sensitive entities. There is a clear need for further research aimed at deceiving automated text identification using noise injection or text perturbation \cite{xu2020differentially}. For instance, we could push automated text classification algorithms to fail to recognize specific identifiers---for instance, transforming the text to show bad grammar, incorrect punctuation, or noisy text.
Future directions also include proposing a generic and standard framework for automatically evaluating and classifying redacted documents based on their privacy-utility levels.

\textbf{Key Properties.} The key properties to consider when selecting an optimal sanitization method for protection against information leakage are: (a) \textit{intelligibility}: the amount of alterations done should be minimal as to not render the document's context unintelligible or unreadable; (b) \textit{Confidentiality and Privacy}: the text should not leak sensitive information; (c) \textit{Pseudonymity}: most use-cases require that the obfuscation be reversible; and, (d) \textit{Multi-level security}: the text is obfuscated based on the user's security clearance. The first two properties (intelligibility and security) are addressed in most text sanitization studies. However, pseudonymity and multi-level security are addressed based on the user requirement. Pseudonymity is applicable to the protection of health information in the medical industry. 

\textbf{Review of Methods.} Currently, the research in text de-identification is driven by the protection of health information \cite{ribaric2016identification}, defined by the 18 HIPAA rules. The de-identification process should, therefore, be reversible and secure for medical records. As shown in Table \ref{tab:text_sanitization_comparison}, we compare the different text sanitization techniques previously discussed. Ad-hoc obfuscations such as blocking and masking provide high protection but compromise the readability of the text and are vulnerable, except for a few variations, to the Naache-Whelan attack. In this context, pixelization and blurring work the same way as blocking, except that they rather provide an aesthetic appeal to the redacted document; however they have been long proven to be ineffective against several attack types \cite{hill2016effectiveness}. Substitution achieves a good privacy-utility balance when done correctly, i.e., when the semantics of the substituted text is preserved. We recommend generalization (irreversible) and pseudonymization (reversible) methods as they achieve the highest utility-privacy balance. Generalization can adopt the differential privacy paradigm to achieve provable security. In \cite{lee2020identification}, other data protection techniques used to protect medical documents are listed: noise addition (random and arbitrary substitution of sensitive entities, or additive semantic noise), swapping, data aggregation, and data reduction, where the latter two apply to numerical values.

\section{Video Sanitization}
\label{video_obfuscation}
Video surveillance systems are today's most ubiquitous source for video capturing people and their activities \cite{sah2017video,ribaric2015overview}. As expected, the public use of video surveillance (including CCTV and drone surveillance) introduces the need for privacy protection, specifically for law-abiding citizens, while providing the authorities the ability to retrieve the original video for specific purposes, e.g., crime investigation. When releasing the video, portions of the footage exposing sensitive information are redacted. Such information include but is not limited to faces, license plates, tattoos, clothing, computer screens, and house numbers \cite{sah2017video}. Videos are a series of time-related image frames and, optionally, include an audio overlay. Hence, the expectation is that image and audio de-identification techniques can be applied separately to videos. However, the process is inherently more complex,  since neighboring frames in a video sequence are closely related. 

Object detection and tracking accuracy play a significant role in identifying and the sanitization of sensitive regions correctly. Any incorrect detection of sensitive areas in a single frame can destroy the protection for the entire video sequence \cite{sah2017video,korshunov2012subjective}. Object detection and tracking in video sequences also face further challenges, such as: variation in the object's localization and size throughout the scenes; bad lighting conditions; variable illumination; cluttered scenes; and, partial occlusions due to obstructing components (e.g., sunglasses, accessories, and facial hair on face regions) \cite{ribaric2016identification,ribaric2015overview}. Additionally, some video surveillance systems require real-time detection and de-identification of moving objects. For that reason, the selection of the obfuscation method is critical since real-time processing should be fast and effective. In general, for most video surveillance systems, the requirements to be satisfied are: (a) real-time processing; (b) efficiency and low cost; (c) utility-preserving obfuscation (preserving gait, physical actions, and facial gestures); (d) reversible obfuscation; (e) adaptive obfuscation (adjustable levels of privacy filters parameters, or selective obfuscation based on preset rules); and, (f) format compliance (decodable by commercial video players).

\subsection{Techniques}
Various approaches have been proposed for privacy protection in videos. The most common solution is applying visual transformations on sensitive image regions. Other approaches, which are out of scope but are defined briefly hereafter, include physical intervention, software intervention, and secure multiparty computation \cite{padilla2015visual}. Physical intervention creates capture-resistant regions using specialized hardware (e.g., adhesive camera blockers, pulsating LEDs directed at camera lenses). In contrast, software intervention is achieved through modifying the firmware of a capturing device to prevent capturing specific scenes \cite{padilla2015visual}. Secure multiparty computation allows image processing tasks (e.g., face detection, object tracking, image segmentation) to be computed anonymously. The traditional approach for visually transforming sensitive regions in videos is manually identifying, annotating, and masking ROIs in each video frame. However, considering a typical surveillance video of 30 frames per second, manual redaction is unpractical being  extremely time- and labor-intensive \cite{ribaric2015overview}. Automatic obfuscation can target specific objects or the entire image frame (aka, general/global obfuscation). The latter assures higher protection yet it removes the contextual information necessary for monitoring people and their behavior. Instead of masking objects or the entire scene, Dai et al. (2015) \cite{dai2015towards}explore whether down‑sampling the video itself can deliver privacy without losing utility. By reducing spatial resolution to as low as one pixel per frame and adjusting frame rates between 2–30 Hz, they measure how gesture‑recognition performance degrades. Their experiments show that even five single‑pixel cameras operating at 30 Hz achieve a 75.70\% classification rate across nine gestures—only 13.9\% lower than the same setup at 10×10 resolution—indicating that ultra‑low‑resolution video may offer a viable privacy–utility compromise. Even though our main focus in this subsection is object obfuscation, we still give a brief overview of the state-of-the-art object detection and object tracking, focusing on face objects, since they are closely related to the obfuscation process. Next, we review image obfuscation methods in terms of how suitable they are for real-time surveillance.

\subsubsection{Object Detection}
This step is needed for the automatic localization of relevant objects in a frame. The detection algorithm identifies the object's bounding box, and segmentation obtains the precise pixels outlining the object within the bounding box. There are four main approaches to object detection: sliding search window, region proposals, DL-based, and pixels-based \cite{sah2017video}. The sliding window approach sweeps a detection template across an image, computes a function with respect to the template, and classifies the image region. It is a computationally demanding method and, therefore, it is not typically used for video redaction. Region proposal methods use classifiers (e.g., SVM, k-nearest neighbor) that obtain a confidence score for each candidate image region using its extracted low-level features (e.g., SIFT, histogram of oriented gradient (HOG), Harris Corners, etc.). The frequently used deep learning models for object detection include Regional CNN (RCNN), Fast RCNN, Mask RCNN, Multibox, and YOLO (You Only Look Once) \cite{sah2017video,asghar2019visual}. YOLO uses regression models to predict the coordinates of the bounding box and its class confidence by applying a single-pass CNN \cite{sah2017video}. Pixel-based methods achieve semantic segmentation and assign a class label to each pixel in a frame. Some approaches use fully convolutional neural networks (FCN) and Conditional Random Field (CRF) to achieve per-pixel classification \cite{sah2017video,qin2017automatic}. 

In terms of face objects, face detection algorithms \cite{zafeiriou2015survey} are based on feature-level or image-level methods. The feature-based approach uses low-level analysis (edges, skin color, motion, gray-scale), active shape models (snakes, deformable templates, PDM), or facial feature searching and extraction (constellations, Viola-Jones, Gabor Feature). The image-based approach is a learning process where a trained model classifies images as face or non-face, e.g., using linear subspace methods (Eigenfaces), statistical methods (GMM, PCA, SVM), and neural networks \cite{ribaric2015overview,chen2009protecting}. A recent survey providing a comprehensive overview of face detection techniques is referred in \cite{kumar2019face}. 

The majority of automated face detection algorithms are limited by head poses, occlusions, image resolution, and so on. To mitigate the inaccuracy of some face detection software in suboptimal conditions, Chen et al. (2018) \cite{chen2018response} proposed ReSPonSe---a system that uses human trace tracking instead of face detection to identify the pixels representing facial information in a more robust way. Human trace tracking gives higher assurance that the face and the body regions are still identifiable by color-based skin detection when faces are not detected. The system architecture has two main stages, the encapsulation stage, and the decapsulation stage. In encapsulation, the video sequence is processed in real-time to remove private information using human trace tracking and image encryption. Human trace tracking consists of: face block extraction to accelerate face detection; in-block face detection to detect faces in the extracted blocks; and, detection rectification. This step also utilizes the distribution of skin regions in image tiles surrounding selected key blocks. The extracted detections are then encrypted. In decapsulation, decryption is performed on the redacted video to selectively recover the original faces based on the viewer's certificate. Human trace tracking does not work well when non-face objects (e.g., walls or cars) possess a color similar to skin colors; it results in the unnecessary redaction of the background, thus reducing the utility of the video. Therefore, Chen et al. (2019) \cite{chen2019fis} proposed the Facial Information Segmentation (FIS) algorithm, which combines color information, Harris Corner, and face detection algorithms to identify the pixels associated with the face region. The authors rely on the observation that the density of Harris Corners is lower on non-face objects than other face objects with skin color; non-face objects tend to be smoother and have fewer features.

\subsubsection{Object Tracking}
The purpose of object tracking is to estimate and predict the size and location of a detected (and tagged) object over time in subsequent video frames. Tracking algorithms are susceptible to noise factors due to variations in illumination, object scale, occlusions, pose or camera relative perspective, and motion blur. There are four main approaches to object tracking: motion-based; appearance-based; tracking by detection; and, deep learning methods \cite{sah2017video}. Motion-based modeling computes an estimation of the stationary scene background and subtracts the estimated background from each frame having a moving object. The differences calculated from the background subtraction between consecutive scenes indicate the presence of objects in motion. Appearance-based matching of hand-crafted or machine-learned image features includes color histogram matching, shape matching, or texture matching. Again, It is limited by dramatic changes in camera perspective, heavy shadows, etc.. Tracking by detection extends the use of object detection algorithms to enable tracking frame by frame  (e.g., using adaptive search windows). However, this method has a high computational overhead and may not be suitable for real-time processing. Lastly, more advanced techniques such as recurrent neural networks exploit the history of an object's location to learn temporal dependencies between frames. 

In the case of face tracking, to preserve naturalness for de-identified videos, deep learning methods for face landmark localization (locating facial key points around the eyes, corners of the mouth, and nose) and motion and pose estimation are considered \cite{ribaric2016identification,asghar2019visual}. Additionally, the algorithm should discriminate and predict the location of all moving faces in the video sequence. Although using temporal information for tracking increases robustness, it is ill-suited for real-time tracking. In some solutions, markers or RFID tags worn by the individuals were used to enhance tracking accuracy \cite{sah2017video,cheung2009protecting}. 

\subsubsection{Obfuscation Methods}
Video surveillance systems follow the same protection techniques used in still images. Masking, pixelization, and blurring are standard methods for video redaction since they do not require an excessive processing power and can be done in real-time. Korshunov et al. (2012) \cite{korshunov2012subjective} performed a subjective evaluation to assess the detection accuracy of abnormal behavior in video surveillance scenes when the previous methods were applied to body silhouettes and faces. The results showed that the recognition of the obfuscated people was still possible to a certain degree, specifically the gender attribute from the shape of the body image in the video, even when the masking filter is applied. 
Pixelization was found to offer higher utility in terms of action recognition than blurring in this context, and masking overall provides the highest privacy protection because it can reduce recognition accuracy to near-zero \cite{korshunov2012subjective,korshunov2014towards}. In \cite{ivasic2014person}, Ivasic et al. (2014) studied the impact of Gaussian Blur filters on human activity recognition and de-identification. The results show that specific activities (e.g., jump in place and wave one hand) reveal more discriminant information that can be used for subject identification than others (e.g., walking and running). The authors suggested increasing the variance of the Gaussian blur for such activities.

There are other methods that are also not as demanding for processing power, while being more utility-preserving. For instance, cartooning \cite{erdelyi2014adaptive}, edge detection \cite{chen2009protecting}, style transfer \cite{brkic2016identifying,brkic2016towards}, false colors \cite{cciftcci2015using}, warping \cite{korshunov2013using}, and morphing \cite{korshunov2013usingfacemorphing}. In the following, we introduce some examples from the literature. In \cite{chen2009protecting}, Chen et al. (2008) applied edge motion history image (EMHI) to obscure body images while still capturing the structure of the bodies and their motion using edges. Korshunov et al. (2014) \cite{korshunov2014towards} evaluated morphing and warping against blurring, pixelization, and masking. 
Warping is the least suitable filter because it affects recognition only at very high distortion levels, whereas it is also more complex than the other simple filters. The authors concluded that the morphing filter seems to be the best choice among the evaluated privacy filters for video surveillance since it is reversible, and recognition accuracy can be varied linearly by varying the intensity strength. False-colorization is a general obfuscation method applicable to video surveillance since it is reversible. It also hides identity information without impacting surveillance-related information, such as the count of people in an area. 

In some video applications—such as sign language—facial dynamics convey linguistic content; naive obfuscation removes meaning together with identity. Recent approaches therefore anonymize identity while explicitly preserving articulation and expression trajectories across frames. This requires temporally consistent synthesis and conditioning on structural cues so that privacy is achieved without degrading linguistic interpretability. A key work in the literature conducted by Xia et al. (2024) introduces DiffSLVA, a novel methodology that uses pre-trained large-scale diffusion models for text-guided sign language video anonymization \cite{xia2024diffusion}. In their project they use text-guided latent diffusion with ControlNet cues, cross-frame attention, and an expression-preservation module to anonymize the signer while retaining the facial dynamics essential for understanding, avoiding pose-estimation dependencies and reducing temporal artifacts.

Lightweight partial encryption and scrambling (i.e., ROI-based) \cite{dufaux2008scrambling,ruchaud2016privacy} are other applicable methods to real-time processing, even though they are less secure than naïve encryption \cite{padilla2015visual}. The videos are obfuscated in a way that they remain format compliant and viewable in the distorted form. Other advanced methods such as object replacement (e.g., avatars \cite{blavzevic2015towards}) or face replacement can be used; however, both require precise position and pose tracking \cite{cheung2009protecting}. Blavezvic et al. (2015) in \cite{blavzevic2015towards} proposed a reversible real-time de-identification pipeline that obfuscates all biometric identifiers in video footage. The system relies on computer vision algorithms, namely human body image detection, segmentation, and tracking of the detections. The detected human body in the original image is replaced with a 3D avatar. The 3D model is rendered precisely on top of the original body with the help of human pose estimation (joint modeling) algorithms. The detection and pose estimation strongly depends on the viewing angle, body shape, clothing, etc.. The de-identified region is encoded and embedded in the carrier image using steganography to enable reversing the concealment. The authors identified some limitations in their 3D modeling; specific body movements are not supported, such as jumping or rotating. In addition, sudden movements are not always accurately tracked. The likelihood of erroneous output and incorrect localization of some joints (and therefore incorrect positioning of the 3D avatar) due to the tracking system is relatively high. To countermeasure the tracking inaccuracies, the authors pixelized and blurred the body sections visible behind the avatar in the original frame. Further works included developing more natural-looking avatars.

\subsection{Attacks and Countermeasures}
Since the same protection methods used for images are used for videos, the same attacks described in Section \ref{image_obfuscation}-B apply. In the following, we explain the major differences in the recognition-based and reconstruction-based attack styles for videos.

\textbf{Recognition-based Attacks.} Recognition-based attacks (or re-identification attacks) are also a significant concern about de-identified videos. Full-body images are more commonly seen in surveillance videos, and even when face masking is used, a person's identity can be exposed using \textit{pairwise constraints} attack \cite{sah2017video,chen2009protecting}, also known as, faceless recognition \cite{oh2016faceless}. A pair-wise constraint is when two masked faces can be determined to belong to the same person by using the variations of clothing, body shape, or other cues, across images of the same person, even though the faces are still hidden. Other soft-biometric features in full-body photos, such as age, gender, race, ethnicity, birthmarks, and tattoos, provide further hints for re-identification \cite{oh2016faceless}; however, such features are more complex to detect. In the same direction, obstructing the entire body provides better protection than face-masking alone; however, this way still results in higher unfavorable utility loss. For instance, facial expressions and body actions become imperceivable in the surveillance footage. The body silhouette and the gait can leak information, therefore, Agrawal et al. (2011) \cite{agrawal2011person} proposed slightly dilating or expanding the silhouette area to be obfuscated to hide the gait information and further increase the privacy.  

\textbf{Reconstruction-based Attacks.} The same attacks applied to obfuscated images apply to redacted videos, except for attacks that utilize temporal dependencies between adjacent frames in the video sequence. Cavedon et al. (2011) \cite{cavedon2011getting} developed a technique that completely recovers the identity of a pixelized face in video streams under certain conditions. When pixelization is applied to video sequences depicting the same subject, there is a high probability that the pixelization squares will change position with respect to the image background, hence averaging different pixels at different times. Unlike super-resolution methods, the technique aims to recover the image with its original quality. The authors used a \textit{Maximum Posteriori} to find an image such that, if shifted and pixelized in multiple frames, it will produce a video sequence as close as possible to the input video. Their method assumes that the same image can be tracked in different frames at different positions. A similar attack described in \cite{padilla2015visual} uses image inpainting for interpolating masked parts in a given frame using information from adjacent video frames. 

The primary solution to this type of attack is to utilize more precise detection mechanisms, perhaps by combining object detection, segmentation, and tracking algorithms. Chen et al. (2008) \cite{chen2009protecting} proposed a bi-directional tracking algorithm that combines background subtraction, face detection, and face tracking. Manual or automatic screening of the redacted video frames can reduce missed face detections. Once a face is detected, it is simultaneously tracked in both previous and subsequent frames to locate any missed (usually non-frontal) faces. Background subtraction is used to segment the foreground, containing people or other moving objects, from the background. Furthermore, for face images, color-based detection can be combined with other forms of detection to achieve better results \cite{chen2018response}. Another countermeasure includes applying general obfuscation (i.e., full-frame obfuscation), but it is not favorable since it degrades the utility of the video. 

\subsection{Discussion}
\textbf{Applications.} Video de-identification is required in scenarios such as public video publishing to protect, among others, the privacy of the individuals who appear involuntarily in the videos. Furthermore, in public surveillance videos, bypassers are captured mainly to record any behavioral anomalies; their actions are relevant but not their identities in real-time. The need for low-cost surveillance has driven the concept of privacy by design into smart camera devices, so specific objects are never stored or transmitted, and complex processing (such as natural face replacement) is moving towards edge computing \cite{sah2017video}. Also, drones are now used for surveillance since they can stealthily access a broader range of locations and closely track the object of interest \cite{ribaric2016identification,bonetto2015privacy}.

\textbf{Challenges and Open Research.} Face de-identification in videos is more challenging than in still images. The challenges are related to the complexity of facial detection and localization in video streams. Although face detection in varying poses and profiles, illumination conditions, and occlusions have been studied, the problem remains a known challenge. Similar to still images, for an identity to be protected, multimodal de-identification must be considered. Sensitive personal identifiers are of many diverse forms (e.g., skin, logos, location information such as house numbers, storefront signs, street signs, and graffiti); therefore, fully automating the redaction process still requires further research. Skin color-based segmentation still faces limitations since skin occurs in many tones. Optical character recognition in outdoor scenes is also limited by suboptimal conditions, such as poor lighting and geometric perspective. 

\textbf{Key Properties.} There are a few more considerations that should be taken into consideration when selecting an obfuscation method: real-time processing, low computational overhead, reversibility, and utility-privacy trade-off in hiding identity information while preserving behavior-related information (gait, actions, facial gestures). Surveillance and monitoring systems that require real-time de-identification should not use heavy processing methods such as inpainting, encryption, and face replacement. Ad-hoc and simple transformations such as blocking, masking, pixelization, and blurring techniques are more commonly used in real-time scenarios, even though they provide lower protection. Nonetheless, they are also not reversible and severely reduce the utility of the video such that user actions and gait are not readily perceivable \cite{ivasic2014person}. Some works recommend using other simple but utility-preserving filters such as cartooning, edge detection, style transfer, false colors, warping and morphing, where only the last three can be reversed if the transformation parameters are known. When secure processing is required, the more advanced transformations discussed above should be used, with the caveat that only encryption and scrambling are reversible. 

\textbf{Review of Methods.} In Table \ref{tab:video_obfuscation_examples}, we list examples of real-time sanitization methods applicable to video surveillance scenarios. We compare the reversibility and the computational complexity of the methods as well---this latter feature expressed in a qualitative appreciation of the required time. As shown, video obfuscation follows the same protection techniques used in still images. The exception is that the obfuscation systems would be slightly modified to ensure the efficiency and speed of object detection, tracking, and object denaturing.

\begin{table}[ht]
\centering
\captionof{table}{Methods for Video Obfuscation in the Literature}
\label{tab:video_obfuscation_examples}
\resizebox{\columnwidth}{!}{%
\begin{tabular}{|l|c|c|c|} 
\hline
\textbf{Method} & \multicolumn{1}{l|}{\textbf{References}} & \multicolumn{1}{l|}{\textbf{Processing Time}} & \multicolumn{1}{l|}{\textbf{Reversibility}} \\
\hline
\begin{tabular}[c]{@{}l@{}}Pixelization\\Blurring (Body)\\Masking\end{tabular} &  \cite{agrawal2011person,yu2008privacy} & Fast & No \\
\hline
Cartooning & \cite{erdelyi2014adaptive} & Fast & No \\ 
\hline
Edge & \cite{chen2009protecting} & Fast & No \\ 
\hline
Style transfer & \cite{brkic2016identifying,brkic2016towards} & Slow & No  \\ 
\hline
False colors & \cite{cciftcci2015using} & Fast & Yes \\ 
\hline
Warping & \cite{korshunov2013using} & Fast & No \\ 
\hline
Face morphing & \cite{korshunov2013usingfacemorphing} & Fast &  No \\ 
\hline
Face replacement & \cite{yu2008privacy,gafni2019live} & Slow & No \\ 
\hline
Scrambling & \cite{dufaux2008scrambling,ruchaud2016privacy} & Fast & Yes  \\
\hline
Avatars (Body) & \cite{blavzevic2015towards} & Fast & Yes \\
\hline
Diffusion models & \cite{xia2024diffusion} & Slow & No \\
\hline
\end{tabular}}
\end{table}

\section{Lesson Learned}
\label{sec:lessons}
In this section we summarize the important insights regarding the current state, challenges, and limitations of the field highlighted in this survey. One of the most evident conclusions is that no single obfuscation technique can universally address all privacy concerns. Methods based on distortion, replacement, or removal each exhibit specific advantages and drawbacks, often shaped by the targeted media type, context of deployment, and expected balance between privacy protection and content utility. For instance, while image inpainting and encryption-based techniques offer strong privacy guarantees, they may severely compromise usability. Conversely, aesthetic transformations such as cartooning or false coloring may preserve semantic information but remain susceptible to modern recognition systems.

Another critical lesson is the intrinsic nature of the privacy-utility trade-off, which manifests consistently across all modalities. Striking a balance between concealing sensitive data and retaining content intelligibility remains one of the central tensions in sanitization research. The effectiveness of an obfuscation technique must be measured not only in terms of information suppression but also with respect to how well it preserves the content’s functional or communicative purpose. In practice, many methods implicitly prioritize one of these objectives, often at the expense of the other.

Furthermore, it has become increasingly clear that adversarial capabilities are advancing at a rapid pace. Traditional techniques such as blurring, pixelation, and masking---once considered sufficient---are now defeated by machine learning models, especially those based on deep neural networks and generative architectures. This observation underscores the need to develop privacy-preserving techniques that are resilient not only to human inspection but also to automated recognition and reconstruction attacks.

The survey has also highlighted a noticeable asymmetry in the maturity of sanitization techniques across different media. While visual sanitization, especially in the context of face de-identification, has seen significant progress, other domains such as audio and multimodal data remain comparatively underexplored. This imbalance raises concerns about the applicability of current methods in real-world scenarios where sensitive information is often conveyed through multiple, overlapping modalities.

An additional concern is the lack of standardized evaluation protocols across the field. Many proposed techniques are validated using custom datasets or loosely defined performance metrics, limiting the ability to compare methods rigorously. There is a need for publicly available benchmarks and reproducible evaluation frameworks that include both privacy leakage estimates and utility assessments. Without such standards, it becomes difficult to assess the generalizability or robustness of existing approaches.

Finally, the integration of sanitization systems into real-world applications reveals practical limitations that are often overlooked in academic research. Systems deployed in surveillance, healthcare, or online platforms must address constraints related to latency, interpretability, legal compliance, and user acceptance. Furthermore, the contextual nature of privacy requires that obfuscation techniques be adaptive to content, task, and cultural norms. Despite growing interest in learning-based and generative techniques, few approaches currently offer guarantees on fairness, transparency, or accountability.

\section{Conclusion}
\label{conclusion}

In this paper, we have presented a comprehensive and up-to-date survey of the major approaches to multimedia sanitization. Specifically, we reviewed existing sanitization methods for images, audio, text, and video; discussed their current applications and associated constraints; identified open key challenges; and, highlighted potential directions for future research. We have also shown that obfuscation-based strategies are susceptible to recognition, reconstruction, and inference attacks, and we discussed relevant research trends and countermeasures aimed at addressing these vulnerabilities.

In conclusion, a central quest in this research domain is to achieve an optimal privacy–utility balance for each sanitization method, although, as our survey underscores, no unified framework exists for evaluating the privacy and utility of multimedia sanitization systems. Nevertheless, by systematizing the vast current body of knowledge and highlighting contemporary research trends and available technologies, our work outlines key research directions to guide the development of sanitization techniques that provide the desired levels of security and privacy while preserving the necessary level of utility.

\balance
\bibliographystyle{IEEEtran}
\bibliography{main}

\begin{IEEEbiography}[{\includegraphics[width=1in,height=1.25in,clip,keepaspectratio]{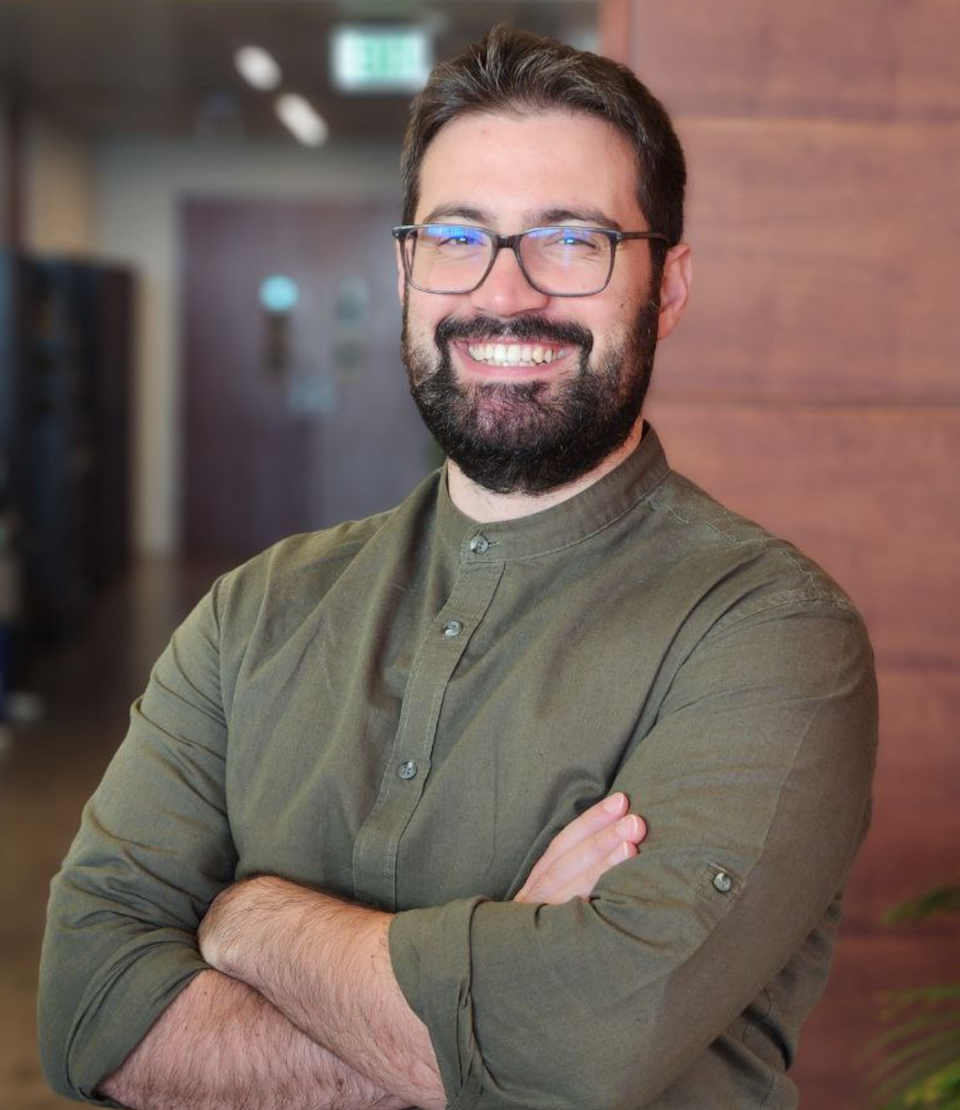}}]{Andrea Ciccotelli}
 is a Ph.D. student in Computer Science at King Abdullah University of Science and Technology (KAUST).
His research focuses on cybersecurity in cyber-physical systems, with particular interests in side-channel, fingerprinting, privacy, and advanced cryptographic protocols. He is also passionate about hardware prototyping, actively developing ESP32- and Raspberry Pi–based platforms.
His approach integrates both theoretical foundations and hands-on experimentation.
\end{IEEEbiography}

\begin{IEEEbiography}[{\includegraphics[width=1in,height=1.25in,clip,keepaspectratio]{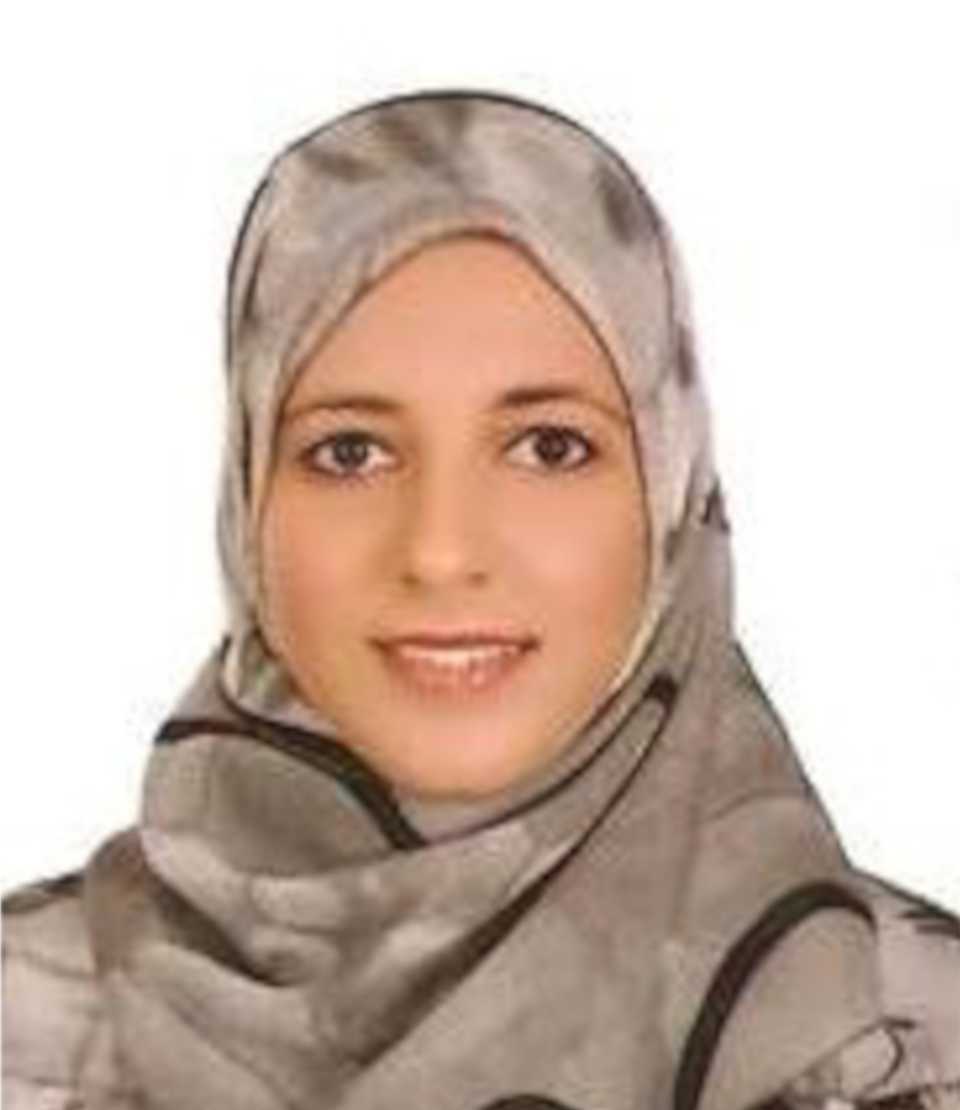}}]{Hanaa Abbas}
holds a B.Sc. in Electrical and Computer Engineering from Texas A\&M University (TAMU-Q) and an M.Sc. in Cybersecurity from Hamad Bin Khalifa University (HBKU), where her research focused on security and privacy challenges in distributed systems. She has professional experience as a Software Engineer working on full-stack development across R\&D projects in domains such as smart grids and defense systems for unmanned vehicles. Currently, she is an AI Engineer specializing in the development and deployment of AI-driven applications, with a focus on large language models (LLMs).
\end{IEEEbiography}

\begin{IEEEbiography}[{\includegraphics[width=1in,height=1.25in,clip,keepaspectratio]{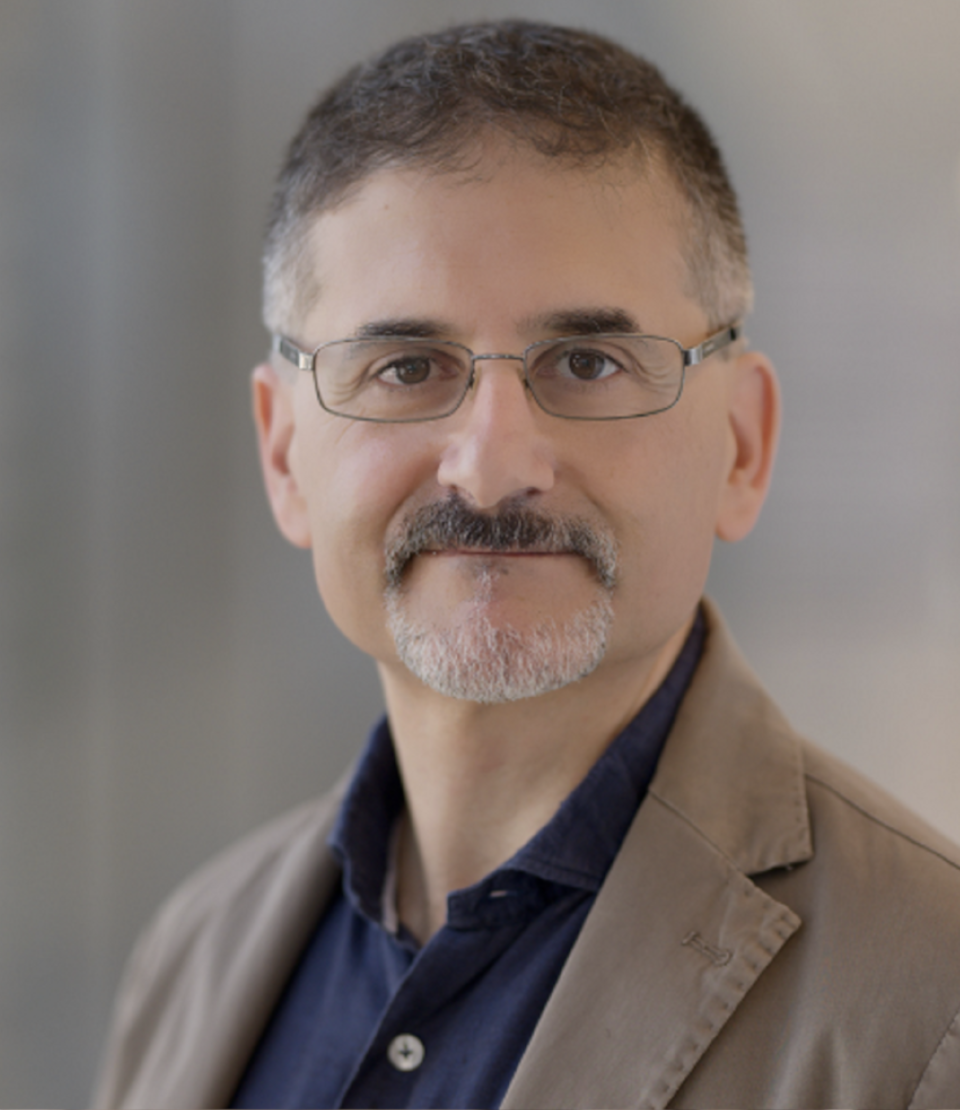}}]{Roberto Di Pietro}
is Full Professor of Computer Science at King Abdullah University of Science and Technology (KAUST). His research focus is on AI-driven  cybersecurity, critical infrastructure protection, and security and privacy of distributed systems---with interests ranging from blockchain and cloud to IoT, UAVs, and applied cryptography. With over thirty years of experience, he has held prominent academic and industrial positions, including Global Head of Security Research at Bell Labs (Alcatel-Lucent/Nokia) and founder of the Cyber-Security Research Innovation Lab. He is an IEEE Fellow, ACM Distinguished Scientist, member of the Academia Europaea and the AAIA, and recipient of the Jean-Claude Laprie Award for pioneering contributions to dependable computing.
\end{IEEEbiography}

\end{document}